\documentclass[11pt, oneside]{Thesis} 

\graphicspath{{Pictures/}} 

\usepackage{hyperref}
\usepackage{setspace}
\usepackage{caption}
\usepackage{textcomp}

\usepackage[usenames,dvipsnames,svgnames,table]{xcolor}
\usepackage[square, numbers, comma, sort&compress]{natbib} 
\usepackage{booktabs}
\hypersetup{colorlinks=true,linkcolor=black,citecolor=black,filecolor=black,urlcolor=black,}

\title{\ttitle} 

\begin{document}

\frontmatter 

\setstretch{1.6} 

\fancyhead{} 
\rhead{\thepage} 
\lhead{} 

\pagestyle{fancy} 

\newcommand{\HRule}{\rule{\linewidth}{0.5mm}} 

\hypersetup{pdftitle={\ttitle}}
\hypersetup{pdfsubject=\subjectname}
\hypersetup{pdfauthor=\authornames}
\hypersetup{pdfkeywords=\keywordnames}


\begin{titlepage}
\begin{center}

%
{\large Graduate Institute of Biomedical Electronics and Bioinformatics}\\[0.3cm]
{\large College of Electrical Engineering and Computer Science}\\[0.3cm]
{\large National Taiwan University}\\[2.0cm]
{\Large Doctoral Dissertation}\\[2.0cm]
{\huge \bfseries \ttitle}\\[2.5cm] 
{\Large Author : Yu-Ting Lin}\\[1.9cm]
{\Large Advisor : Jenho Tsao, Ph.D.}\\[4.0cm]
%
{\Large February 2015}\\[0.01cm]

%
%
 
\vfill
\end{center}

\end{titlepage}

\pagestyle{empty} 

\null\vfill 

\textit{``All composite things are not constant. Work hard to gain your own enlightenment."}

\begin{flushright}
Siddh\={a}rtha Gautama
\end{flushright}

\vfill\null 

\clearpage 

%
%
%
%

\addtotoc{Abstract} 
%
\begin{center}
\setstretch{1.5}
\null\vfill
{\Huge Abstract} \\[1cm]
\end{center}
{
Variations of instantaneous heart rate appears regularly oscillatory in deeper levels of anesthesia and {less regular} in lighter levels of anesthesia. It is impossible to observe this ``rhythmic-to-non-rhythmic" phenomenon from raw electrocardiography waveform in current standard anesthesia monitors. To explore the possible clinical value, I proposed the \emph{adaptive harmonic model}, which  fits the descriptive property in physiology, and provides adequate mathematical conditions for the quantification. Based on the \emph{adaptive harmonic model}, multitaper Synchrosqueezing transform was used to provide time-varying power spectrum, which facilitates to compute the quantitative index: ``Non-rhythmic-to-Rhythmic Ratio" index (NRR index). I then used a clinical database to analyze the behavior of NRR index and compare it with other standard indices of anesthetic depth. The positive statistical results suggest that NRR index provides addition clinical information regarding motor reaction, which aligns with current standard tools. Furthermore, the ability to indicates the noxious stimulation is an additional finding. Lastly, I have proposed an real-time interpolation scheme to contribute my study further as a clinical application.}\\[1cm] 

\null\vfill
\clearpage 


\setstretch{1.2} 

\acknowledgements{\addtocontents{toc}{\vspace{1em}} 

First of all, I would like to thank Professor Jenho Tsao for all thoughtful lessons, discussions, and guidance he has provided me in the last five years. His immense knowledge in electrical engineering has also been a great addition to assisting me in my research. I thank Professor Hsiao-Lung Chan, Professor Kung-Bin Sung, Professor Nien-Tsu Huang, and Professor Hau-tieng Wu for joining my dissertation defense committee. I would also like to thank Professor Chih-Ting Lin for his great advice in my proposal defense.

I would like to thank Professor Yi Chang and all my colleagues in the department of anesthesia, Shin Kong Wu Ho-Su Memorial Hospital, for supporting me on seeking the answer to my clinical question. I am particularly grateful to Cheng-Hsi Chang M.D. for his constant assistance and advice in regards to my clinical practice and my research.

I would like to thank Professor Hau-tieng Wu for his inspiration and collaboration. It has been a great blessing that we were both able to work together and understand the problems from the perspective as a scientist, engineer, and physician. This has helped us drawn wonderful and meaningful results. I am very grateful for the valuable assistance and discussion from all the members in Signal Processing Laboratory. I am particularly grateful to Pei-Feng Lin, Ming-huang Chen, Zongmin Lin for all their valuable assistance and inspiring discussions during my stay at the NTU campus.

I would like to express my sincere gratitude to Hui-Hsun Huang M.D. in the department of anesthesia, National Taiwan University Hospital. His pioneer work on using time-frequency analysis to attack the IHR problem in anesthesia hugely inspired me since my residency.

I would also like to thank Professor Charles K. Chui for his guidance in seeking a solution on real-time interpolation.

My acknowledgements would be incomplete without recognition of Shu-Shya Hseu M.D. and Huey-Wen Yien M.D. for their extraordinary mentorship since I was a medical student.

My greatest appreciation are reserved for my family. My parents and brother have been the constant support of my life. My daughter and my son have always brought me joy. Lastly, I am most grateful to my wife's consistent dedication to me since we have met.
}

\clearpage 


\pagestyle{fancy} 

\lhead{\emph{Contents}} 
\doublespacing
\tableofcontents 

\lhead{\emph{List of Figures}} 
\listoffigures 

\lhead{\emph{List of Tables}} 
\listoftables 


\clearpage 

\setstretch{1.6} 

\lhead{\emph{Abbreviations}} 
\listofsymbols{ll} 
{
\textbf{NRR} & \textbf{N}on-\textbf{R}hythmic-to-\textbf{R}hythmic  \\
\textbf{HR} & \textbf{H}eart \textbf{R}ate \\
\textbf{IHR} & \textbf{I}nstantaneous \textbf{H}eart \textbf{R}ate \\
\textbf{HRV} & \textbf{H}eart \textbf{R}ate \textbf{V}ariability \\
\textbf{ECG} & \textbf{E}lectro\textbf{C}ardio\textbf{G}raphy \\
\textbf{$\PK$} analysis & \textbf{P}rediction probability analysis\\
\textbf{BIS} index & \textbf{Bis}pectral index\\
\textbf{PS} & \textbf{P}ower \textbf{S}pectrum \\
\textbf{tvPS} & \textbf{t}ime-\textbf{v}arying \textbf{P}ower \textbf{S}pectrum \\
\textbf{STFT} & \textbf{S}hort \textbf{T}ime  \textbf{F}ourier \textbf{T}ransform \\
\textbf{CWT} & \textbf{C}ontinuous \textbf{W}avelet \textbf{T}ransform \\
\textbf{TF} & \textbf{T}ime-\textbf{F}requency  \\
\textbf{RRI} & \textbf{R}-peak to \textbf{R}-peak \textbf{I}ntevral \\
\textbf{EDR} & \textbf{E}CG-\textbf{D}erived \textbf{R}espiration \\
\textbf{HF} power & \textbf{H}igh \textbf{F}requency power \\
\textbf{LF} power & \textbf{L}ow \textbf{F}requency power \\
\textbf{tvHF} power & \textbf{t}ime-\textbf{v}arying \textbf{H}igh \textbf{F}requency power \\
\textbf{tvLF} power & \textbf{t}ime-\textbf{v}arying \textbf{L}ow \textbf{F}requency power \\
}


%
%

%
%
%
%
%


\setstretch{1.3} 

\pagestyle{empty} 

\dedicatory{To My Family and to the Memory of My Grandparents.} 

\addtocontents{toc}{\vspace{2em}} 


\mainmatter 

\pagestyle{fancy} 



\chapter{Introduction} 

\label{Chapter1} 

\lhead{Chapter 1. \emph{Introduction}} 


\section{Field of Project}


The human body undergoes unique conditions when under anesthesia. Anesthesiologists administer anesthetics, also known as anesthetic medicine, dependent on the patient's physiological information. The physiological information may include the patient's heart rate, blood pressure, and electroencephalography (EEG) as this information reflects the inner dynamics of the human body. My study will present a new perspective on the physiological information that could contribute to the clinical practice of anesthesia.

\section{An Obscure Phenomenon}

\begin{figure}[htbp]
\centering
\includegraphics[width=1\textwidth]{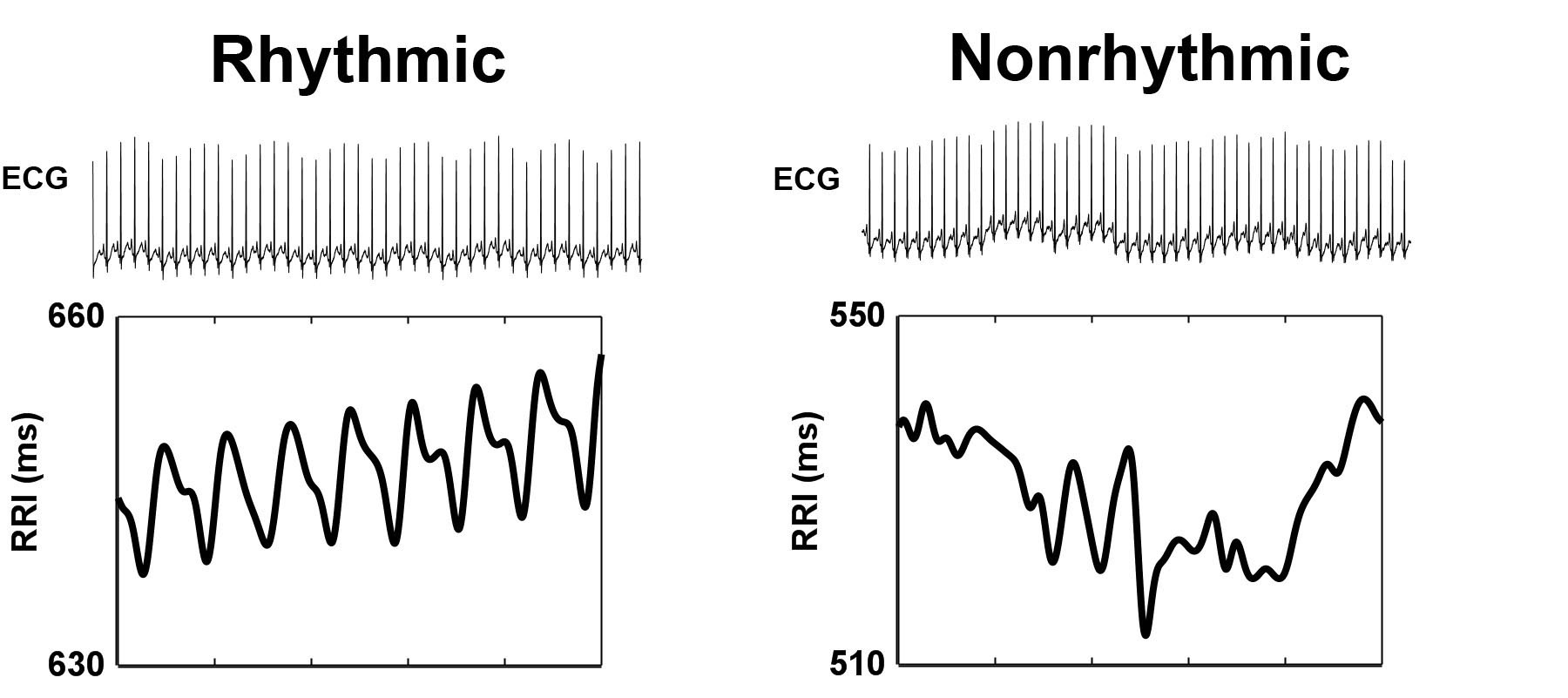}
\caption[Rhythmic-to-Non-rhythmic phenomenon]{

The tracing of R-R peak interval (RRI) represents instantaneous heart rate (IHR) exhibiting ``rhythmic" or ``non-rhythmic" patterns during anesthesia. This phenomenon is difficult to be read from the raw electrocardiography (ECG) waveform. Interestingly, the ECG waveform is mandatory in standard anesthesia monitor, but IHR is not.

}
\label{fig:NRR}
\end{figure}

As a practicing anesthesiologist, I have noticed that the instantaneous heart rate (IHR) fluctuates dependent on the various levels of anesthesia -- it became regularly oscillatory in deeper level of anesthesia, whereas it became less regularly in lighter level.(see Fig.\ref{fig:NRR}) This phenomenon, I refer to as ``\textit{rhythmic-to-non-rhythmic phenomenon}", has been described qualitatively in the past. However, to the best of my knowledge, no quantitative method has been proposed. Also, the underlying physiological mechanism has not been addressed. 

\section{Quantification for Depth of Anesthesia}


This phenomenon has induced three questions -- 1) how to quantify this phenomenon into a useful index, 2) whether the quantitative index can reflect the level of anesthesia, and 3) what are the underlying physiological mechanisms.

For an anesthesiologist, information from this phenomenon is unavailable in the clinical perspective. From Fig.\ref{fig:NRR}, it is apparent that the ``rhythmic-to-non-rhythmic" phenomenon can be seen from the tracing of instantaneous heart rate, but not from raw electrocardiographic (ECG) waveform. Although real-time monitoring of ECG is mandatory in the modern anesthesia monitor, the tracing of instantaneous heart rate is not displayed. Even though an anesthesiologist was aware of this phenomenon, he cannot obtain ``rhythmic-to-non-rhythmic" phenomenon information from modern anesthesia monitor to improve the anesthetic management.

If NRR index can provide additional clinical value, I believe that it could help in future developments of standard anesthesia monitor.

\section{Project Motivation and Goals}

My clinical observation leads to an intuition that this phenomenon arises from the inner physiologic dynamics under surgery and anesthesia. Hence, the quantification could provide valuable information to help anesthesiologists in administering anesthetics.

Pursuing these questions involves model construction based on knowledge of three distinct fields: signal processing, clinical anesthesiology and physiology. Then, I will propose a quantitative index, referred to as ``non-rhythmic-to-rhythmic ratio''(NRR) to quantify the phenomenon. The computation of NRR index requires multitaper Synchrosqueezing transform, a recently developed time-frequency analysis technique. 

I used a clinical database to explore the property of NRR index. Meanwhile, NRR index was compared with standard depth-of-anesthesia indices used in current clinical practicing to understand its performance. 

\section{Thesis Outline}
My thesis will begin with a literature review of the clinical anesthesiology, neural physiology, and time-frequency analysis. 
Bringing multidisciplinary knowledge up to date leads to a theory for the possible mechanism of rhythmic-to-non-rhythmic phenomenon. 
I proposed a model as the foundation of further analysis using time-frequency analysis. The \emph{adaptive harmonic model} is compatible in both ways: fitting the descriptive property in physiology, and providing adequate mathematical conditions for the quantification. Based on the adaptive harmonic model and multitaper Synchrosqueezing transform, NRR index was developed to quantify rhythmic-to-non-rhythmic phenomenon. I used a clinical database to analyze the behavior of NRR index, and compare it with other current standard indices of anesthetic depth. The statistical result shows positive value in clinical: NRR index provides information regarding motor reaction earlier and better than all the other depth-of-anesthesia indices. The result suggests a possible anatomic area and functional part in the brain that mediate the rhythmic-to-non-rhythmic phenomenon. To benefit each individual patient undergoing surgery and anesthesia, NRR index should be implemented in real-time. An optimal spline interpolation technique is proposed for the development of ``real-time NRR index".


\chapter{Background and Previous Work} 

\label{Chapter2} 

\lhead{Chapter 2. \emph{Background}} 


\section{Historical Background in Anesthesia}


More than seventy years ago, Guedel made a remark on the connection between anesthesia and respiratory patterns\cite{Guedel:37}. It appears that there is a direct relation between anesthetic depth and respiratory patterns. A deeper level of anesthesia results in regular respiratory patterns while a lighter level of anesthesia results in irregular respiratory patterns. During that era, depth of anesthesia were concluded from physical signs, such as the changes in pupil size and movement of the eyeballs. However, the introduction of muscle relaxant in clinical anesthesia has hindered the observation of muscle movement and Guedel's assessment method.


Even though Guedel's observations became less known, this ``rhythmic-to-non-rhythmic" phenomenon has been documented in other literature. About twenty years ago, Kato used power spectrum to analyze IHR in anesthesia\cite{kato1992spectral}. His study showed that the spectrum appears to be concentrated during deeper level of anesthesia; whereas the spectrum is dispersing in lighter level of anesthesia. Kato also mentioned the respiratory effect on this phenomenon. It is possible that the power spectrum technique helps to capture the obscure phenomenon in IHR at that era.

\section{Current View on Anesthesia}
Today, we would like to further explain this phenomenon through the advancement of various scientific fields in signal processing, anesthesiology, and neural physiology.

Because of the advancement in anesthesiology, we now understand that anesthetics exert effect mainly on the brain. Distinct anatomic location mediates different functions, with differential susceptibility to same anesthetics. Furthermore, each patient has his own unique genetic profile that renders diverse response to anesthetics. The inter-individual genetic variability, plus the dynamic influence of surgical procedure, make it difficult to determine appropriate anesthetic levels based on the dosage of anesthetics or single index of anesthetic depth. There is strong evidence that anesthetic management can improve the long-term health of patients, it is important to have monitoring instruments facilitating a comprehensive anesthetic management specific for each unique patient. Hence, the quantification of ``rhythmic-to-nonrhythmic" phenomenon should be worthwhile.


Current advancements in signal processing has resulted in the development of time-frequency analysis. In particular, reassignment and Synchrosqueezing technique were not readily accessible in the past. Furthermore, current advancements in neural physiology has also provided more insight into the underlying mechanism of the ``rhythmic-to-non-rhythmic" phenomenon, which has not been addressed according to my best knowledge. A more detail literature review regarding three main disciplines is as follows.


\section{Perspective from Anesthesiology}

Anesthesia comprises several components, including hypnosis, analgesia, immobility, amnesia, and autonomic nerve system stability\cite{miller2014miller,seitsonen2005eeg,vakkuri2004time,storm2013nociceptive}. For each patient, the dosage for different  anesthetics is tailored to the patient's specific surgical needs. Therefore, anesthesiologists need to continuously monitor the patient's response to the anesthesia. Bispectral Index\textsuperscript{\textregistered} (BIS), an index computed from  electroencephalography(EEG), is the current standard tools used to monitor depth-of-anesthesia for either the hypnotic component or the level of sleepiness. However, analgesia component, commonly known as the inhibition of pain sensation in anesthesia, requires an indirect interpretation of physiologic information such as heart rate (HR) and blood pressure\cite{velly2007differential,soehle2008comparison,ekman2004comparison,schmidt2004comparative,wennervirta2008surgical}. It is impossible to achieve optimal anesthesia using only one kind of anesthetics. Also, anesthesiologists are unable to monitor the depth of anesthesia by using merely one index, consequently inadequately measuring anesthetic effects. For a more comprehensive administration of anesthesia, it may be helpful to integrate different types of information to come to a decision.

It is important to understand what is wrong with simply relying on BIS monitor. BIS is a processed EEG, which measures the activity of the frontal cortical area only. On the other hand, amnesia is the suppression of memory, which is governed by medial temporal cortical area. Analgesia is the suppression of pain sensation, which is governed by various subcortical areas. Immobility is the inability of movement which is typically controlled by the thalamus and spinal cord\cite{velly2007differential, antognini1993exaggerated, antognini2000isoflurane}. Lastly, the entire autonomic nerve system is controlled by the brainstem\cite{eckberg2009point,poyhonen2004effect,hall2010guyton}. Although BIS index is the current gold standard of anesthetic depth monitor, BIS sensor is unable to measure the activities from all above listed anatomical areas except the frontal cortex. Literature in recent years indicates that anesthetics exert differential effects on different brain areas\cite{velly2007differential, miller2014miller, detsch2002differential}. Thus, there is a considerable risk measuring frontal cortex to conclude the anesthetic effect on other brain areas. For example, researchers have reported that the BIS monitor cannot reduce the rate of intra-operative awareness, which requires adequate memory function suppression\cite{avidan2011prevention}.

\section{Anesthesia and Long-Term Mortality}

Anesthesia is more than merely falling asleep. Evidences support the connection between anesthesia and long-term mortality. Too deep of hypnotic index (BIS index) for too long in surgery is associated with worse one-year mortality\cite{monk2005anesthetic}. It has also been reported that different anesthetics cause different immune responses, which affect tumor metastasis\cite{melamed2003suppression}.

Evidence suggests incorporating ``stress reduction" strategy into anesthetic management leads to longer life expectancy. First, infants who needed surgery for their congenital heart disease had lower long term mortality rate if they received adequate analgesics\cite{anand1992halothane}. Second, more adult patients who underwent major surgeries survived one year later if they receive adequate medication to stabilize autonomic nerve system\cite{wallace2004effect,mangano1996effect}. Lastly, when epidural anesthesia, a regional anesthesia technique, is combined with general anesthesia for patients undergoing major surgeries, studies have showed that epidural anesthesia reduces stress and pain, leads to faster recovery, and overall better outcome\cite{bardram1995recovery}. There are more compelling studies that support the benefit of anesthesia management on long term outcome. These evidences suggest that a comprehensive optimization of anesthesia is worthwhile.

In summary, anesthesia comprises several components. A comprehensive anesthesia management that covers several components could benefit patient for many years. Liteurature survey also brings hope that quantitative analysis of ``rhythmic-to-non-rhythmic" phenomenon could be useful to monitor the anesthetic effect on subcortical regions.

%


\section{Perspective from Physiology}


From a physiological perspective, the underlying mechanism of the ``rhythmic-to-non-rhythmic" phenomenon could be partially explained by breathing mechanism and cardiopulmonary coupling\cite{eckberg2009point}.
It is known that the neural respiratory control comprises of two systems, 1) the involuntary automatic control system and 2) the voluntary control system. The involuntary automatic control system is controlled by the respiratory center in the brainstem. The voluntary control system is controlled by the forebrain\cite{mitchell1975neural}. The anatomies of these two systems are distinct from one another. The respiratory neural signals from these two systems compete with each other but are both integrated at the spinal cord to control the respiratory motor neuron. During spontaneous respiration, the involuntary automatic respiratory pattern generated in the preB\"otzinger complex \cite{mitchell1975neural,elsen2005postnatal,rekling1998prebotzinger,ramirez1998hypoxic,shea1996behavioural}, is rhythmic. On the contrary, the breathing pattern of the voluntary respiratory motor control is non-rhythmic, which involves larger cerebral areas, including cortical processing and thalamic integration\cite{mitchell1975neural,mcfarland2002thalamic, ramsay1993regional,mckay2003neural,murphy1997cerebral,simonyan2007functional}. 

The suppression effect of anesthetics on the forebrain, including the cortex and the thalamus, is stronger than that on the brainstem\cite{sloan1998anesthetic}. Meanwhile, studies have shown that the preB\"otzinger complex, the involuntary respiratory control center, is less susceptible to the anesthetics\cite{kuribayashi2008neural,takita2010effects}. These relations suggest that under deeper level of anesthesia, the non-rhythmic respiratory is more suppressed than the rhythmic respiratory center. Hence, the IHR exhibits oscillation that is more rhythmic during deeper anesthetic level. Contrarily, the non-rhythmic respiratory center, being less suppressed during lighter level of anesthesia, causes activity more non-rhythmic in IHR. This concludes that IHR exhibits more nonr-hythmic. In summary, literature in physiology supports the correlation that the differential anesthetic effects between the involuntary control system and voluntary control system induces the ``rhythmic-to-non-rhythmic" phenomenon in IHR.
Thus, I hypothesize that IHR exhibits the central respiratory activity via cardiopulmonary effect\cite{eckberg2009point}, and it reflects the integration of rhythmic and non-rhythmic respiratory activities. Although more evidence is necessary to clarify this hypothesis, I propose using the NRR quantification methodology as a potential tool to evaluate the depth of anesthesia from a different aspect than using EEG-based monitoring.

\section{Physiology of Amplitude and Frequency Modulation}

The human body constantly regulates the respiration to meet the requirement of metabolism. In other words, the human body has to inhale oxygen (O$_\text{2}$), and expel carbon dioxide (CO$_\text{2}$) adequately. The respiratory control center in the brainstem receives the feedback information of oxygen concentration and the acidity to modulates the volume and the rate of breath. The acidity is due to the fact that CO$_\text{2}$ exists as hydrogen ion (H$^\text{+}$) and carbonic acid in human body\cite{fink1963central,hall2010guyton}.

Since CO$_\text{2}$ are the metabolic products of human body, the H$^\text{+}$ concentration and O$_\text{2}$ concentration in the human body are gradually, but consistently, changing. Thus, we can describe that the respiration signal in the brainstem is oscillatory with amplitude modulation (AM) and frequency modulation (FM). The AM and FM are continuous and slow change in time. These physiologic features can be depicted in mathematical language for subsequent modeling and quantification in the next chapter.

In summary, literature from physiology supports an integrated \textit{rhythmic} and \textit{non-rhythmic} neural activity reflecting the level of anesthesia. From a clinical anesthesiology perspective, technological advancements on monitoring various aspects of anesthetic depth can help improve the quality of overall patient care.

\section{Instantaneous Heart Rate and Heart Rate Variability}

\begin{figure}[htbp]
\centering
\includegraphics[width=1\textwidth]{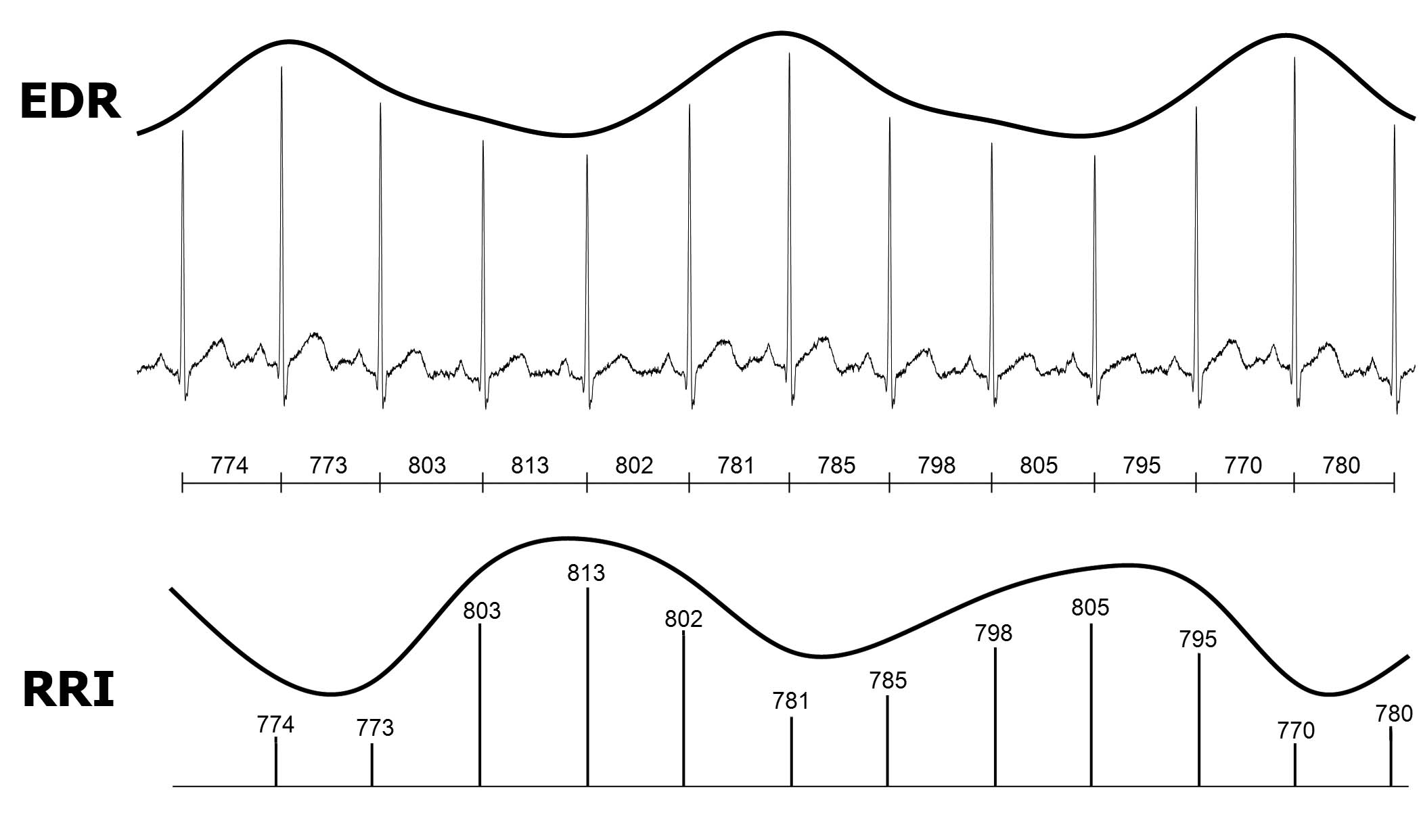}
\caption[R-R peak interval and instantaneous heart rate]{A schematic diagram for the R-R interval (RRI) signal derived from the raw lead III electrocardiographic (ECG) signal. We refer to RRI signal as instantaneous heart rate. The diagram also showed the ECG-derived respiration (EDR)}
\label{fig:RRI}
\end{figure}

The term \textit{heart rate} is important to all anesthesiologists and is usually referred to as the number of heartbeats within one minute. To an anesthesiologist, an ongoing heart rate indicates the stability of the heart condition, or the stress responses to surgical stimulation. The heart rate is mainly controlled by both sympathetic nerve and parasympathetic nerve system from the brainstem to the heart\cite{hall2010guyton}. Due to the dynamic nature of patient's under surgery and anesthesia, we do not feel comfortable with the heart rate information at an one-minute interval. Instead, a standard requirement in anesthesia practice includes a continuous real-time display of the heart rate. In this thesis, the ``rhythmic-to-non-rhythmic" phenomenon is only observable at the \textit{instantaneous} level. Conversely, the average heart rate of a normal routine excludes the information of ``rhythmic-to-non-rhythmic" phenomenon.

Both the \textit{Instantaneous heart rate} (IHR) and \textit {ECG derived respiration} (EDR),(Fig.{\ref{fig:RRI}}) can be derived from the raw ECG waveform\cite{moody1985derivation,houtveen_groot:2006,flaherty:1967}.
Although we can say that the ``rhythmic-to-non-rhythmic" phenomenon in IHR is a kind of ``generalized" heart rate variability (HRV), a medical guideline has set up the standard method of HRV analysis\cite{task1996north}. Unfortunately, the standard analysis techniques stated in this guideline cannot reveal the information of the ``rhythmic-to-non-rhythmic" phenomenon. In particular, the SDNNx, pNNx or the power spectrum technique measures information in a specified period, which inevitably neglects the dynamic and transient changes within this period. The guideline discusses but does not resolve the potential issue of \textit{stationarity} in the data. This also supports my desire to rely on the current scientific advancements on signal processes to resolve the current medical shortcomings in my project.

\section{Integration of Multidisciplinary Backgrounds}

Integration of multidisciplinary backgrounds provides partially theoretical background for the mechanism of ``rhythmic-to-non-rhythmic" phenomenon. Below is a list of key facts from various disciplines: 

\begin{enumerate}
\item In respiratory control, non-rhythmic activity originates from the forebrain, whereas rhythmic activity originates from the brainstem.
\item The forebrain has far more neurons and synapses when compared to the brainstem.
\item Neural structures with more synapses are more susceptible to anesthetics
\item Respiration have an coupling effect on IHR

\end{enumerate}

Fact 1 and 2 are knowledge in neurology and neural anatomy. Fact 3 is from anesthesiology, and fact 4 is physiology knowledge.
Each single fact seems ordinary when considered in its own disciplines. However, when I combine them together, the theoretical ground is justified. Then, we can proceed to the modeling problem in the next chapter.


\chapter{Theory and Modeling} 

\label{Chapter3} 

\lhead{Chapter 3. \emph{Modeling}} 

\section{Proposal}

The ``rhythmic-to-non-rhythmic" phenomenon in IHR\cite{kato1992spectral,Lin_Hseu_Yien_Tsao:2011} suggests that IHR during anesthesia consists of two components: one more rhythmic, less affected by anesthesia, and one more non-rhythmic, suppressed by deep anesthesia, and the strength of these two components vary according to the anesthesia depth.

I hypothesize that quantifying the "rhythmic-to-non-rhythmic" phenomenon will reflect the level of anesthesia. In order to quantify this change in IHR, I will introduce a novel index which I referred to as \textit{Non-rhythmic to Rhythmic Ratio} (NRR) index.

\section{Description of Proposed Methodology}

My hypothesis will be tested through a four step method. I will capture the changes in rhythm through IHR using the adaptive harmonic model. Next, I will use the multitaper Synchrosqueezing transform to extract the dynamic behavior from the ECG signals. In order to quantify ``rhythmic-to-non-rhythmic" phenomenon, I will use the NRR index. Lastly, I will use a clinical database to compare the NRR index with other anesthetic depth indices.

\section{Adaptive Harmonic Model}

It is important to consider which appropriate model to utilize to adequately describe and quantify this rhythmic-to-non-rhythmic phenomenon. Specifically, it is important to recall the high difficulty on describing the inner dynamic of the human body through mathematical formulas. This difficulty is due to the fact that the human body is a complex system creating non-linear and unpredictable interactions with the outer environment. The drastic influences of surgery and anesthetic medication complicates the difficulty further. It is fairly simple to ruin the capability of a model to describe the reality by imposing unrealistic conditions. That is the most challenging task currently at hand when we try to model the biological phenomenon of our study. In order to do so we have to be extremely cautious on what conditions we choose and impose on our study model.

After careful observation from a time-varying spectrum, I propose a  relatively conservative model to analyze the \textit{rhythmic} part of IHR. As is shown in Fig.\ref{fig:NRR}, the definition of \textit{rhythmic} implies that the oscillation exhibits time-varying frequency and time-varying amplitude. However, the modulations are in a relatively slow rate. This presents the notion that the relative predictable oscillatory activity can be perceived as a \textit{rhythmic} movement. Hence, a more conservative phenomenological modeling, also known as \textit{adaptive harmonic model}, will be proposed as follows.

The adaptive model integrates with the Synchrosqueezing transform creating a whole new scheme starting from the theoretical grounds to the clinical phenomenon. We start from motivating the technical model we consider to analyze $\IHR(t)$. It is commonly observed in clinical settings that stronger anesthetic levels are associated with more regular patterns in the instantaneous heart rate $\IHR(t)$, which is caused by the cardiopulmonary coupling effect under anesthesia \cite{kato1992spectral,Lin_Hseu_Yien_Tsao:2011}.
Based on this major observation, our treatment of the $\IHR(t)$ signal will be purely phenomenological; that is, the parameters and indices we will derive from the $\IHR(t)$ signal will be based solely on these signals themselves, and not on explicit, quantitative models of the underlying mechanisms. Since we mainly extract IHR from ECG signal as RRI, we will consider the following {\it phenomenological model} for $\RRI(t)$:
\begin{equation}\label{decomp1}
\RRI(t) = T(t)+A(t)\cos(2\pi\phi(t)),
\end{equation}
%
%
%

where $T(t)>0$, $A(t)>0$ and $\phi'(t)>0$.
We call $T(t)$ the trend, which captures the ``average heart rate'' over a long period; we require the trend to be positive, but not to be constant; we allow it to vary in time, as long as the variations are small enough. We call the derivative $\phi'(t)$ of the function $\phi(t)$ the {\it instantaneous frequency} (IF) of $\RRI(t)$; we require IF to be positive, but not to be constant; we allow it to vary in time, as long as the variations are slight from one period to the next, i.e. $|\phi''(t)|\leq \epsilon\phi'(t)$ for all time $t$, where $\epsilon$ is some small, pre-assigned positive number. Likewise, We call $A(t)$ the {\it amplitude modulation} (AM) of $R(t)$, which should be positive and smaller than $T(t)$, but is allowed to vary slightly as well, i.e. $|A'(t)|\leq \epsilon\phi'(t)$ for all time $t$. In summary, we have the following conditions for the model \ref{decomp1} \cite{Daubechies_Lu_Wu:2011,Chen_Cheng_Wu:2013}:
\begin{align}\label{decomp_cond}
&0<A(t)<T(t),~\phi'(t)>0,\\
&~|A'(t)|\leq\epsilon \phi'(t),~|\phi''(t)|\leq \epsilon \phi'(t),~|T'(t)|\leq \epsilon ~ \mbox{ for all } t.
\end{align}
If a RRI signal satisfied the model (\ref{decomp1}), we call it {\it rhythmic}; otherwise we call it {\it non-rhythmic}. It is possible that a function can be composed of both the rhythmic and non-rhythmic components.

This seeming complicated model is actually a direct generalization of the Fourier harmonic model \cite{Daubechies_Lu_Wu:2011, Chen_Cheng_Wu:2013}. Close inspecting the waveform of IHR, we can see that the conditions on $\phi''(t)$ and $|A'(t)|$ are reasonable. In other words, the \textit{rhythmic} RRI exhibits an oscillation with ``small enough" frequency modulation and amplitude modulation. Without adequate knowledge of underlying mechanism, the adaptive harmonic model fulfills the phenomenological behavior of rhythmicity feature.

\section{Non-rhythmic to Rhythmic Ratio}\label{NRRtheory}

The amount of variance inside the RRI signal are known as Heart Rate Variability (HRV). HRV has been proven to be closely related to the autonomic activity \cite{task1996north}. The most commonly used tool to quantify HRV is the power spectrum {(PS)} \cite{task1996north}. As useful as the PS is, however, in this study the PS cannot capture the momentary dynamics in the RRI time series. This issue limits the application to human under highly dynamic situation -- in particular, the ``rhythmic-to-non-rhythmic" phenomenon in anesthesia. The main technical focus of this paper is to resolve this limitation.

In this paper I will quantify the \textit{rhythmic-to-non-rhythmic} pattern of the RRI series by using a novel index referred to as \textit{Non-rhythmic to Rhythmic Ratio} (NRR). NRR is motivated by the clinical finding that deeper anesthetic levels are associated with the appearance of more regular oscillatory patterns in IHR. Thus, under deeper anesthesia, IHR oscillates more regularly.
We call this regular oscillatory IHR ``rhythmic'', which appears as a sharp peak on the PS. On the other hand, when the subject is under lighter anesthesia, IHR varies irregularly, and we call this IHR ``non-rhythmic''. The non-rhythmic IHR exhibits irregular and random-like behavior that appears as a plateau on the PS. As a result from this observation, the RRI time series is composed of a \textit{non-rhythmic component} and a \textit{rhythmic component} with various ratios.  
Based on the above facts, we hypothesize that the time-varying ratio change of the non-rhythmic and rhythmic components may reflect the anesthetic effect on the brain.

To quantify the ratio of the rhythmic component and non-rhythmic component, we will use a new {signal processing} tool to fully capture the momentary RRI time series change. We will replace the PS with the {\it time-varying power spectrum} (tvPS) by applying a recently developed time-frequency analysis technique \cite{Lin_Hseu_Yien_Tsao:2011,huang1997time}, which is referred to as multitaper Synchrosqueezing transform \cite{Daubechies_Lu_Wu:2011,Chen_Cheng_Wu:2013,xiao2007multitaper}. The tvPS is a non-parametric generalization of the PS providing instantaneous PS information. We will then be able to trace the momentary strength of the rhythmic component and non-rhythmic component through analyzing the tvPS.

By using the tvPS we will further extend the definition of the classical frequency domains HRV parameters.
Recall that traditionally, the high frequency (HF) power, the low frequency (LF) power, and the low frequency to high frequency power ratio (LHR), are determined from the PS of the RRI time series. Vagal activity mainly contributes to HF power, meanwhile, sympathetic activity influences both LF power and LHR \cite{task1996north,mazzeo2011heart}. By using the tvPS of RRI, we are able to obtain time-varying HRV parameters, including the time-varying high frequency power (tvHF), the time-varying low frequency power (tvLF) and the time-varying low frequency to high frequency power ratio (tvLHR). All these parameters are suitable to provide a better understanding of the dynamical situations of anesthesia. 

\begin{figure}[htbp]
\centering
\includegraphics[width=0.9\textwidth]{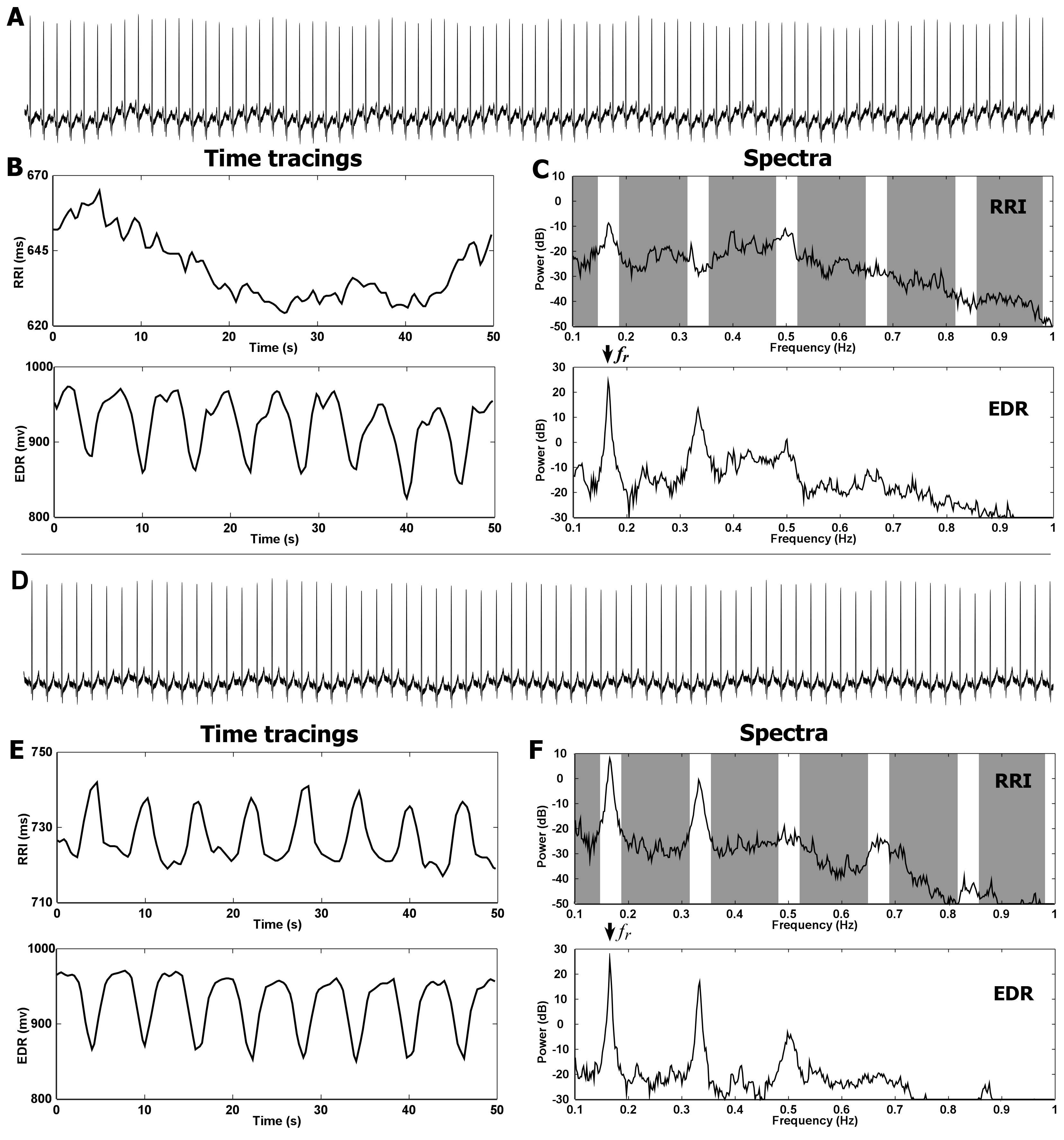}
\caption[The Calculation of Rhythmic Power and Non-rhythmic Power]{Fifty-second electrocardiography (ECG) waveforms, time tracings of R-R interval (RRI), ECG-derived respiration (EDR), and the corresponding spectra in lighter level of anesthesia (A, B, C) and deeper level of anesthesia (D, E, F). Whether in lighter (A) or deeper level of anesthesia (D), it is hard to read the IHR information from raw ECG waveform by naked eyes. Time tracings (B, E) of RRI and EDR and their power spectra (C, F) show that RRI is more non-rhythmic in lighter level of anesthesia then in deeper level. The frequency bands around the instantaneous frequency $fr$ and its multiple frequencies (the unshaded area) constitute the rhythmic component. The rest (shaded area) constitutes the non-rhythmic component. dB = decibel}
\label{fig:nrrcal}
\end{figure}

The main index NRR is defined as the ratio between momentary non-rhythmic power to momentary rhythmic power. 
In mathematical terms, we compute the rhythmic component power by identifying the location of peak on tvPS. The non-rhythmic power is then defined as tvHF subtracted by the momentary rhythmic power. 
Finally, NRR is defined as 
\[
\NRR = \log_{10} \left(\frac{\text{Non-rhythmic power}}{\text{Rhythmic power}} \right),
\]
where the value of NRR is high with non-rhythmic RRI and the value of NRR is low with rhythmic RRI. As previously discussed, the dynamical behavior of $\RRI(t)$ is captured by tvPS. Please see Figure \ref{fig:nrrcal}

In practical calculation, the non-rhythmic power is calculated as the rhythmic power subtracted from the high frequency power of RRI. The rhythmic component power is calculated as the sum of the power near the IF of the rhythmic component within the range of high frequency on the tvPS. 
We define the IF of the rhythmic component, denoted as $f_r(t)$, as the time varying position of the maximal power within the high frequency range on the tvPS of $\RRI(t)$:
\begin{align}\label{curve_extraction}
f_r(t) :=  \argmax_{0.1\leq\xi \leq1}M_{\RRI,K}(t, \xi),
\end{align}
which is implemented by the curve extraction algorithm considered in \cite{brevdo_fuckar_thakur_wu:2012}.
Next, we will calculate the rhythmic component power, denoted as $P_r(t)$, as the sum of the power inside the bands around $f_r(t)$ on the tvPS of $\RRI(t)$. In particular,
\[
P_r(t) := \int_{n f_r(t) - 0.01 Hz}^{n f_r(t) + 0.01 Hz}  M_{\RRI,K} (t, \xi) \ud \xi.
\]
Here the width of the band is chosen to be $0.01$Hz in an ad hoc way. We mention that when the signal is of broadband spectrum, there is no dominant curve that can be visualized, as shown in Fig.{\ref{fig:nrrcal}}. In this case, the most dominant curve defined in (\ref{curve_extraction}) from the time-frequency representation is viewed as the rhythmic component. 

The non-rhythmic component power is defined as the power within the high frequency subtracted by the rhythmic component power. Lastly, NRR is defined intuitively as the ratio of the non-rhythmic component power to the rhythmic component power:
\[
\NRR(t) = \log_{10} \left(\frac{\HFMR(t)-P_r(t)}{P_r(t)} \right) .
\]
Through defining power within absolute boundary as $P_r(t)$, NRR can differentiate the level between ``rhythmic" and ``non-rhythmic". In this thesis, the variables affixing $(t)$ mean that they are functions of time. It is important to notice that these sorts of time-varying parameters describe the momentary behavior of the signal. Since this instantaneous information cannot be captured by the classical power spectral analysis, we would expect to gain more insight into the IHR signal by considering our definitions. 

\section{Non-rhythmic Component vs. Stochastic Process}
For the purpose of this thesis, the term \textit{non-rhythmic} is defined as any signal that appears ``irregular", ``noisy", or ``as random". This poses two interesting questions. First, does this mean that we should treat all non-rhythmic signals as a stochastic process? Second, will we be able to model the non-rhythmic signal as an auto-regressive moving-average (ARMA) model or other stochastic process in order to analyze it using parametric methods? However, I would like to also remind that we should consider these issues using our medical knowledge.

One of my hypotheses in the present study is that the non-rhythmic signal reflects highly structured and complex activity of the human brain. Although it is too early to determine that the human brain is a ``deterministic" or ``stochastic" system, when awake, the human brain can retain memory for longer periods of time to execute complex tasks. The thought or idea in the brain lasts longer too. Therefore, the brain exhibits \textit{non-rhythmic} activity when its inner state is longer lasting. However, the above statements would contradict the stochastic process behavior that presents \textit{no memory effect} in a Gaussian random process. That is, the autocorrelation function is a Dirac's delta function:

\begin{displaymath}
R_{yy}(l) = \sum_{l\in Z}  { y(n) \bar{y}(n-l)} ,
\end{displaymath}

 $\mathbb{E}[ RR_{yy}(0)]=\sigma$ and $\mathbb{E}[ RR_{yy}(k)]=0$ for $k>0$ when $y$ is a Gaussian process.

Although \textit{non-rhythmic} signals may appear similar to a stochastic process, the human brain that shows \textit{non-rhythmic} activities should result in a more structured state. Therefore, a \textit{non-rhythmic} signal in its natural state is not completely ``stochastic". This is why I would like to use the term "non-rhythmic".
 
My second hypothesis is that there are two components, non-rhythmic and rhythmic, that are contained in the "rhythmic-to-non-rhythmic" phenomenon. It is unknown how these two components coexist but one simple way to model their co-existence would be \textit{superposition}. Regardless if we fully understand how the two components coexist, the fact that stochastic process is fundamentally a ``white-noise-driven" process means it doesn't match the ``rhythmic component plus non-rhythmic component" concept.

The above technical issues regarding parametric modeling for ``rhythmic-to-non-rhythmic" phenomenon needs to be carefully addressed in the future work. To narrow down the problems I have to solve in my thesis, I decided to choose a non-parametric method as a starting point. Still, statistical signal processing technique is a viable direction for my future studies.

\section{Summary}

In summary, the proposed \textit{adaptive harmonic model} (\ref{decomp_cond}) describes a fundamental physiologic property: the rhythmic component possesses frequency modulation and amplitude modulation. However, the law of physiology and chemistry sets a limit to these modulations.

From the previously mentioned theory to the modeling and quantification of the NRR index, I keep seeking the \textit{time-varying} ability to capture the momentary dynamics, which is essential to the dynamic nature of clinical anesthesia. This feature is also fundamentally distinct to theories of other signal analysis methods. After establishing the theoretic part, I will focus on the next problem: the time-varying spectrum.


\chapter{Time-Frequency Analysis, Reassignment, and Synchrosqueezing} 

\label{Chapter4} 

\lhead{Chapter 4. \emph{Synchrosqueezing Transform}} 


\section{Time-Frequency Analysis}

Chapter 3 provided strong evidence that in order to execute the quantification of NRR index within anesthesia the \textit{adaptive harmonic model} requires a strong time-varying spectrum. A recently developed time-frequency analysis technique is called the multitaper Synchrosqueezing transform. I believe that this technique is the most useful for my proposed study.

Time-frequency analysis is a type of signal processing tool that generalizes the power spectrum to provide \textit{time-varying} ability. This type of analysis generates a series of power spectra that has the ability to represent instantaneous spectra. Short-time Fourier transform (STFT) and continuous wavelet transform (CWT) are two types of well known tools for biomedical signal analysis\cite{Flandrin:99,daubechies:1992,Lin_Hseu_Yien_Tsao:2011,Chui:1997,Chui:1992}. Both of these tools are able to capture the instantaneous frequency information of the biological signal. Huang and Chan were the first to employ time-frequency analysis to analyze the dynamic changes in IHR during anesthesia \cite{huang1997time}.

Both STFT and CWT require window functions to ``focus" or ``localize" the signal. Take STFT as example:

\begin{equation}\label{STFT}
V_{X}^{(h)}(t,\xi)=\int^\infty_{-\infty} X(s)h(t-s)e^{-i2\pi \xi s}\ud s
\end{equation}

where $h$ is a window function chosen by the user, for example, the Gaussian function $h(t)=e^{-t^2}$. Note that $X(\tau)h(t-\tau)$ is the piece of signal around $t$ chosen by the window function. We then define the spectrogram of $X(t)$ \cite{Flandrin:99} as

\begin{equation}\label{spectrogram}
G_{X}^{(h)}(t, \xi) =\big|V_{X}^{(h)}(t, \xi)\big|^2.
\end{equation}

From equation \ref{STFT}, it is apparent that choosing different window functions causes different results. The influence of the chosen window function in relation to the STFT is no less than the signal per se. In particular, the window function imposes a resolution limit which relates to the well-known Heisenberg's uncertainty principal. The larger the time window $h$ is, the higher the spectrum resolution is, but with low temporal resolution, and vise versa. 

The window function also influences other TF methods, such as CWT, and Smoothed Pseudo Wigner-Ville Distribution, to name a few. Rather than fiddling with various TF methods, various window functions, and different sizes, it is highly advised to choose a technique that alleviates the above issues.

\section{Reassignment}

\begin{figure}[htbp]
\centering
\includegraphics[width=1\textwidth]{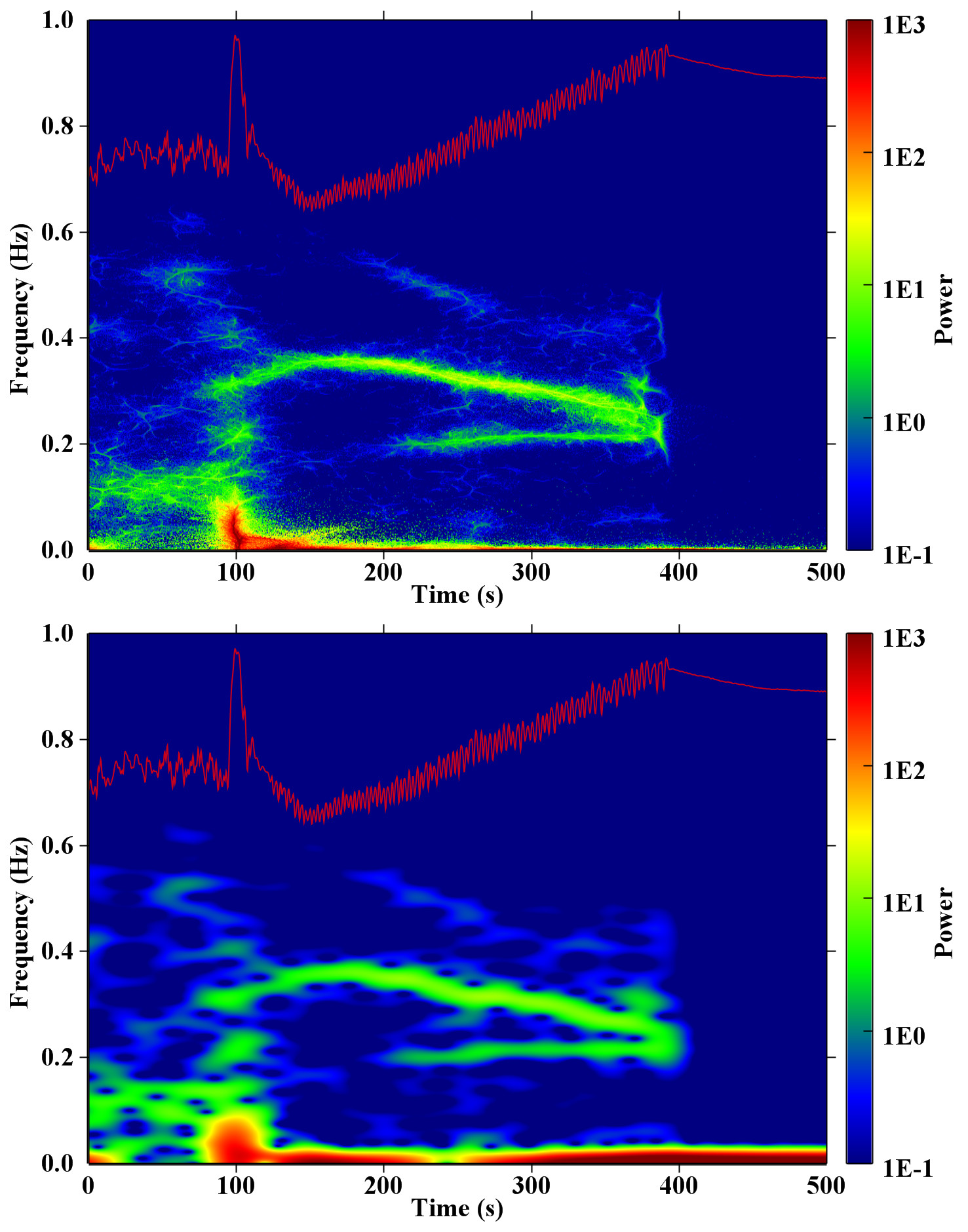}
\caption[Color Graph: STFT spectrogram vs. Reassigned spectrogram]{A side-by-side comparison between STFT spectrogram (A) and Reassignment spectrogram (B). The tracing of original R-R interval signal recorded in anesthesia is denoted as red line.}
\label{fig:tfcomp}
\end{figure}


To cope with the resolution limitation of time-frequency analysis, it is possible to further use the phase information $\varphi(t,\xi)$, which is discarded in the classical STFT spectrogram (\ref{spectrogram}) and scalogram\cite{Flandrin:99, daubechies:1992}. Kodera et al. first proposed this technique\cite{kodera_gendrin_villedary:1978} by calculating the instantaneous frequency $\partial_t \varphi(t,\xi)$ and group delay $\partial_\xi \varphi(t,\xi)$. Auger and Flandrin coined  this technique as the \textit{reassignment} and further generalized it to ``Fourier transform based" time-frequency analysis\cite{Flandrin:99}. The reassignment technique provides a sharper representation, which improves the readability of the STFT spectrogram. Reassignment is helpful in visualizing the dynamics of the biological signal that contains multiple nonstationary components in the TF plane (Fig. \ref{fig:tfcomp}).

The phase information hidden in STFT contains information that explains why time-frequency domains are spreaded out.\cite{auger_flandrin:1995}. To further eliminate this issue, we can shift the STFT coefficients accordingly to the reassignment rules determined by the phase information. The temporal reassignment rule and frequency reassignment rule are as follows:

\begin{align}
&\widehat{t}_{t,\xi}:=t+\Re\left\{ \frac{V_{X}^{(Th)}(t,\xi)}{V_{X}^{(h)}(t,\xi)}\right\}\nonumber\\
&\widehat{\xi}_{t,\xi}:=\xi-\Im\left\{ \frac{V_{X}^{(Dh)}(t,\xi)}{V_{X}^{(h)}(t,\xi)}\right\}\nonumber,
\end{align}

where $\Re$ denotes the real part and $\Im$ denotes the imaginary part and $h$, $Th$, and $Dh$ are the three related window functions. Here $Th$ is the time-ramped window used to determine the temporal reassignment rule and $Dh$ is the differential window used to determine the frequency reassignment rule. 
The reassignment is then achieved by applying the reassignment rules:
\[
RS_{X}^{(h)}(t,\xi):=\int^\infty_{-\infty}\int^\infty_{-\infty} V_{X}^{(h)}(s,\eta)\delta(s-\widehat{t}_{s,\xi})\delta(\eta-\widehat{\xi}_{s,\xi})\ud s\ud \eta,
\]
where we move the spectrogram coefficient $V_{X}^{(h)}(s,\eta)$ at time $s$ and frequency $\eta$ to a new location $\widehat{t}_{s,\xi}$ and $\widehat{\xi}_{s,\xi}$ according to the reassignment rules.


The reassigned spectrogram is a valuable tool for signals that contain dynamic information of anesthesia in my current study\cite{Lin_Hseu_Yien_Tsao:2011}. However, aside from the visual information represented in the TF plane, it is also desirable to execute further analysis of the signals, such as isolating a component in TF plane, or performing reconstruction in time domain. \textit{Synchrosqueezing transform} is a recently developed reassignment technique that can provide this possibility.

\begin{figure}[htbp]
\centering
\includegraphics[width=1\textwidth]{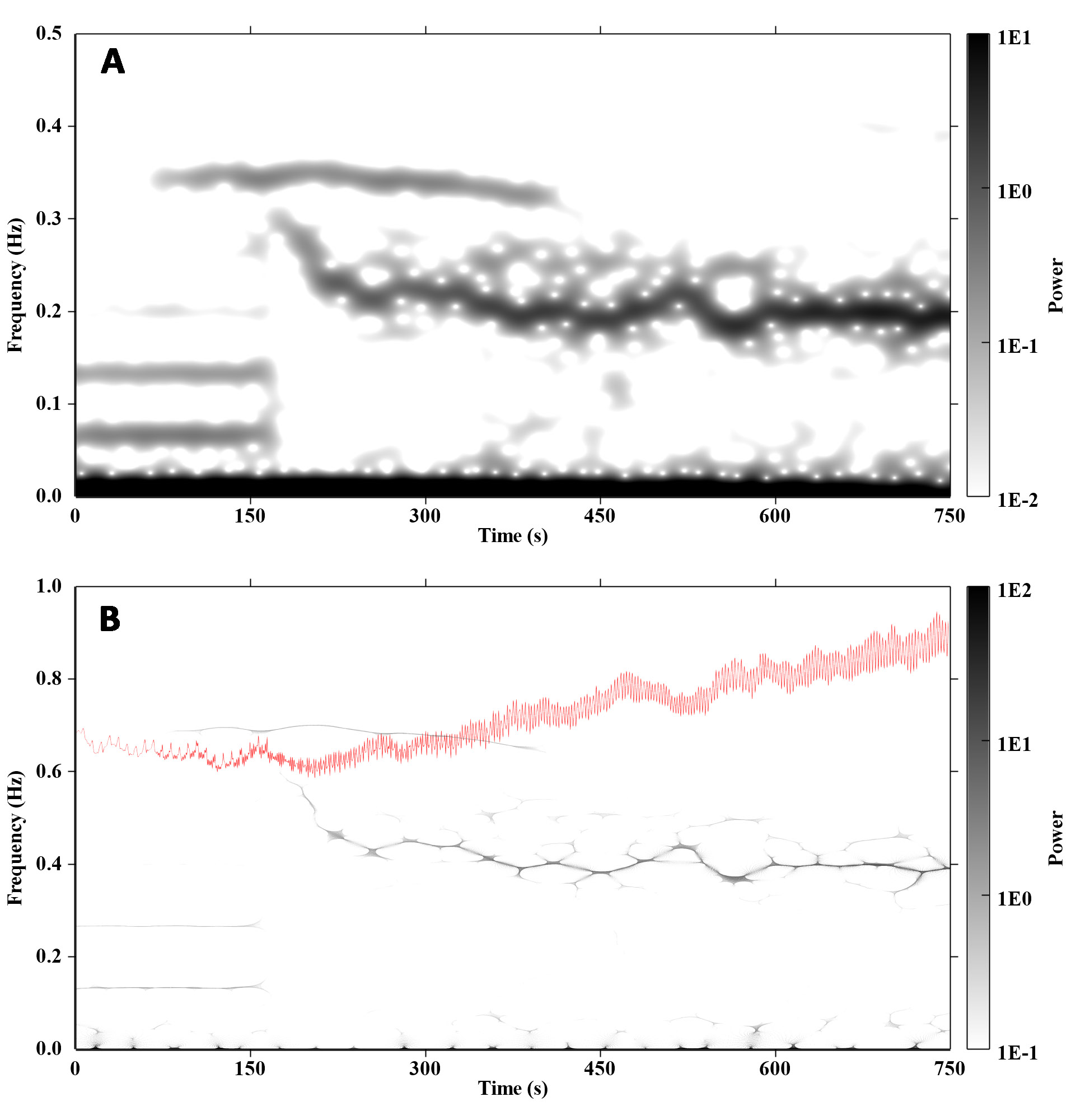}
\caption[STFT spectrogram vs. Reassigned spectrogram]{A side-by-side comparison between STFT spectrogram (A) and Reassignment spectrogram (B). The tracing of original R-R interval signal recorded in anesthesia is superimposed as red line in panel B}
\label{fig:3comcomp}
\end{figure}

\section{Synchrosqueezing Transform}

Synchrosqueezing transform is a special kind of reassignment technique\cite{daubechies_maes:1996,Daubechies_Lu_Wu:2011,Chen_Cheng_Wu:2013}. This technique uses a light-weight mathematical model ({\ref{sstdecomp_cond}}).  The light-weight mathematical model helps with further processing of signals, such as isolating a component in the TF plane or reconstruction in time domain.

For my present study, Synchrosqueezing transform can be used to present non-parametric generalizations of the power spectrum. The Synchrosqueezing transform is able to provide the time-varying power spectrum (tvPS) which reveals the instantaneous changes that happen in IHR. In addition, comparing with traditional STFT and wavelet, it is visually more informative, more resistant to several kind of noise, and more resistant to the influence of chosen window function \cite{Daubechies_Lu_Wu:2011, Chen_Cheng_Wu:2013, brevdo_fuckar_thakur_wu:2012}. Mathematical works have supported these abilities, as well as the ability of inverse transform. Luckily we do not need to impose too much effort on choosing parameters as using the Synchrosqueezing transform technique can easily capture the oscillatory activity.

The detailed information on the algorithm leading to tvPS is as follows. It has been known that the phase information hidden in STFT contains more information that explains why the STFT spectrogram is spreaded out. A solution that could help remedy this issue is by shifting the STFT coefficients according to reassignment rules determined by the phase information. The frequency reassignment rule is as follows:

\begin{equation}\label{freassign}
\Omega(t,\xi):= \frac{-i\partial_t V_{X}^{(h)}(t,\xi)}{2\pi V_{X}^{(h)}(t,\xi)}
\end{equation}

where $\partial_t$ means the partial derivative with related to $t$. 
The Synchrosqueezing transform is then achieved by applying the reassignment rule:
\[
S_{X}^{(h)}(t,\xi):=\int^\infty_{-\infty}\int^\infty_{-\infty} V_{X}^{(h)}(t,\eta)\delta(|\xi-\Omega(t,\eta)|)\ud \eta,
\]
where for each fixed time $t$, we move the STFT coefficient $V_{X}^{(h)}(t,\eta)$ at frequency $\eta$ to a new location according to the reassignment rules $\Omega(t,\xi)$.

One way that the Synchrosqueezing transform is unique to the reassignment technique is that the Synchrosqueezing transform has the ability to reconstruct back to the time domain. Suppose the signal $X(t)$ analyzed by Synchrosqueezing transform is: 

\begin{equation}\label{sstdecomp1}
X(t) = T(t)+A(t)\cos(2\pi\phi(t)),
\end{equation}
where $T(t)>0$, $A(t)>0$ and $\phi'(t)>0$. 

In other words, $X(t)$ comprises of an oscillatory component and a trend component.

In addition, the amplitude modulation and frequency modulation of the oscillatory component is under the condition as follows:
\begin{align}\label{sstdecomp_cond}
&0<A(t)<T(t),~\phi'(t)>0,\\
&~|A'(t)|\leq\epsilon \phi'(t),~|\phi''(t)|\leq \epsilon \phi'(t),~|T(t)'|\leq \epsilon ~ \mbox{ for all } t.
\end{align}

Relevant results of the Synchrosqueezing transform can be found in literatures\cite{Wu_Flandrin_Daubechies:2010,brevdo_fuckar_thakur_wu:2012,Thakur_Wu:2011,Wu:2013,Wu_Chan_Lin_Yeh:2013}. A recent review article discussed in depth on the relation of the TF analysis, the reassignment and Synchrosqueezing technique\cite{Auger_Flandrin_Lin_McLaughlin_Meignen_Oberlin_Wu:2013}.

\section{Multitaper Estimation}

From a non-parametric spectral estimation perspective, Thompson proposed a powerful multi-window technique that improves spectral estimation of the random process\cite{kaymodern,thomson1982spectrum}. Thompson uses the \textit{discrete prolate Spheroidal sequence} to define a series of window function, which are orthogonal to each other, that have the same length in time domain. The average of the spectra from multiple windows reduces the spectrum \textit{variance} but also prevents the \textit{bias} of spectral estimation induced by varying window size. One strong advantage of Thompson's spectrum is that we are able to estimate the spectrum of a stationary random process without resort to parametric estimations.

The principle of Thompson's technique has been extended to the field of nonstationary signal. Multitaper in time-frequency analysis is designed to be used with Hermite functions as the sequence of multiple windows that are orthogonal to each other in time-frequency plane. The final spectrogram estimation can then be obtained from the average of multiple realizations. Taking the STFT spectrogram as an example, the multitaper spectrogram can be defined as follows:

\[
Q_{{X}, K}(t, \xi) := \left|\frac{1}{K}\sum_{k = 1}^K V_{X}^{(h_k)}(t, \xi)\right|^2 .
\]

The multitaper spectrogram and multitaper scalogram were shown to make success for nonstationary stochastic signal\cite{bayram2000multiple}. To achieve both high resolution for deterministic nonstationary signal and low variance for stochastic nonstationary signal, Xiao and Flandrin combined the reassignment technique and multitaper technique{\cite{xiao2007multitaper}}.

From the background information mentioned in Chapter {\ref{Chapter2}} and the model in Chapter {\ref{Chapter3}}, it can be seen that the IHR under anesthesia should contains both non-stationary component and stochastic component. We have reported the advantage of using multitaper time-frequency reassignment in extracting these dynamic features{\cite{Lin_Hseu_Yien_Tsao:2011}}.

\section{Multitaper Synchrosqueezing Spectrogram}
Synchrosqueezing transform responds well to signals from different types of stochastic process. One dilemma we face is understanding the true mechanism in biological signal. Since the quantification of the \textit{non-rhythmic} component is an important part of our current model(\ref{NRRtheory}), we have decided to apply the multitaper method to further optimize the performance of the Synchrosqueezing transform $X(t)$.
We have chosen the \textit{multitaper Synchrosqueezing spectrogram} to achieve the tvPS as follows: 

\[
M_{{X}, K}(t, \xi) := \left|\frac{1}{K}\sum_{k = 1}^K S_{X}^{(h_k)}(t, \xi)\right|^2,
\]
where $K>0$ is the number of windows chosen. We choose the the $k$-th Hermite function as $h_k$. For my current study, I will use $K=10$ and the first ten Hermite window functions will be $h_k$, $k=1,\ldots,10$ when deriving tvPS.

\begin{figure}[htbp]
\centering
\includegraphics[width=1\textwidth]{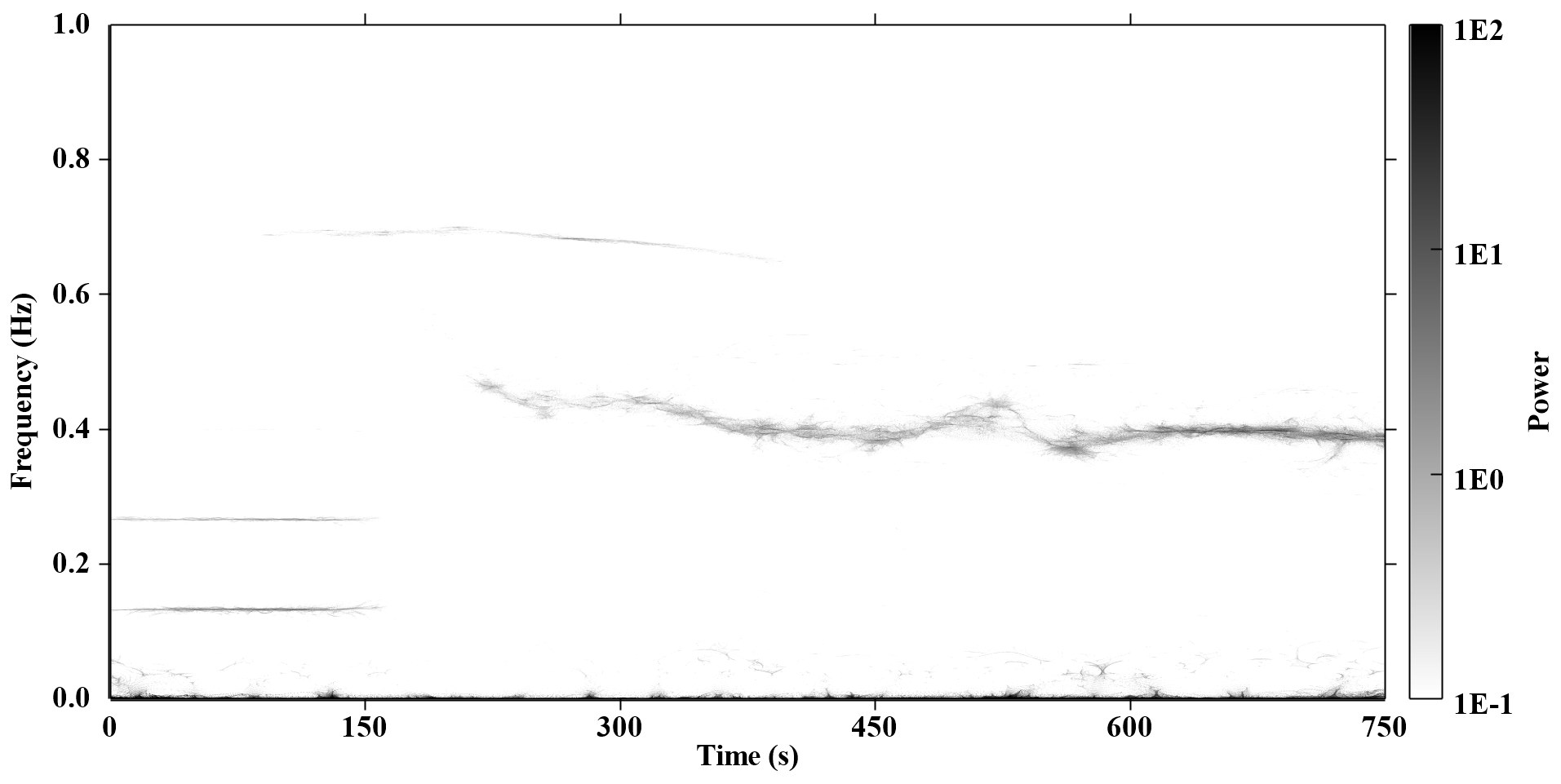}
\caption[The time-varying power spectrum: multitaper Synchrosqueezing spectrogram]{The time-varying power spectrum of instantaneous heart rate (IHR) as data used in figure {\ref{fig:3comcomp}}, realized by multitaper Synchrosqueezing spectrogram, showing rich dynamic features of human body in anesthesia. As shown in figure {\ref{fig:3comcomp}}, the readability is better than STFT spectrogram, and also better than the reassigned spectrogram. }
\label{fig:3comsstm2}
\end{figure}

The advantage of using a multitaper Synchrosqueezing spectrogram is summarized as follows:

\begin{enumerate}
\item Robust to several different kinds of noise, which might be slightly nonstationary.
\item Visually informative. When the signal is rhythmic, that is, satisfying the model (\ref{decomp1}), a dominant curve following the IF can be seen on the multitaper Synchrosqueezing spectrogram; otherwise the multitaper Synchrosqueezing spectrogram is blurred.
\item Adaptive to the data so that its dependence on the chosen window function is weak.
\end{enumerate}

\begin{figure}[htbp]
\centering
\includegraphics[width=1\textwidth]{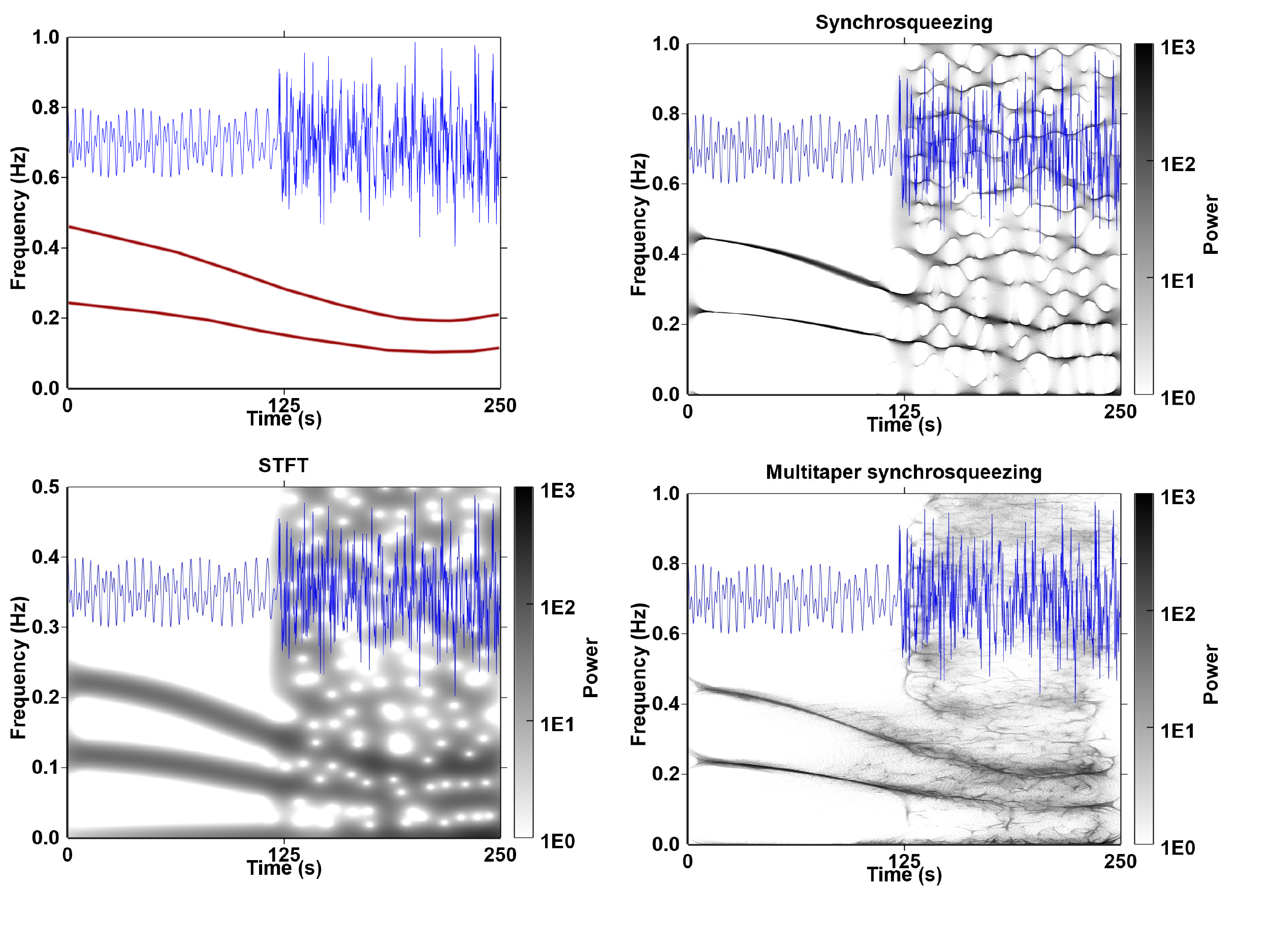}
\caption[Demonstration of multitaper Synchrosqueezing using artificial data]{An artificial data, as a summation of two time-varying components and a Gaussian white noise, and its TF analysis realized by STFT spectrogram, Synchrosqueezed spectrogram and the multitaper Synchrosqueezed spectrogram. From 125 s to 250 s, it is apparent that Synchrosqueezing technique improves the readability of the two components, and multitaper technique refines further.}
\label{fig:simusst}
\end{figure}

We compared various methods of TF analysis in Table{\ref{table:comparetf}} and in simulated data analysis(Fig.\ref{fig:simusst}).

%
%
%

\begin{table}[h]
\centering
\caption[Comparison of TF analysis methods]{Summary of abilities of TF methods. Details can be found in references cited in this chapter. X: unable; V: feasible; ?: questionable}
\renewcommand{\arraystretch}{2.3}
\label{table:comparetf}

\begin{tabular}{lllll}
\hline
                                   & Reassignment & SST  & STFT & CWT  \\ \hline
Instantaneous frequency estimation & good         & good & fair &      fair \\
Visual readability                 & good         & good & fair & fair \\
Component reconstruction           & X            & V    & V    & V    \\
Robust to noise                    & ?            & good & fair & fair \\ \hline
\end{tabular}
\end{table}

\section{Adaptive Filter and Parametric Estimation}

Another kind of techniques based on parametric modeling for time series data analysis is the adaptive filter, including the famous Kalman filter, and the parametric spectral estimation\cite{Chui_Chen:2009x}. The adaptive filter is designed to use a cost function to represent the weighted mean square error. The goal is finding a set of parameters $a_i$ to minimize the cost function.
 \[
 a_i= \argmin \mathbb{E} \{e^2[n]\}
 \]
If we have adequate knowledge about the signal, we can obtain the optimal point on the performance surface of the cost function by using partial differential.
 \[
 \frac{\partial}{\partial a_i} \mathbb{E} \{e^2[n]\} = 0 
 \]
It is apparent that TF reassignment is fundamentally different from adaptive filter. TF reassignment is a non-parametric deterministic signal processing technique, whereas the adaptive filter is a parametric technique for stochastic signal processing. Regardless of the difference, both techniques use the same mathematical tools to achieve their goals.

Kalman filter in essence is a real-time prediction-correction algorithm that provides parametric estimation in the sense of weighted least-square fitting\cite{kaymodern,Chui_Chen:2009x}. As the physiologic mechanism underlying rhythmic-to-non-rhythmic phenomenon is not adequately understood, a technique which requires both elaborate model construction and careful parameter selection is not my first-line choice for my research problem. Besides, From Chapter {\ref{Chapter3}}, it shows that the NRR index has a high relation to oscillation. Therefore, for my study, the output of the Kalman filter as a matrix of state variables may not be as intuitive as the spectrum technique for data exploration and interpretation at this stage. 

Parametric spectral estimation has a complete theory of modeling based on stochastic process{\cite{kaymodern}. One critical issue of the parametric estimation is that the mechanism of the signal we are facing in the present study is unknown, and is constantly changing. The parametric spectral estimation has a lower tolerance to incorrect modeling. In other words, it is crucial to ensure that the correct model are constructed to generate reasonable parametric spectral estimation. In addition, parametric spectral estimation does not behave well when the signal is a superposition of multiple components. The rigidity of model construction contrasts well with the flexibility of Synchrosqueezing: flexible in window selection and adaptive to signal. Certainly, it is possible that the technique realizing time-varying power spectrum in the future may belong to the family of parametric spectrum. 

\section{Common Condition in Physiology and Mathematics} 
The most unique characteristic of the Synchrosqueezing transform that is relevant to this project is the mathematical condition {\ref{decomp_cond}} which shows that the frequecy modulation and amplitude modulation are limited. The condition used on the Synchrosqueezing transform is exactly the same condition as the \textit{adaptive harmonic model} in order to describe this \textit{rhythmic} physiological behavior. Therefore, the Synchrosqueezing transform matches perfectly with the adaptive harmonic model for us to quantify the ``rhythmic-to-non-rhythmic" phenomenon that was previously discussed.


\chapter{Performance Evaluation using Clinical Data} 

\label{Chapter5} 

\lhead{Chapter 5. \emph{Performance Evaluation}} 


\section{Clinical Database}
In my hopes to investigate the clinical value of NRR index, I have collected and analyzed a clinical database. The prospective observational study spanned the whole period of typical surgical anesthesia. ECG, BIS index, and the end-tidal concentration of sevoflurane were all recorded continuously. Anesthetic events, including loss of consciousness (LOC), skin incision, first reaction of motor movement during emergence period, and return of consciousness (ROC), were registered with precise time stamps.

After receiving approval from the Institutional Review Board (Taipei Veterans General Hospital, Taipei, Taiwan), we enrolled 31 female patients. All patients were \textit{American Society of Anesthesiologists} (ASA) physical status of I or II and were scheduled for laparoscopic gynecological surgery undergoing general anesthesia. We obtained informed consent in paper from each of our patients. All ASA I or II status patients are generally under healthy conditions. It is important to investigate the changes of the NRR index under the normal physiological conditions. We are aware that some diseases can affect HRV (and IHR) untowardly. Thus, patients that are excluded were those younger than 20 years old or older than 55 years old, have a body mass index below 19 or higher than 30 and have either cardiovascular disease, arrhythmia, intake of medication that affects the neurological and cardiovascular system, potential cancer, diabetes mellitus, or anticipated difficult airway. All these surgeries were performed between the hours of 8 a.m. and 7 p.m.

\section{Anesthesia Protocol}

The standard anesthetic monitoring (HP agilent patient monitoring system) was applied to all patients. This included ECG, pulse oximeter (SpO$_\text{2}$) and non-invasive blood pressure. A BIS electrode (BIS-XP sensor) and the ECG recorder (MyECG E3-80; Micro-Star Int’l Co., New Taipei City, Taiwan) were applied to the patients prior to anesthetic induction. After the initial step of cleaning the skin with rubbing alcohol, the BIS sensor is applied to the right side of the forehead. As a standard practice, all anesthetic managements, including the administration of medications, and airway management were under the discretion of the anesthesiologist. After intravenous infusion of lactated Ringer's solution begins, we started anesthetic induction with fentanyl 2.5 \--- 3 \textmu g/kg. Next, right after preoxygenation, hypnosis was induced with propofol 2 \--- 2.5 mg/kg. LOC was then assessed by no response to verbal command. Cisatracurium 0.15mg/kg was used to facilitate tracheal intubation. Subsequently, mechanical ventilation was started from volume control mode with oxygen-gas-sevoflurane mixture as low flow anesthesia. The respiratory rate and tidal volume of the ventilator was adjusted to maintain an end-tidal carbon oxide (ETCO$_\text{2}$) that ranges between 35 and 40 mmHg, and to keep the peak airway pressure lower than 25 mm Hg.

The laparoscopic skin incision was made by a surgical blade. The adequacy of anesthetic depth (during the whole period of surgery) was determined by the BIS index ($<60$) and based on the anesthesiologist's judgment. If the anesthesiologist determined an inadequate level of analgesia, a bolus dose of fentanyl would be given to the patient. Toward the end of the surgery, the peritoneal gas was deflated and the patient's body position recovered from Trendelenburg's position to level-supine position. As the wound closure began, the anesthetic gas was on a consistent decrease. Muscle relaxation was reversed with a combination of neostigmine 0.05 mg/kg and glycopyrrolate 0.01 mg/kg during controlled ventilation. Once the patient regained spontaneous breathing the controlled ventilation was halted.
 Patients that exhibited inadequate spontaneous breathing (ETCO$_\text{2}>$50 mmHg or SpO$_\text{2}<$95\%) were assisted with manual positive ventilation, otherwise the patients continued to breath spontaneously until regaining consciousness without the interference of positive-pressure ventilation.ECG and BIS are the main electrophysiological signals we analyze for this project. When we observed the emergence period, physical contact to the patient was avoided and minimized if necessary to prevent the interference on the sensors of ECG and BIS monitor. Any unnecessary interference to patient's respiration from the manual (bag) ventilation was avoided also. When the patient showed adequate consistency in their spontaneous breathing, we removed the endotracheal tube. If in any case there was inadequate ventilation or upper airway obstruction, corrective actions were taken that include either mask ventilation or nasopharyngeal airway insertion. During the emergence period, I carefully assessed first reaction as any first visible motor reaction such as movement of the arms or legs, coughing, or grimace. ROC was assessed through the patient's ability to open their eyes and follow simple procedural commands.

\section{Data Acquisition}
 The BIS index was continuously recorded from an Aspect A-2000 BIS monitor (version XP, Host Rev:3.21, smoothing window: 15 seconds; Aspect Medical Systems, Nattick, CA, USA) connected to a laptop computer (Asus Corp., Taipei, Taiwan). During surgery, corrective measures were taken to improve signal quality of the BIS sensor when the signal quality index was lower than 50; otherwise the BIS data during this period was discarded. The raw limb lead ECG sampled at 1000Hz and 12-bit resolution, was recorded for off-line analysis. The clocks on the laptop and ECG recorder were synchronized with a time accuracy of $\pm 1$ second. Time stamps of BIS record were provided by the laptop. The inhaled and end-tidal concentrations of anesthetic gas, which was sampled from the connection piece close to the endo-tracheal tube and detected by the gas analyzer on a Datex-Ohmeda S/5 anesthesia machine (GE Health Care, Helsinki, Finland), were also recorded on the laptop. All data was recorded without any interruption until the patient regained ROC. All data collected for analysis of sevoflurane concentration correlation and anesthetic events, including skin incision, first reaction and ROC, are under single ventilation mode. There were no transitions from the mechanical ventilation to the spontaneous ventilation or vice versa in the data. However, during LOC,  the transition from the spontaneous ventilation to mechanical ventilation is possible.

The offline ECG signal was analyzed in several steps. The R peak detection was automatically determined by taking lead I, II and III ECG signals into account to improve accuracy. The ectopic beats were removed and interpolated from the data to eliminate incorrect R peaks and ectopic beats through visual verification. When electrocauterization is severely interfered by the ECG, the data segment were also discarded. Whenever electro-cauterization severely interfered the ECG, this data segment was discarded. IHR was then derived from the R peaks through the cubic spline interpolation. RRI were resampled to be equally spaced at 4 Hz from IHR for subsequent analyses (Fig.\ref{fig:RRI}). 

In order to correct the hysteresis between the end-tidal sevoflurane concentration and the anesthetic effect on the brain, the estimated effect-site sevoflurane concentration ($C_{\textup{eff}}$) was derived from the end-tidal sevoflurane concentration ($C_{\textup{et}}$) by the following first order differential equation. This equation is based on the pharmacokinetic-pharmacodynamic modeling\cite{sheiner1979simultaneous}:


$$
\frac{\ud C_{\textup{eff}}}{\ud t} = K_{\textup{e0}} ( C_{\textup{et}}-C_{\textup{eff}} )
$$
where the constant $K_{\textup{e0}}$ was assumed to be constant for all patients and defined as 0.20 /min according to previous studies \cite{soehle2008comparison}.


\section{Statistical Analysis}
We treated BIS and ECG-derived continuous indices, including NRR, $\HFMR$, $\LFMR$, $\LHRMR$ and HR, as anesthetic depth indices. The performances of the indices were evaluated in two ways: the ability for these indices to predict anesthetic events and their correlations with effect-site sevoflurane concentration during emergence period. Because the concentration of anesthetic gas consistently decreased and the influence from surgery was relative minor, we considered the emergence period for correlation analysis.  A $p$ value less than 0.05 would be considered as statistically significant. Multiple significant tests were calculated by using Bonferroni correction. Statistical results are expressed as mean (standard deviation [SD]). 

Prediction probability ($\PK$) analysis is a versatile statistical method for measuring the performance of an anesthetic depth index \cite{smith1996measuring}. A value of one means that the observation is always correctly predicted, a value of 0.5 indicates that the observation is predicted at a 50/50 chance. The ability to predict anesthetic events can be investigated by using the serial prediction probability analysis\cite{smith1996measuring}. The correlation with sevoflurane concentration employed $\PK$ analysis and Spearman rank correlation. Estimation of $\PK$ and its standard error was obtained through the Jackknife method. Prior to using the null hypothesis, any $\PK$ value less than 0.5 would be changed into one minus the $\PK$ value\cite{smith1996measuring}.

\section{Serial Prediction Probability Analysis in Predicting Anesthetic Events}
To evaluate the performance of the anesthetic indices in predicting the anesthetic events, serial $\PK$ analysis was designed as follows. The following timestamps are used as the baseline: one minute before LOC, five seconds before skin incision, three minutes before the first reaction and three minutes before ROC. The serial $\PK$ analysis is performed as successive $\PK$ analyzes of data pairs: index on baseline timestamp versus indices on subsequent successive timestamps spaced by five seconds. As a result, the out of the serial $\PK$ analysis is a sequence of \textit{time-varying} $\PK$ values which reveals the temporal relation between anesthetic events and the indices. By plotting the serial $\PK$ value and its standard error bar, it becomes possible for us to do hypothesis test simply with the naked eye. If there is significant difference ($p<0.05$), it can be established if $1.5$ times of their standard error bars do not overlap \cite{smith1996measuring,wolfe2002if}.

\section{Correlations with Sevoflurane Concentration}

Correlation between the above indices and sevoflurane concentration was investigated in the emergence period. We chose the time interval where the sevoflurane concentration is monotonically and continuously decreased. The period of spontaneous breathing is defined from the start of adequate spontaneous breath to ROC. We employed the $\PK$ analysis and Spearman rank correlation to analyze performance of indices sampled every 4 seconds. The results were tabulated as weighted averages according to each patient's data length. The bootstrap method was used to calculate the 95\% confidence interval of Spearman correlation based on 10,000 samplings. A value of Spearman correlation closer to $1$ is better since we expect that the indices will increase to correlate with the decrease of sevoflurane concentration.

\section{Algorithm of Serial Prediction Probability (s$\PK$) Analysis}

In this section we detail a variation of the Prediction probability ($\PK$) analysis \cite{smith1996measuring}, referred to as {\it Serial Prediction Probability} (s$\PK$) analysis. s$\PK$ is employed in this study to evaluate the prediction performance of the BIS and ECG-derived continuous indices, including $\NRR(t)$, $\HFMR(t)$, $\LFMR(t)$, $\LHRMR(t)$ and HR. 

We consider $K$ different measurement times, which are the times when we record the anesthetic indicators. Then we determine two special timestamps, one is referred to as the {\it event time}, indicating the happening of the event, for example, the first reaction, and one is referred to as the {\it base time}, which is $T>0$ seconds before the event time. Here $T$ depends on the event we are studying. Collect $N$ subjects and record the measurement times and the event time of each individual. Note that the measurement times and the event time and are different among individuals, and we denote the $k$-th measurement time of the $n$-th subject as $t_{k,n}$, $k=1,\ldots,K$, and the base time as $t_{0,n}$. 
Denoted as $x^k_n$  the anesthetic indicator (e.g. BIS) of the $n$-th subject recorded at time $t_{k,n}$, and denoted as $y^k_n$ the associated outcome (e.g. first reaction) at time $t_{k,n}$. 

To evaluate the prediction power of an anesthetic indicator at the $k$-th measurement time, we take the following two datasets. At the measurement time $t_{k,n}$, we record the sample anesthetic indicators $\{x^k_1,\ldots,x^k_N\}$ and the sample outcomes $\{y^k_1,\ldots,y^k_N\}$, where each pair $(x^k_n,y^k_n)$ is recorded from the $n$-th subject. Denote $\mathcal{Z}_k:=\{(x^k_1,y^k_1),\ldots,(x^k_N,y^k_N)\}$ as the measurement data at the $k$-th measurement time. Moreover, at the base time $t_{0,n}$, we record the sample anesthetic indicators $\{x^{0}_1,\ldots,x^{0}_N\}$ and the sample outcomes $\{y^{0}_1,\ldots,y^{0}_N\}$, where each pair is recorded at the base time. Denote $\mathcal{Z}_0:=\{(x^0_1,y^0_1),\ldots,(x^0_N,y^0_N)\}$ as the data at the base time.

For a given individual, there are five possible relationships between pairs of anesthetic indicator and outcome at the $k$-th measurement time and those at the base time: 
\begin{enumerate}
\item $(x^k_n,y^k_n)$ and $(x^{0}_m,y^{0}_m)$ are concordant if they are rank ordered in the same direction; 
\item $(x^k_n,y^k_n)$ and $(x^{0}_m,y^{0}_m)$ are disconcordant if they are rank ordered in the reverse direction.
\item $(x^k_n,y^k_n)$ and $(x^{0}_m,y^{0}_m)$ tie in $x$ if $x^k_n=x^0_m$ while $y^k_n\neq y^0_m$; 
\item $(x^k_n,y^k_n)$ and $(x^{0}_m,y^{0}_m)$ tie in $y$ if $y^k_n=y^0_m$ while $x^k_n\neq x^0_m$;
\item $(x^k_n,y^k_n)$ and $(x^{0}_m,y^{0}_m)$ tie in both $x$ and $y$ if $x^k_n=x^0_m$ and $y^k_n= y^0_m$.
\end{enumerate} 
Usually the anesthetic indicator is of finer scale and the outcome is of coarser scale. Thus, unlike the usual notion of correlation, in the $\PK$ analysis the tie in $x$ is undesirable but we tolerate the tie in $y$ and the tie in both $x$ and $y$. Based on this notion, we define the {\it serial $\PK$ index} in the following way. 
Denote $P_c(k)$, $P_d(k)$ and $P_{tx}(k)$ the respective probabilities that two pairs of $(x,y)$ independently drawn from $\mathcal{Z}_k$ and $\mathcal{Z}_0$ with replacement are concordant, disconcordant and tie in $x$ at the $k$-th measurement time. The serial $\PK$ index at the $k$-th measurement time is denoted as 
$$
\PK(k)=\frac{P_c(k)+\frac{1}{2}P_{tx}(k)}{P_c(k)+P_d(k)+P_{tx}(k)}.
$$
The $\PK(k)$ value is interpreted in the same way as that in the $\PK$ analysis. A value of one means that the indicator always correctly predicts the observed depth of anesthesia, a value of $0.5$ means that the indicator predicts no better than $50/50$ chance, and a $\PK(k)$ value less than $0.5$ means that the indicator predicts inversely. To be more precise, a zero value also means a perfect prediction but in a reversed direction. 

The estimation of the serial $\PK$ and its standard error was obtained by the Jackknife method. Before the null hypothesis, the serial $\PK$ value less than 0.5 was converted into one minus the serial $\PK$ value. 

\chapter{Results from Clinical Database} 

\label{Chapter6} 

\lhead{Chapter 6. \emph{Results}} 

\section{Visual Information in TF Plane}
Abundant information can be visually read from the tvPS of IHR. Various features were determined to reflect the physiologic dynamics of human body responding to various anesthetic events (Fig. {\ref{fig:dynamicind}}).

\begin{figure}[htbp]
\centering
\includegraphics[width=1\textwidth]{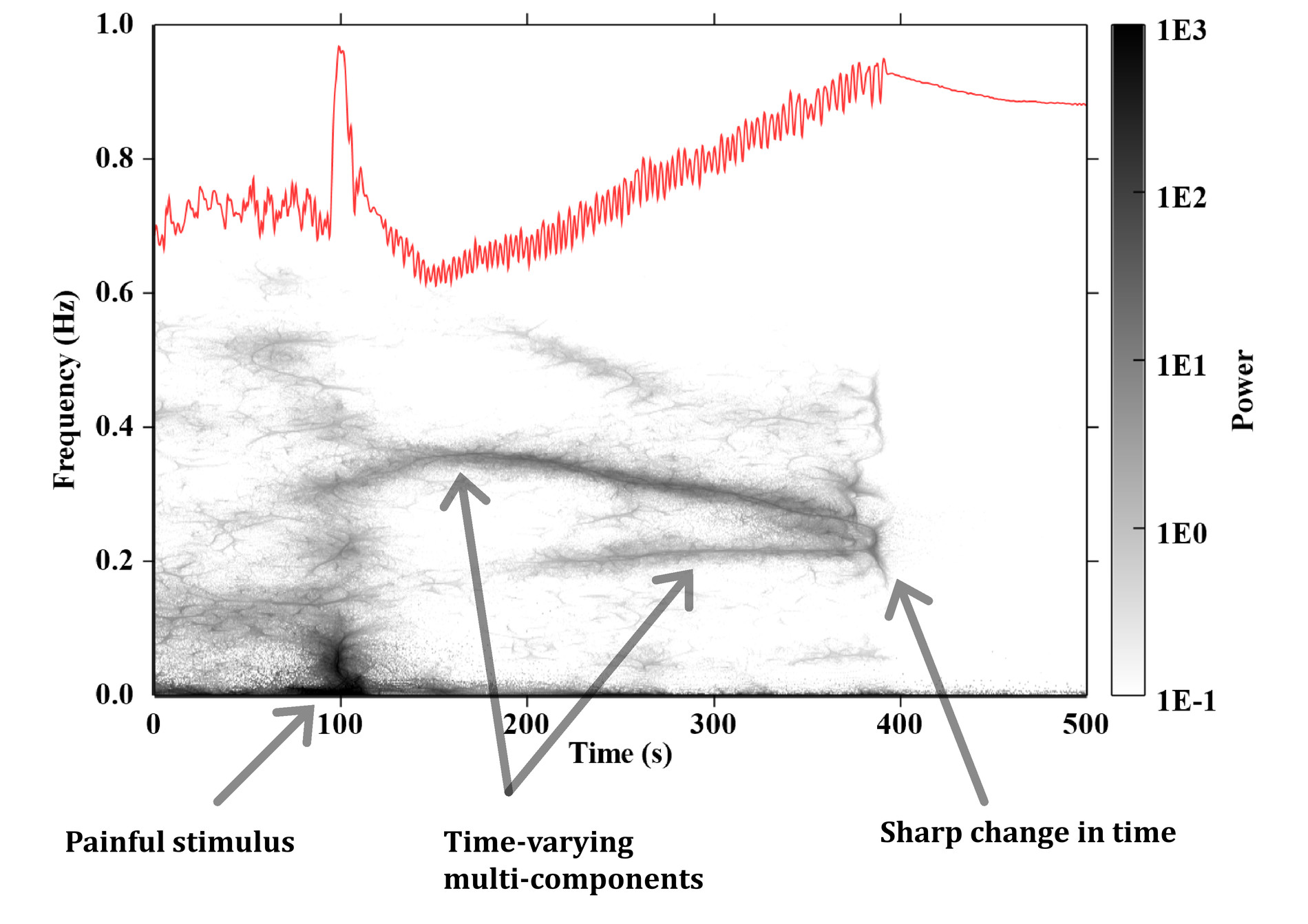}
\caption[Dynamic Feature in Anesthesia]{The time-varying spectrum of R-R peak interval (RRI) recorded during anesthesia showing rich dynamic features of human body in anesthesia. The RRI tracing is superimposed as red line}
\label{fig:dynamicind}
\end{figure}

\begin{figure}[htbp]
\centering
\includegraphics[width=0.9\textwidth]{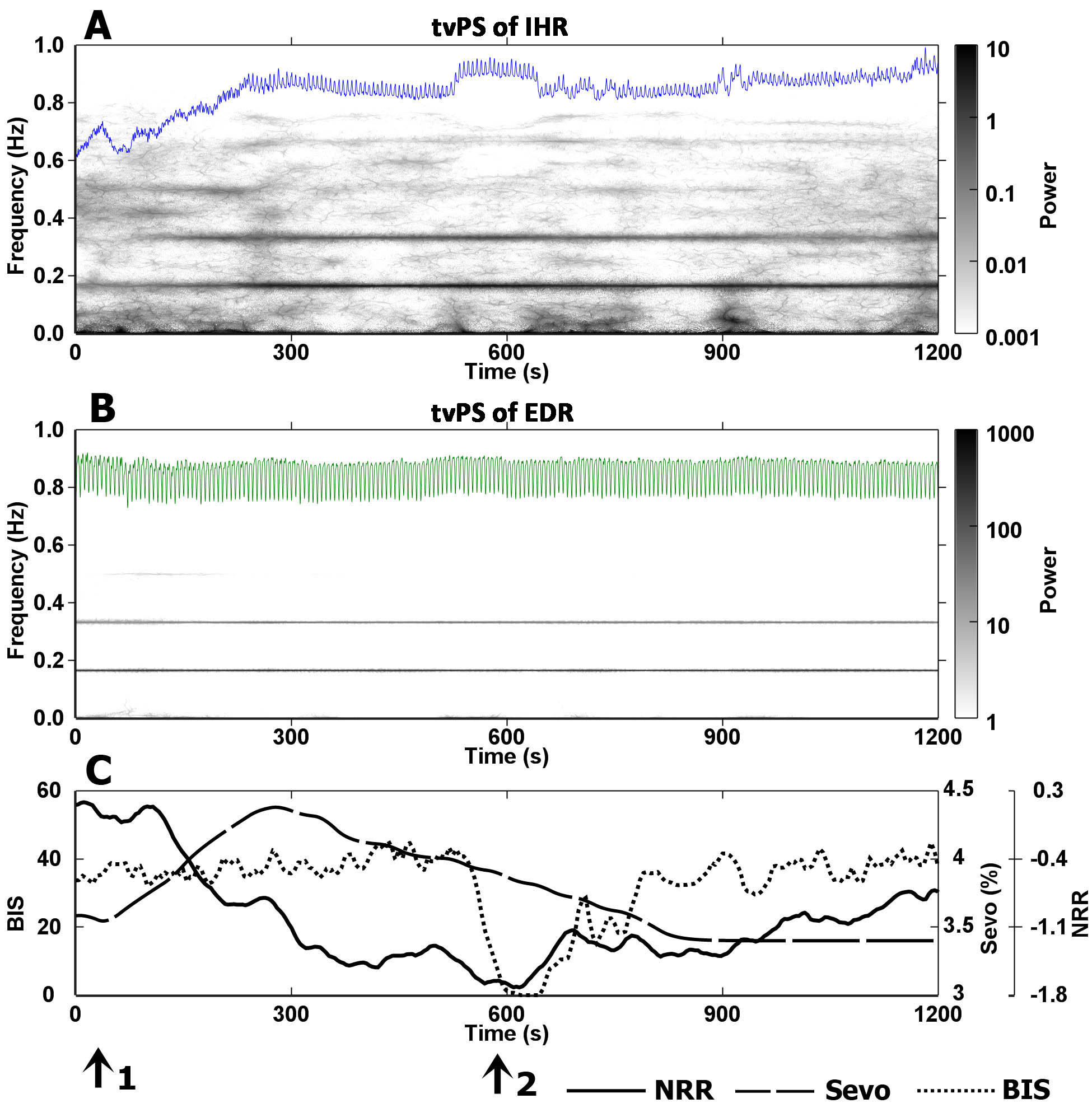}
\caption[Rhythmic-to-Non-rhythmic phenomenon, NRR index and depth of anesthesia varying during controlled ventilation]{A representative data from a patient under anesthesia and controlled ventilation. The 1200 seconds R-R interval (RRI) is superimposed on its time-varying power spectra (A) as a blue solid line. The corresponding electrocardiography derived respiration (EDR) is superimposed on its time-varying power spectra (B) as a green line. Panel C shows simultaneous recorded Bispectral index (BIS), effect-site sevoflurane concentration (Sevo) and non-rhythmic to rhythmic ratio (NRR). Arrow 1: inadequate level of anesthesia was noted and the sevoflurane concentration was raised. Sevoflurane concentration reached top level after 300-second. Arrow 2: both BIS index and NRR reached minimum. There is a dominant straight line around 0.167 Hz in the panel A, whose fundamental frequency is the same as the rate of the ventilator, representing the entrainment of RSA by the controlled ventilation. The time-varying power spectra of RRI showed rhythmic-to-non-rhythmic transition, corresponding to varying anesthetic depth, whereas the time-varying power spectra of EDR was almost unchanged under controlled ventilation. }
\label{fig:nrrmv}
\end{figure}

It is demonstrated (Fig. {\ref{fig:nrrmv}},\ref{fig:nrrsb}) that the tvPS is more concentrated on deeper anesthetic levels and more scattered on lighter anesthetic levels. This then created the hypothesis on trying to understand the ``rhythmic-to-non-rhythmic" phenomenon. This phenomenon is consistently seen in most study cases. During spontaneous breathing (Fig.\ref{fig:nrrsb}), the dominant curves can be seen in the first portion (usually on the left hand side) of the tvPS graph. Meanwhile, less dominant curves can be identified in the second portion (usually the right hand side). The dominant curve represents the typical rhythmicity feature. It is also important to observe that the tvPS becomes more non-rhythmic prior to the appearance of first reaction. (Fig. \ref{fig:nrrsb})

\begin{figure}[htbp]
\centering
\includegraphics[width=0.9\textwidth]{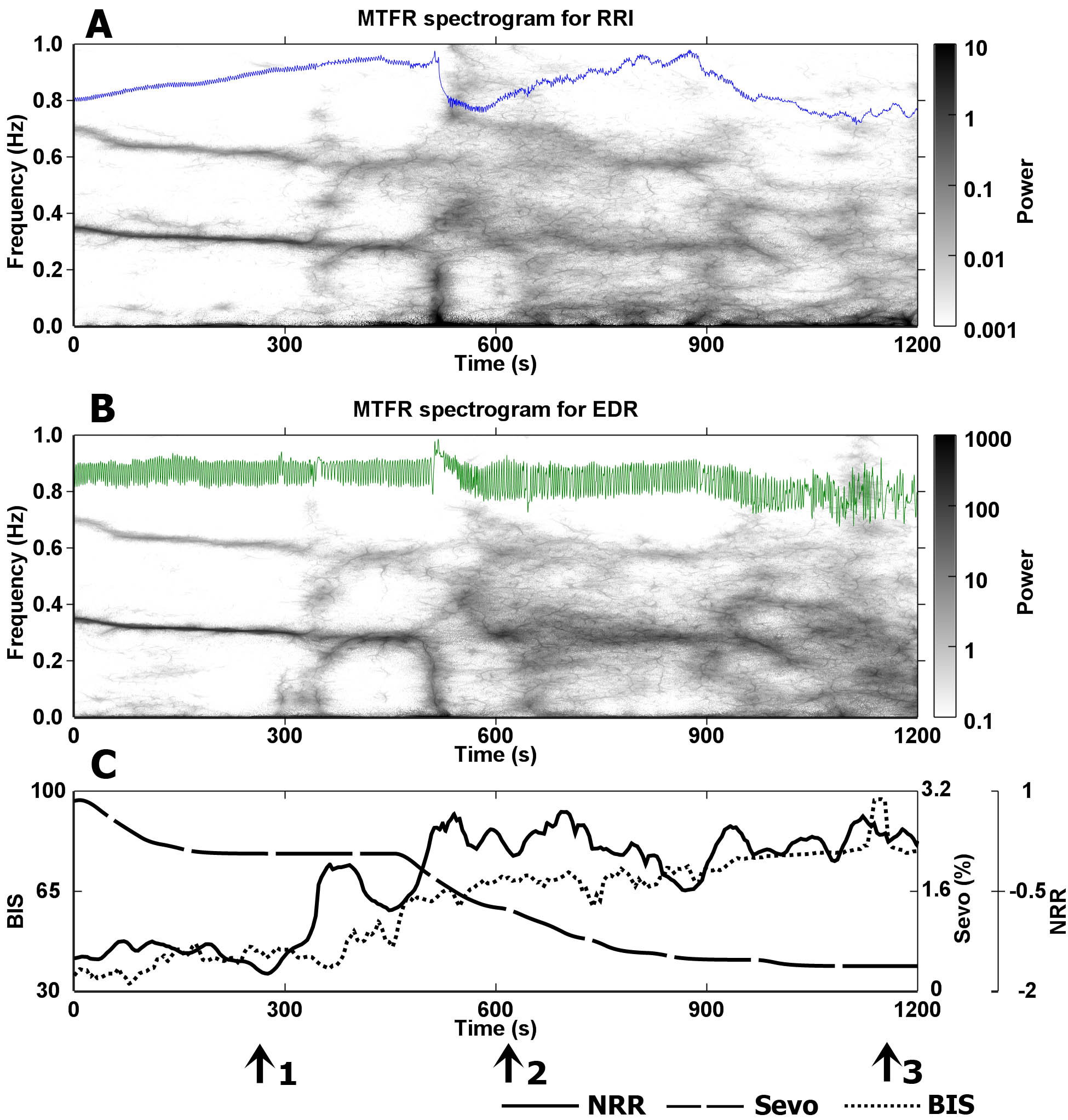}
\caption[Rhythmic-to-Non-rhythmic phenomenon, NRR index and depth of anesthesia during spontaneous breathing]{A representative data obtained from the same patient under spontaneous breathing in emergence period. The 1200-second R-R interval (RRI) is superimposed on its time-varying power spectra (A) as a green solid line. The corresponding electrocardiography derived respiration (EDR) is superimposed on its time-varying power spectra (B) as a blue solid line. Panel C shows simultaneous recorded Bispectral index (BIS), effect-site sevoflurane concentration (Sevo) and non-rhythmic to rhythmic ratio (NRR). Arrow 1: endotracheal suction and extubation, Arrow 2: first reaction, Arrow 3: regain of consciousness. Under spontaneous breath, RRI and EDR are similar in high frequency region of spectrograms. The rhythmic component and its harmonics are time-varying in both time-varying power spectra of RRI and EDR. With the overall decrease of sevoflurane, both RRI and EDR exhibited a rhythmic to non-rhythmic transition. The RRI and EDR became non-rhythmic before the first reaction.(From ref.\cite{Lin_Wu_Tsao_Yien_Hseu:2013})}
\label{fig:nrrsb}
\end{figure}

\begin{figure}[htbp]
\centering
\includegraphics[width=1\textwidth]{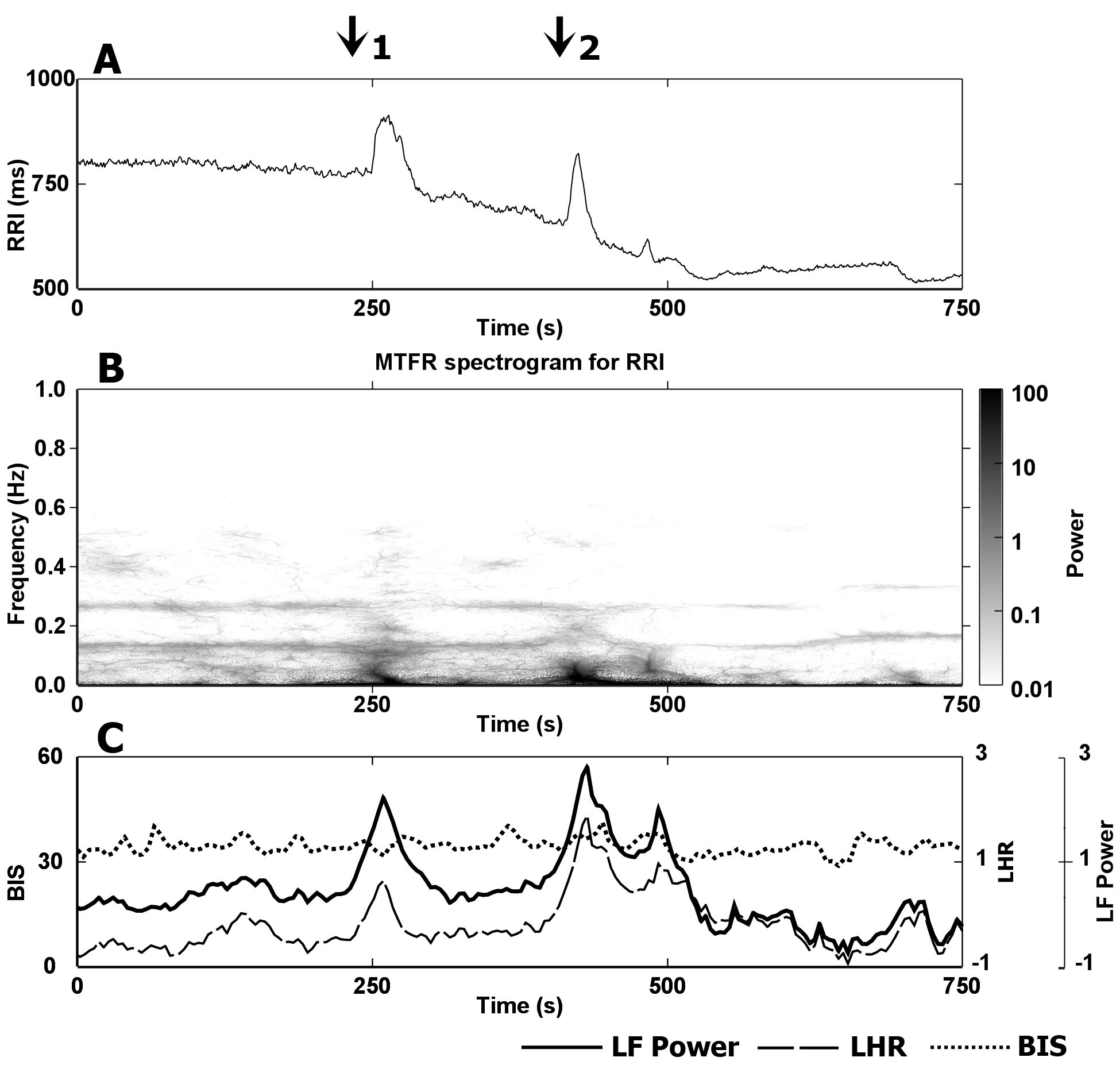}
\caption[Skin Incision during Anesthesia]{A representative data from a patient under anesthesia receiving skin incision for her surgery. The 750-second R-R interval (RRI) is superimposed on its time-varying power spectra (A) as a green solid line. The corresponding electrocardiography derived respiration (EDR) is superimposed on its time-varying power spectra (B) as a blue solid line. Panel C shows simultaneous recorded Bispectral index (BIS), effect-site sevoflurane concentration (Sevo) and non-rhythmic to rhythmic ratio (NRR).}
\label{fig:skinincision}
\end{figure}

Painful surgical stimulations during anesthesia are referred to as \textit{noxious stimulation} because an unconscious patient under anesthesia cannot report pain sensation. Two of the most intense noxious stimulationthat occur in anesthesia and surgery are \textit{endotracheal intubation} and \textit{skin incision}. Clinical anesthesiologist uses the increase of heart rate and blood pressure as indicators for the level of noxious stimulation. In my study, we have found the visual information of tvPS that reveals the influence of noxious information where the low frequency (LF) power increases immediately after noxious stimulation, and wanes soon. The ``\textit{LF surge phenomenon}" appeared in all study cases. Although the main goal of my study is on investigating NRR index, it can be seen that the potential value of LF power as an ``noxious index" deserves to be investigated too.

These visual findings are quantified by NRR and HRV indices, such as tvHF, tvLF and tvLHR for subsequent analyses.

\begin{figure}[htbp]
\centering
\includegraphics[width=1\textwidth]{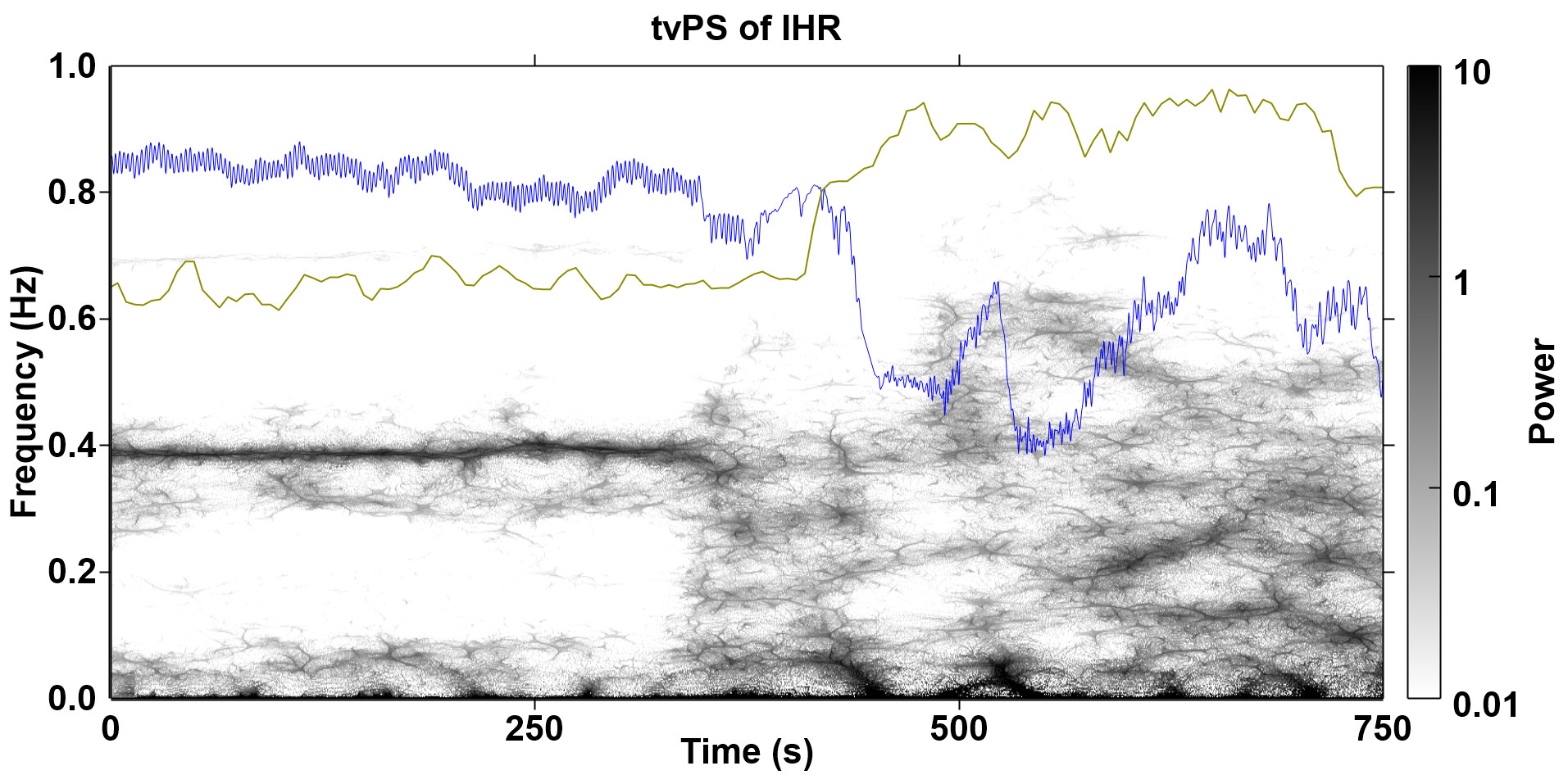}
\caption[Rhythmic-to-Non-rhythmic Phenomenon Demo no.1]{tvPS of IHR during emergence period from the last three consecutive cases (no.1) showing rhythmic-to-non-rhythmic phenomenon. BIS index (golden yellow tracing) and IHR (blue tracing) are superimposed.}
\label{fig:rnrdemo1}
\end{figure}

\begin{figure}[htbp]
\centering
\includegraphics[width=1\textwidth]{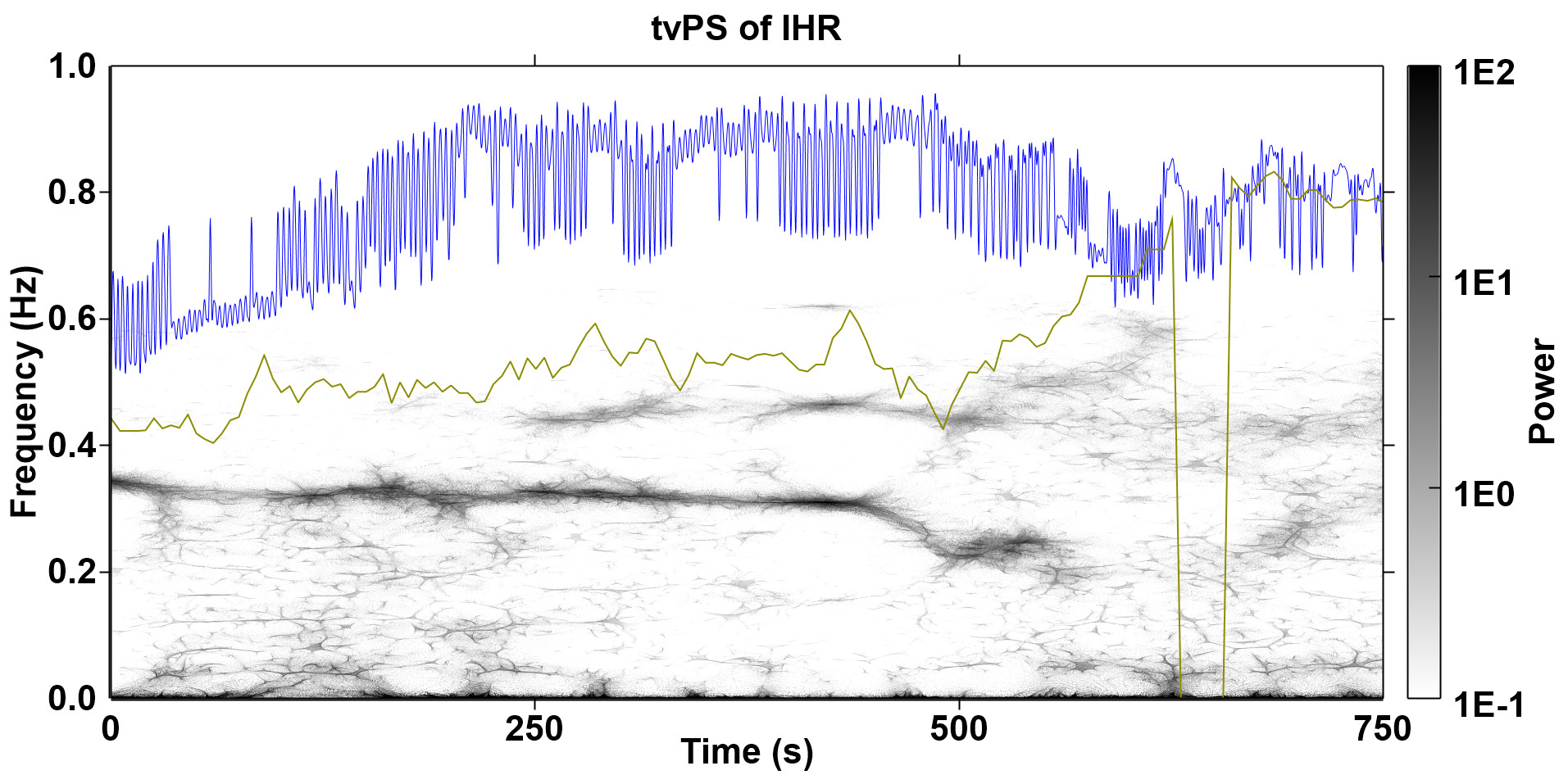}
\caption[Rhythmic-to-Non-rhythmic phenomenon Demo no.2]{tvPS of IHR during emergence period from the last three consecutive cases (no.2) showing rhythmic-to-non-rhythmic phenomenon. BIS index (golden yellow tracing) and IHR (blue tracing) are superimposed.}
\label{fig:rnrdemo2}
\end{figure}

\begin{figure}[htbp]
\centering
\includegraphics[width=1\textwidth]{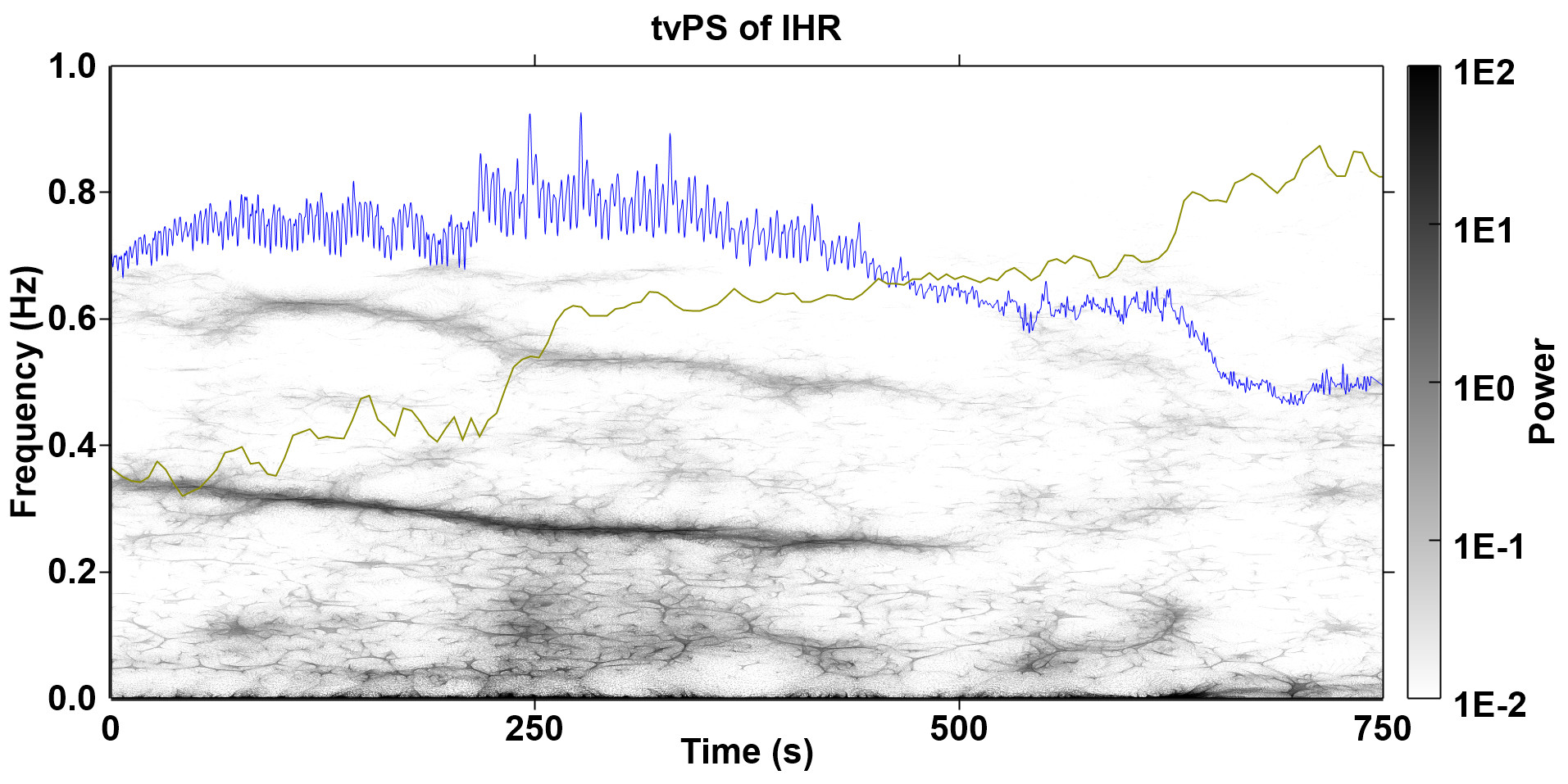}
\caption[Rhythmic-to-Non-rhythmic phenomenon Demo no.3]{tvPS of IHR during emergence period from the last three consecutive cases (no.3) showing rhythmic-to-non-rhythmic phenomenon. BIS index (golden yellow tracing) and IHR (blue tracing) are superimposed.}
\label{fig:rnrdemo3}
\end{figure}


\section{Multiple Component Phenomenon}
Occasionally, multiple independent oscillatory components hidden in IHR can be seen from the tvPS. To further expand on this statement, the term \textit{independent} is defined as the frequency of one component of which the modulation is unrelated to other components. Therefore, this is definitely not the harmonics (Fig.\ref{fig:3comsstm}, \ref{fig:multicomp2}, \ref{fig:rnrmulti}, \ref{fig:rnrmulti2}). The ``multiple component phenomenon" are an interesting finding, which demonstrates the dynamic physiology of human body undergoing anesthesia.

\begin{figure}[htbp]
\centering
\includegraphics[width=1\textwidth]{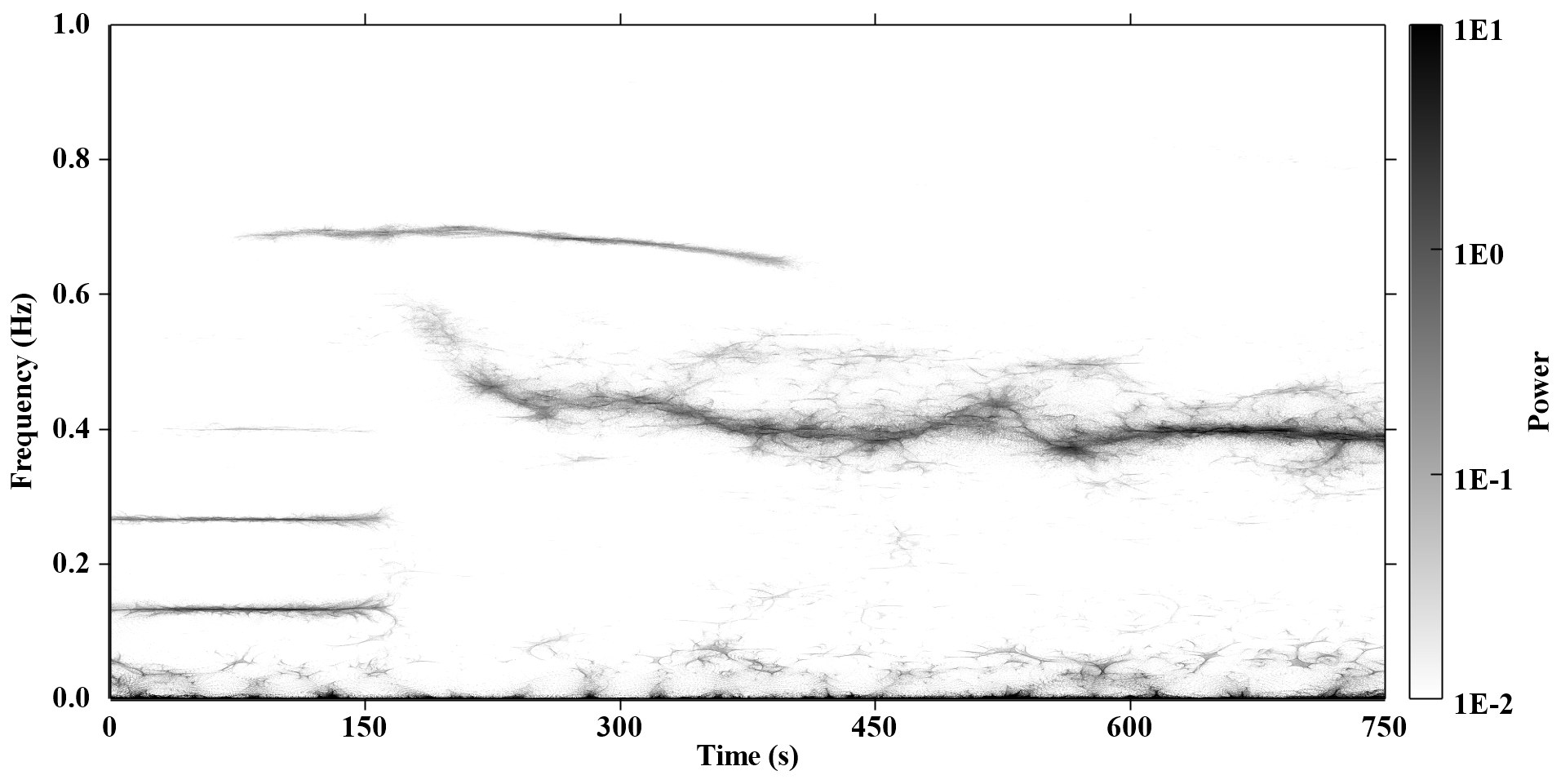}
\caption[Multiple Components in IHR during Anesthesia]{The time-varying power spectrum of instantaneous heart rate (IHR) recorded during anesthesia showing three independent oscillatory components, revealing rich dynamic features of human body in anesthesia. }
\label{fig:3comsstm}
\end{figure}

\begin{figure}[htbp]
\centering
\includegraphics[width=1\textwidth]{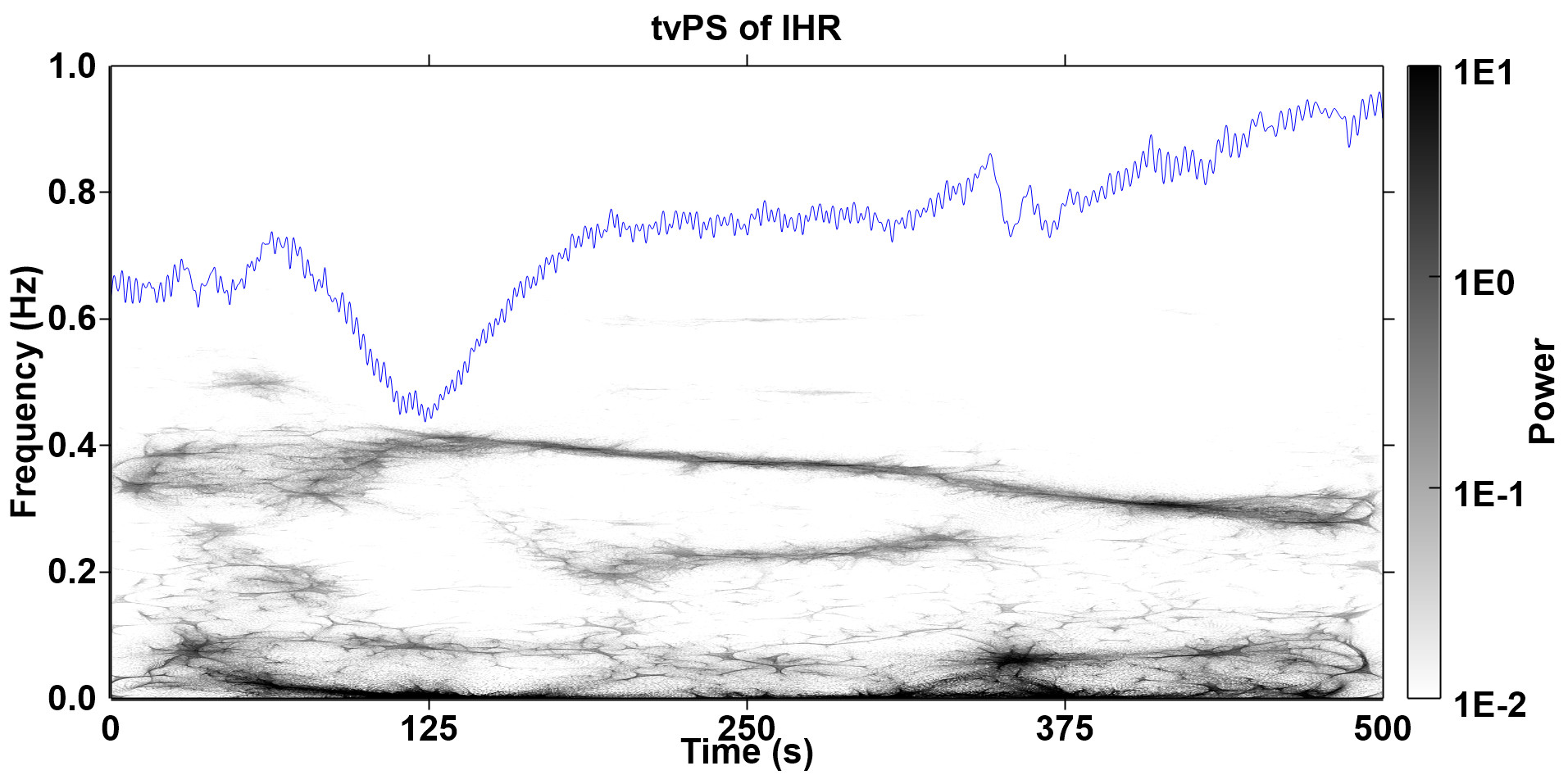}
\caption[Two Components in IHR during Anesthesia]{The IHR (blue tracing) was recorded during spontaneous breathing. The second component appeared shortly (200--300 s).}
\label{fig:multicomp2}
\end{figure}

\begin{figure}[htbp]
\centering
\includegraphics[width=1\textwidth]{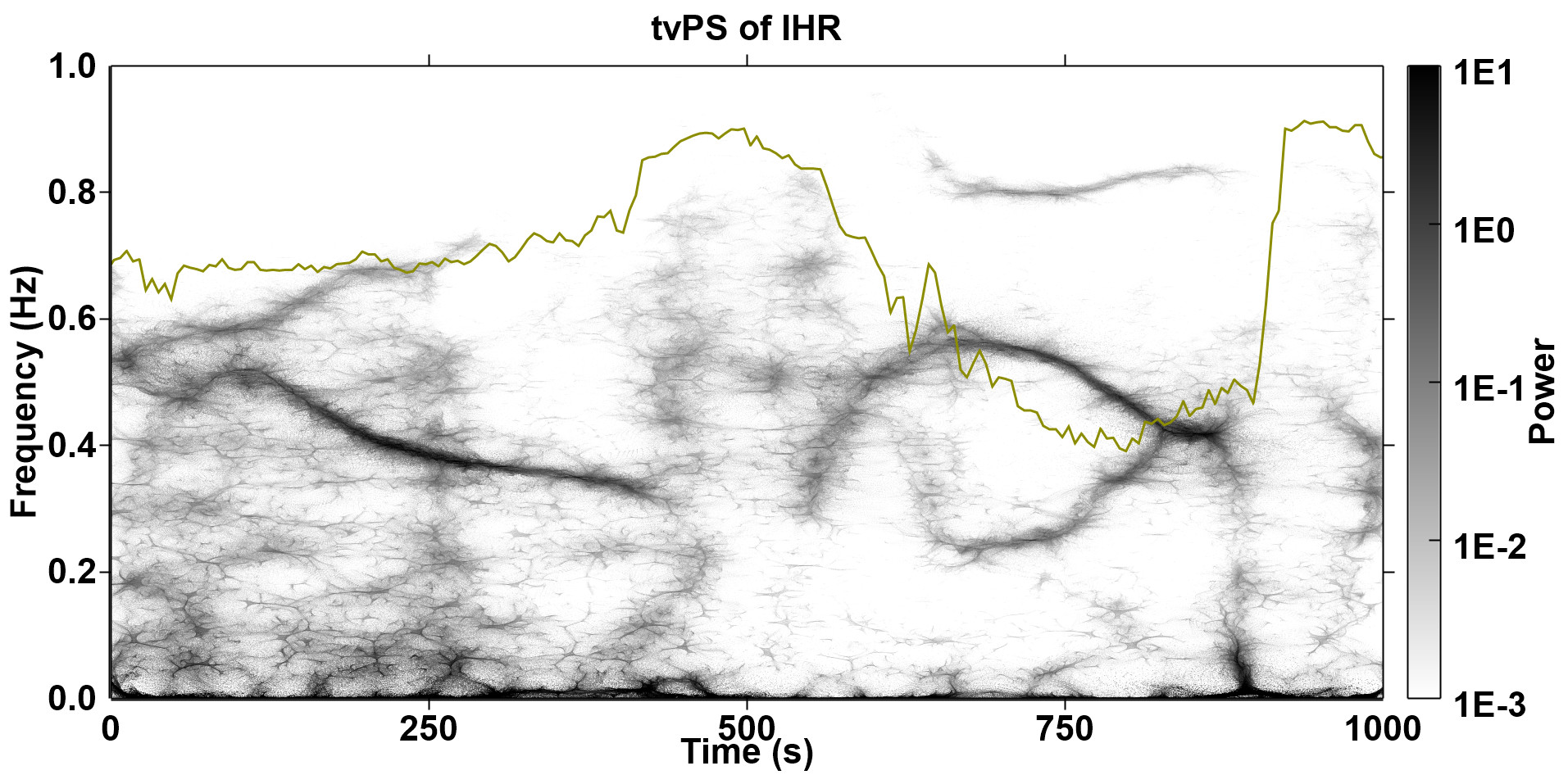}
\caption[Multiple components and rhythmic-to-non-rhythmic phenomenon]{The IHR during spontaneous breathing showed ``rhythmic-to-non-rhythmic" changes with BIS index (golden yellow tracing), accompanied with multiple components during ``rhythmic" period.}
\label{fig:rnrmulti}
\end{figure}

\begin{figure}[htbp]
\centering
\includegraphics[width=1\textwidth]{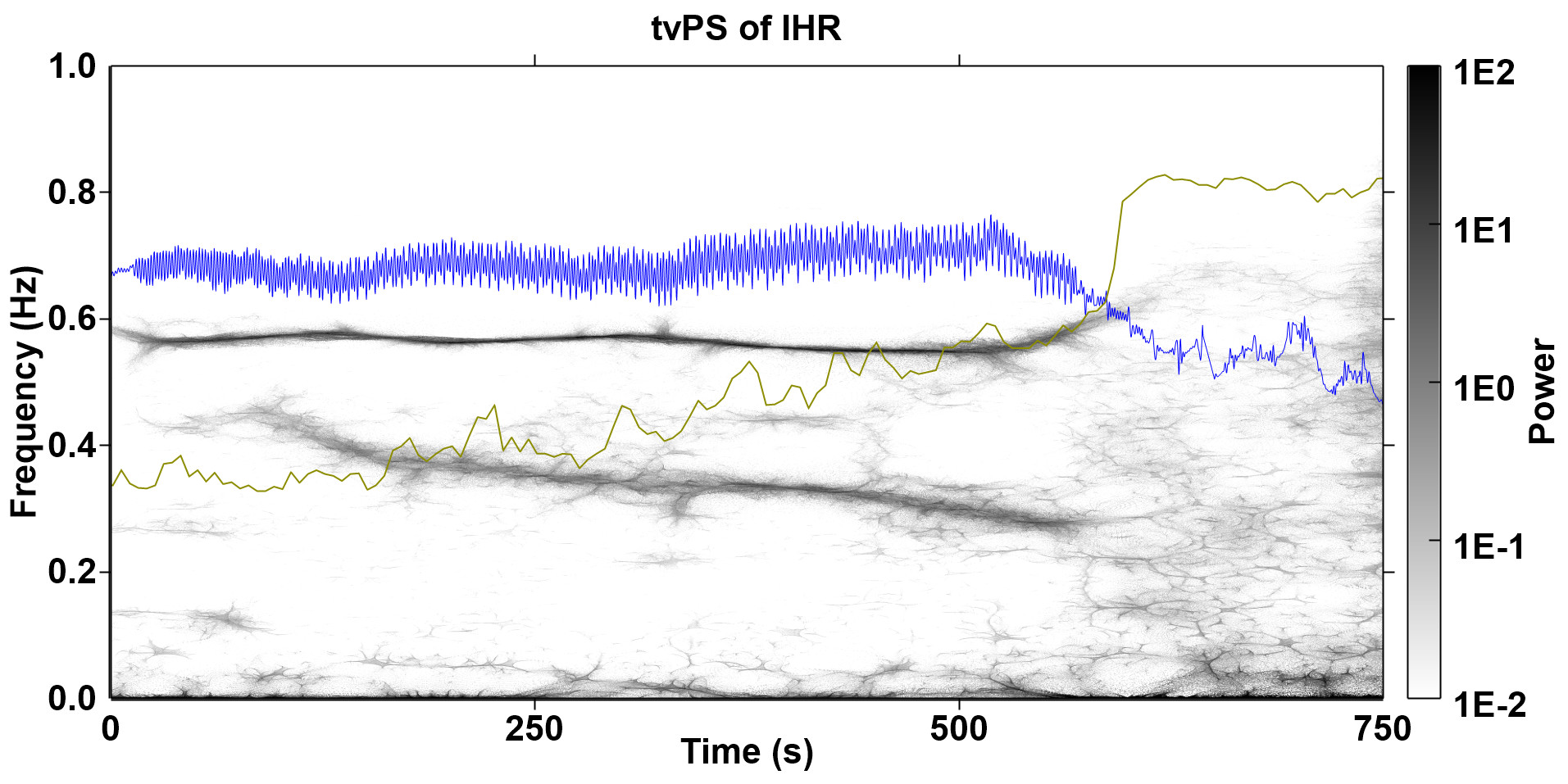}
\caption[Multiple components and rhythmic-to-non-rhythmic phenomenon no.2]{Another IHR data (blue tracing) during spontaneous breathing showed ``rhythmic-to-non-rhythmic" changes with BIS index (golden yellow tracing), accompanied with multiple components during ``rhythmic" period.}
\label{fig:rnrmulti2}
\end{figure}

Apparently, it is nearly impossible to identify the appearance of \textit{multiple component} from raw IHR waveform using naked eye. The tvPS is good at presenting ``multiple component phenomenon", but it is a difficult task for classical power spectrum. The ``multiple component phenomenon" appeared in one thirds of the study cases. The moment that it appeared and disappeared seems random and unpredictable. Even so, this phenomenon is related to specific anesthetic events like only appearing when the patient is undergoing anesthesia. Although mechanism underlying this phenomenon is beyond the goal of my present study, it still presents an interesting research topic for the future.

\section{Ability to Predict Anesthetic Events}
In the serial $\PK$ analysis for the first reaction {(Fig. {\ref{fig:spkfr}})}, the NRR is ahead of BIS ($p<0.05$ for {20} seconds). The NRR was able to predict the first reaction 30 seconds in advance ($\PK >0.90$). At the instance of first reaction (0 seconds), both the NRR and BIS (both PK maxima $>0.95$) were significantly better than other parameters. The time-varying HRV indices and HR ($\PK <0.83$) provides significantly worse results than the NRR.
 
In the serial $\PK$ analysis for skin incision (Fig. {\ref{fig:skinincision}}), tvLF reaches maximum ($\PK >0.95$) roughly at 30 seconds after skin incision. It reflects skin incision best, followed by tvLHR ($\PK <0.85$). The tvLF is significantly better than BIS ($\PK <0.55$) and NRR ($\PK <0.65$). 

In the serial $\PK$ analysis for LOC (Fig. {\ref{fig:spkloc}}), BIS reflects perfectly ($\PK=1$) 50 seconds after LOC. LOC is also associated with a decrease in tvLF, tvHF and HR. Nowever, NRR does not reflects LOC well. In the serial $\PK$ analysis for ROC (Fig. {\ref{fig:spkroc}}), BIS is the best index, and surpasses NRR, HR and the time-varying HRV indices significantly. 

Due to the inadequate data quality, 1 patient has been eliminated from our data in regards to the serial Pk analysis of the initial reaction. For our study, when we analyzed the serial Pk of skin incision, LOC and ROC, we also eliminated 3, 5, and 4 patients respectively.

\begin{figure}[htbp]
\centering
\includegraphics[width=1\textwidth, angle=0]{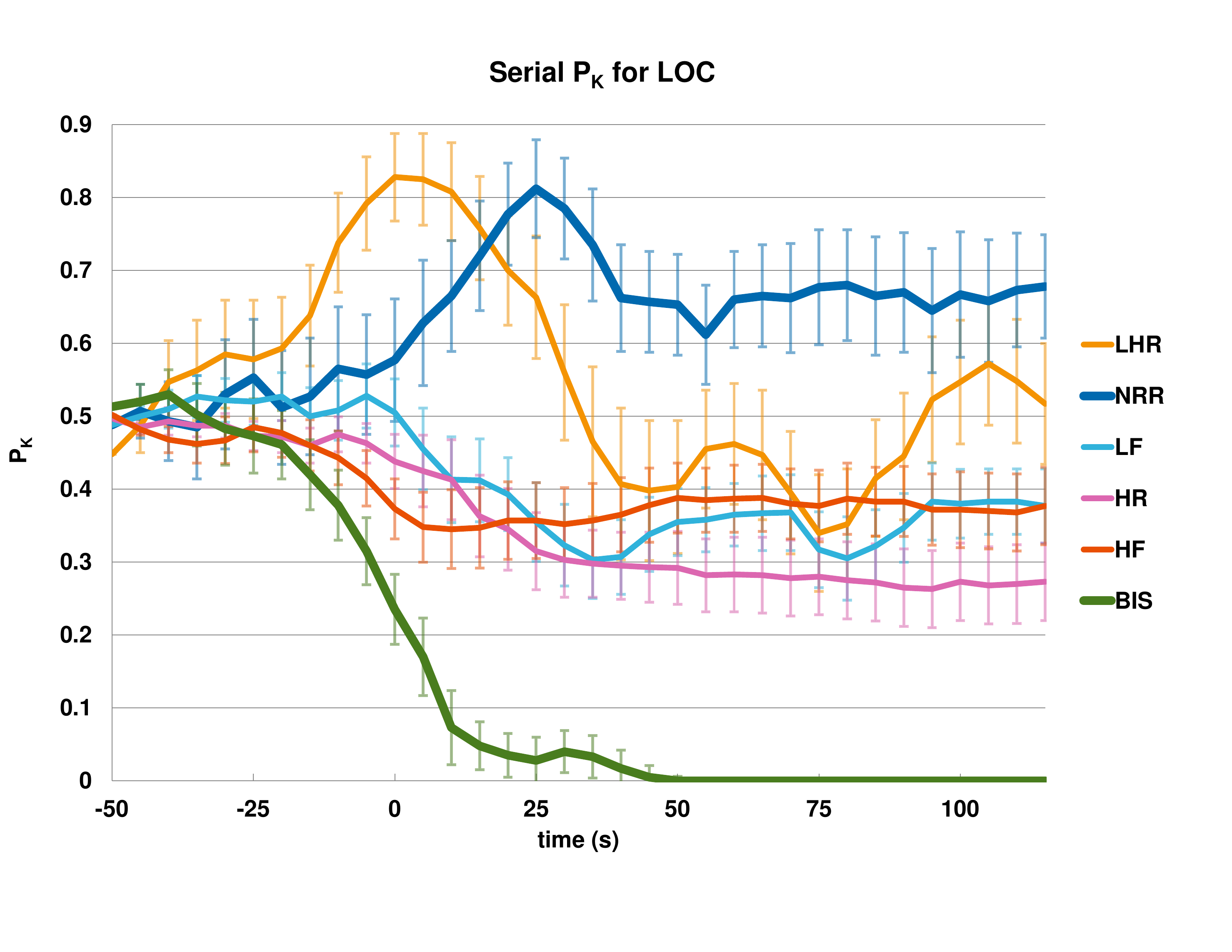}
\caption[Serial $\PK$ Analysis: Loss of Consciousness]{Tracings and their faded error bars about the prediction probability ($\PK$) values of the parameters and their standard errors in serial $\PK$ analysis for  loss of consciousness (LOC). The baseline is one minute before LOC. NRR = non-rhythmic to rhythmic ratio; HF = high frequency power; LF = low frequency power; LHR = LF to HF ratio; BIS = Bispectral Index; HR = heart rate. (From ref.\cite{Lin_Wu_Tsao_Yien_Hseu:2013}) }
\label{fig:spkloc}
\end{figure}

\begin{figure}[htbp]
\centering
\includegraphics[width=1\textwidth, angle=0]{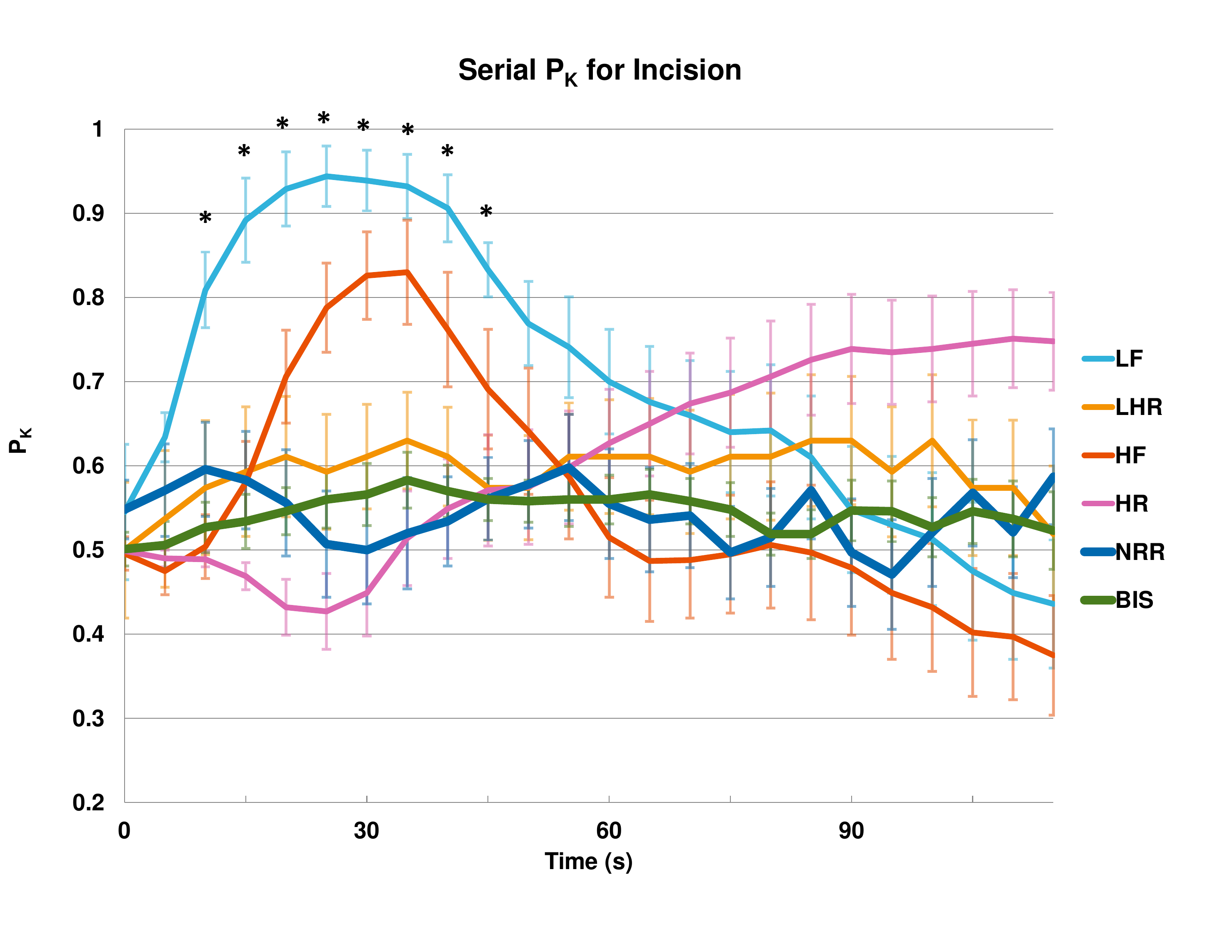}
\caption[Serial $\PK$ Analysis: Skin Incision]{Tracings and their faded error bars about the prediction probability ($\PK$) values of the parameters and their standard errors in serial $\PK$ analysis for skin incision. The baseline is 5 seconds before skin incision. NRR = non-rhythmic to rhythmic ratio; HF = high frequency power; LF = low frequency power; LHR = LF to HF ratio; BIS = Bispectral Index; HR = heart rate. *$ P< 0.05$, LF versus BIS and LF versus NRR. (From ref.\cite{Lin_Wu_Tsao_Yien_Hseu:2013})}
\label{fig:spksi}
\end{figure}

\begin{figure}[htbp]
\centering
\includegraphics[width=1\textwidth, angle=0]{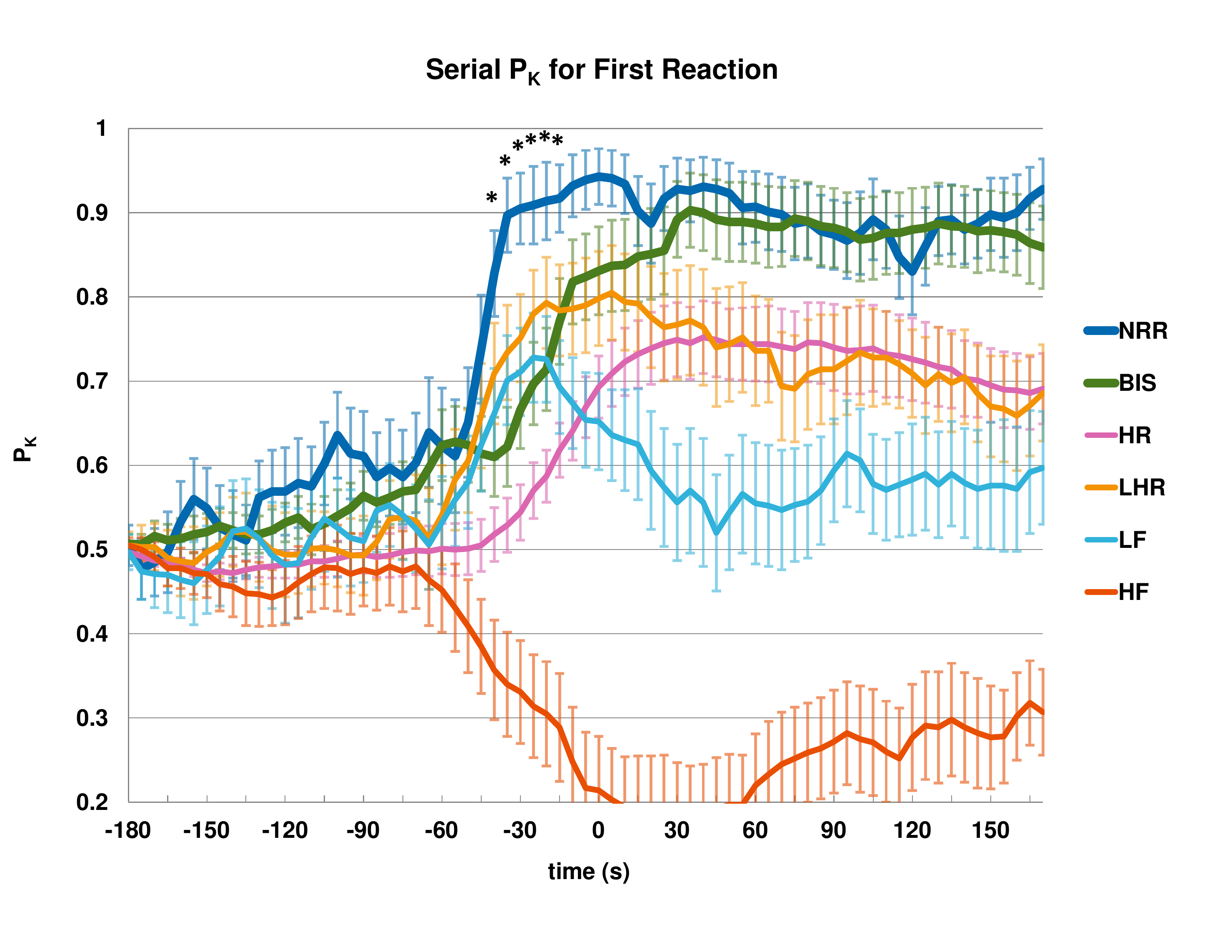}
\caption[Serial $\PK$ Analysis: First Reaction]{Tracings and their faded error bars about the prediction probability ($\PK$) values of the parameters and their standard errors in serial $\PK$ analysis for first motor movement reaction. The baseline is 3 minutes before first motor movement reaction. NRR = non-rhythmic to rhythmic ratio; HF = high frequency power; LF = low frequency power; LHR = LF to HF ratio; BIS = Bispectral Index; HR = heart rate. * $P< 0.05$, NRR versus BIS. (From ref.\cite{Lin_Wu_Tsao_Yien_Hseu:2013})}
\label{fig:spkfr}
\end{figure}

\begin{figure}[htbp]
\centering
\includegraphics[width=1\textwidth, angle=0]{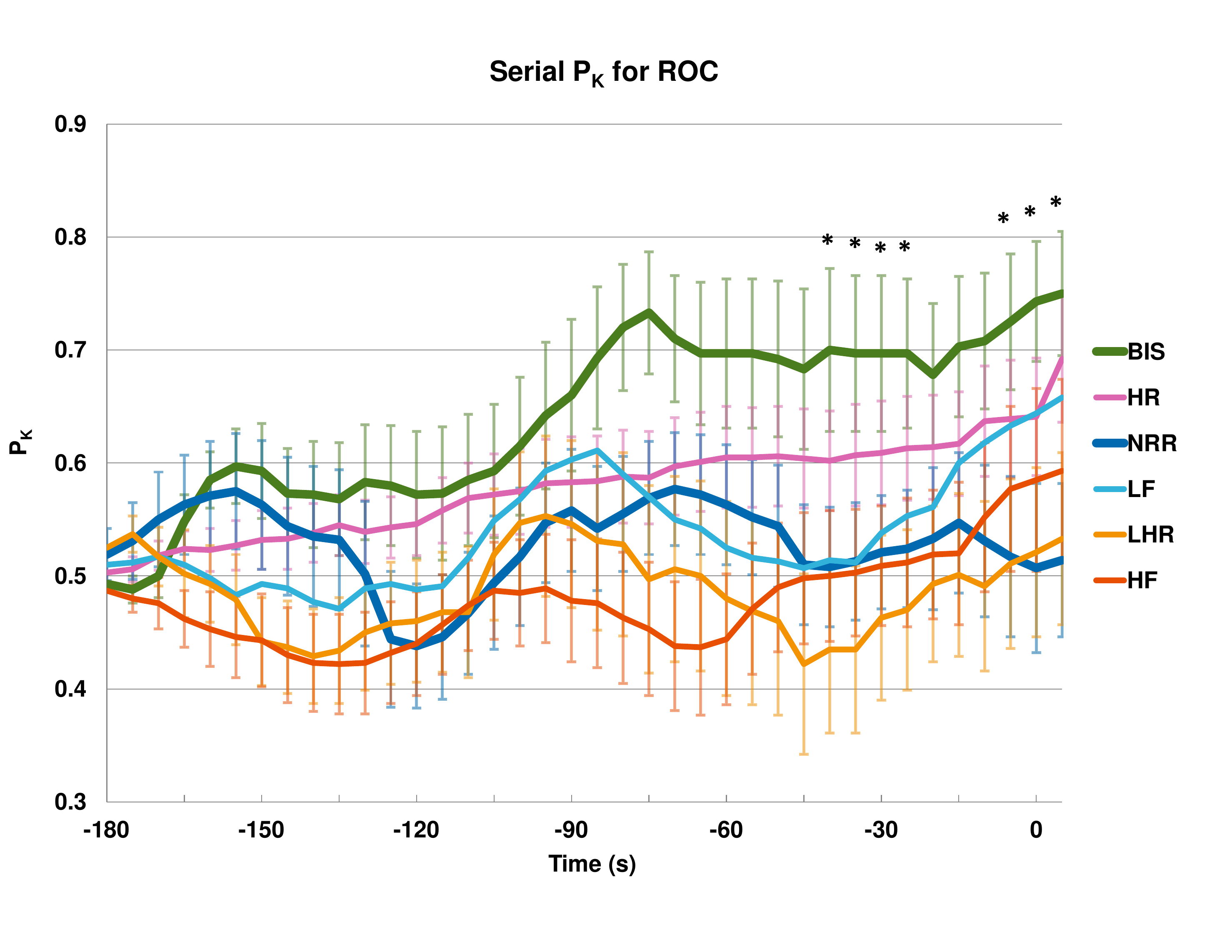}
\caption[Serial $\PK$ Analysis: Return of Consciousness]{Tracings and their faded error bars about the prediction probability ($\PK$) values of the parameters and their standard errors in serial $\PK$ analysis for return of consciousness (ROC). The baseline is 3 minutes before ROC. NRR = non-rhythmic to rhythmic ratio; HF = high frequency power; LF = low frequency power; LHR = LF to HF ratio; BIS = Bispectral Index; HR = heart rate. * $P< 0.05$, BIS versus NRR. (From ref.\cite{Lin_Wu_Tsao_Yien_Hseu:2013})}
\label{fig:spkroc}
\end{figure}

\begin{figure}[htbp]
\centering
\includegraphics[width=1\textwidth, angle=0]{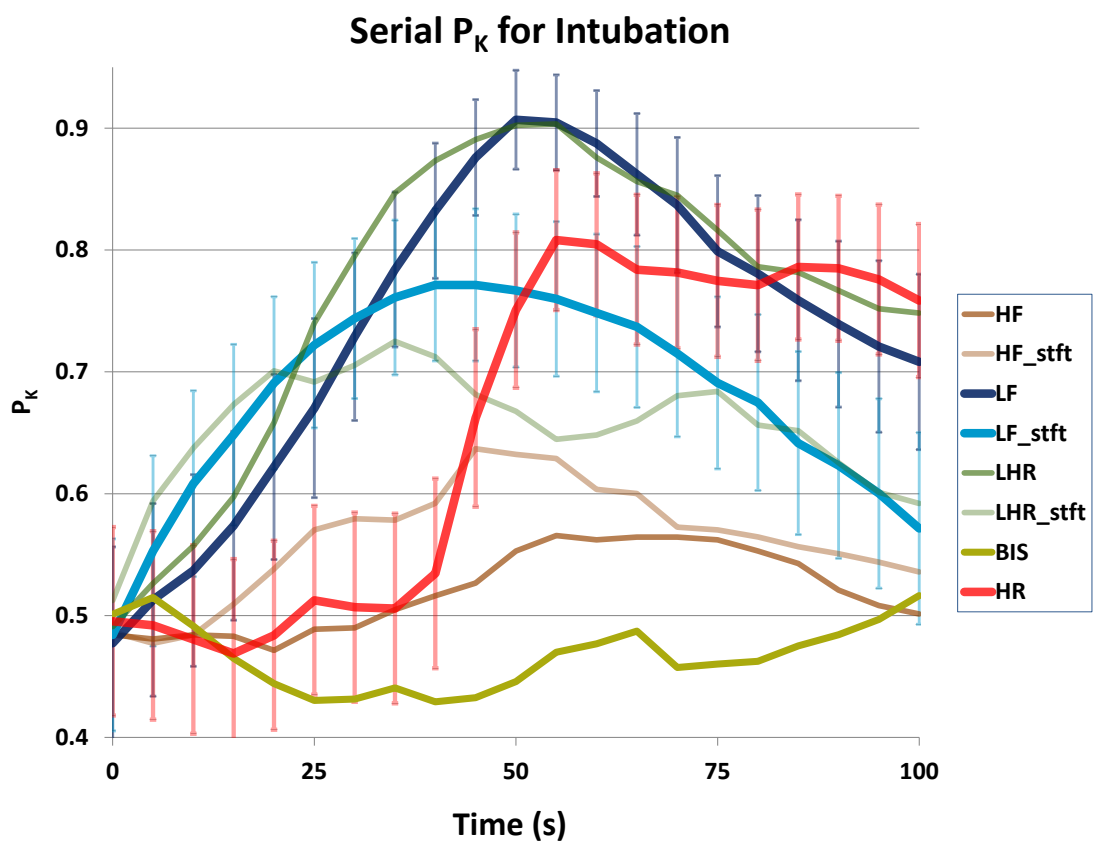}
\caption[Serial $\PK$ Analysis: Endotracheal Intubation]{Tracings and their faded error bars about the prediction probability ($\PK$) values of the parameters and their standard errors in serial $\PK$ analysis for endotracheal intubation. The baseline is 5 seconds before endotracheal intubation. HF = high frequency power calculated by multitaper reassigned spectrogram; LF = low frequency power calculated by multitaper reassigned spectrogram; LHR = LF to HF ratio calculated by multitaper reassigned spectrogram; HF\_stft = high frequency power calculated by STFT spectrogram; LF\_stft = low frequency power calculated by STFT spectrogram; BIS = Bispectral Index; HR = heart rate. }
\label{fig:spkint}
\end{figure}


\begin{table}[h]\scriptsize
\caption[Representative Value of Indices]{Representative values of the Indices during important anesthetic events. Values are presented as median values (lower and upper quartiles). HF, LF, and LHR are expressed in common logarithm (10 base) of millisecond square. Heart rate is expressed in beat per minute. BIS = Bispectral Index; NRR = non-rhythmic to rhythmic ratio; HF = high frequency power; LF = low frequency power; LHR = LF to HF ratio; LOC = loss of consciousness; ROC = return of consciousness; T0= 60 s before LOC; T1= 60 s after LOC; T2= 5 s before incision; T3= 30s after incision; T4= 180 s before first reaction; T5= 5 s after first reaction; T6= 180 s before ROC; T7= 5 s after ROC. (From ref.\cite{Lin_Wu_Tsao_Yien_Hseu:2013})}
\renewcommand{\arraystretch}{2.3}
 \begin{tabular}{  p{0.7cm}  p{1.1cm}  p{1.1cm}  p{1.1cm}  p{1.1cm}  p{1.1cm}  p{1.1cm}  p{1.1cm}  p{1.1cm} }
\toprule
    & \multicolumn{2}{c}{LOC}                    & \multicolumn{2}{c}{Skin incision}        & \multicolumn{2}{c}{First reaction}         & \multicolumn{2}{c}{ROC}                    \\ \midrule
    & T0                    & T1                   & T2                   & T3                  & T4                    & T5                   & T6                   & T7                    \\
\hline
BIS & 97 (94,98)           & 54 (42, 68)         & 39 (36,44)          & 41 (37,44)         & 61 (70, 47)          & 73 (67,80)          & 80 (77, 83)         & 86 (81,94)           \\
NRR & -0.26 (-0.42, 0.21)  & 0.09 (-0.09, 0.31)  & -0.14 (-0.51, 0.02) & 0.04 (-0.19, 0.21) & -0.80 (-1.1, -0.38)  & 0.47 (0.29, 0.61)   & 0.36 (0.12, 0.51)   & 0.43 (0.19, 0.62)    \\
HF  & 2.37 (2.05, 2.76)    & 1.50 (0.97, 1.66)   & 1.21 (0.94, 1.55)   & 1.42 (1.08, 1.81)  & 1.44 (1.09, 1.83)    & 1.07 (0.78, 1.23)   & 1.11 (0.91, 1.40)   & 1.25 (0.93, 1.60)    \\
LF  & 1.84 (1.50, 2.27)    & 1.37 (1.01, 1.60)   & 0.84 (0.46, 0.94)   & 1.59 (1.25, 2.25)  & 0.09 (-0.37, 0.56)   & 0.79 (0.75, 1.31)   & 0.68 (0.01, 1.27)   & 0.91 (0.37, 1.26)    \\
LHR & -0.42 (-0.84, -0.27) & -0.03 (-0.29, 0.19) & -0.50 (0.46, 0.84)  & 0.24 (0.01, 0.50)  & -1.33 (-1.67, -0.66) & -0.25 (-0.73, 0.06) & -0.44 (-0.79, 0.10) & -0.34 (-0.66, -0.12) \\
HR  & 74 (65, 91)          & 64 (58, 72)         & 65 (58, 72)         & 70 (60, 76)        & 71 (62, 80)          & 79 (69, 85)         & 76 (69, 83)         & 84 (75, 91)          \\ \bottomrule
\end{tabular}
\label{table:represent}
\end{table}

\section{Correlation with Sevoflurane Concentration}

\begin{table} 
\caption[Correlation to Estimated Sevoflurane Effect-site Concentration]{Prediction Probability and Spearman Rank Correlation for Indices vs. Estimated Sevoflurane Effect-site Concentration in Controlled Ventilation and Spontaneous Breathing. $\star$ $p<0.05$ between non-rhythmic to rhythmic ratio (NRR) and other indices. Indices with prediction probability ($\PK$) value smaller than 0.5 was corrected by $1-\PK$ before comparison.
BIS = Bispectral Index; $\HFMR$ = high frequency power; $\LFMR$ = low frequency power; $\LHRMR$ = LF to HF ratio; HR = heart rate. (From ref.\cite{Lin_Wu_Tsao_Yien_Hseu:2013})
}
\centering
\renewcommand{\arraystretch}{1.7}
\begin{tabular}{ p{2cm}  p{4cm}  p{4cm} }

\hline
  \multicolumn{3}{c}{Controlled Ventilation}\\
\hline
  & $\PK$ (SE) & Spearman rank correlation (95\% CI) \\
\hline
BIS &0.716 (0.020)& -0.575 (-0.661, -0.476)\\
NRR &0.670 (0.025)& -0.467 (-0.583, -0.337)\\
$\HFMR$& 0.479 (0.027)$\star$& 0.073 (-0.069, 0.211)$\star$\\
$\LFMR$& 0.582 (0.025) &-0.233 (-0.359, -0.101)$\star$\\
$\LHRMR$& 0.581 (0.025) &-0.233 (-0.364, -0.096)$\star$\\
HR& 0.423 (0.024)$\star$& 0.152 (0.034, 0.269)$\star$\\
\hline
\hline
  \multicolumn{3}{c}{Spontaneous Breathing}\\
\hline
  & $\PK$ (SE) & Spearman rank correlation (95\% CI) \\
\hline
BIS index &0.839 (0.014)$\star$ &-0.836 (-0.881, -0.778)$\star$\\
NRR &0.736 (0.019) & -0.619 (-0.688, -0.536)\\
$\HFMR$ & 0.489 (0.023)$\star$ & 0.001 (-0.117, 0.113)$\star$\\
$\LFMR$ & 0.669 (0.022)$\star$ & -0.416 (-0.506, -0.317)$\star$\\
$\LHRMR$ & 0.666 (0.022)$\star$ & -0.420 (-0.514, -0.319)$\star$\\
HR & 0.532 (0.022)$\star$ & 0.027 (-0.075, 0.133)$\star$\\
\hline
\end{tabular}
\label{table:corr}
\end{table}

The correlation with the estimated effect-site sevoflurane concentration is evaluated for the spontaneous breathing (SB) period (table \ref{table:corr}). The data number of SB is $5810$ ($23240$ s) and the average duration for each patient is  $750\pm322$ s. The overall indices ranking of the $\PK$ analysis and Speareman rank correlation ($R$) are consistent. While BIS correlates with effect-site sevoflurane concentration best (SB: $\PK =0.839$, $R=-0.836$, $p<0.0001$), the NRR index is second ($\PK =0.736$, $R=-0.619$, $p<0.0001$). The $\LFMR$ is the best HRV indices ( $\PK =0.669$, $r=-0.416$, $p<0.0001$). Compared to $\LFMR$, the NRR index is significantly better ( $p<0.01$). BIS surpasses NRR significantly in SB ($p<0.001$). The NRR index is significantly better than HR ($p <0.0001$).

\section{Quantitative Results of Noxious Stimulation}
When observing the serial $\PK$ analysis for skin incision (Fig. {\ref{fig:skinincision}}), tvLF reaches the maximum amount ($\PK >0.95$) 30 seconds after skin incision. This shows that the tvLF is the best index of the skin incision, followed by the heart rate ($\PK <0.85$). Furthermore, tvLF is significantly better than BIS ($\PK <0.55$) and the NRR index ($\PK <0.65$). 

In the serial $\PK$ analysis for endotracheal intubation (Fig. {\ref{fig:spkint}}), I compared different indices calculated from \textit{multitaper reassignment spectrogram} (HF, LF, LHR) with the indices calculated from \textit{STFT spectrogram} (HF$_\text{STFT}$ , LF$_\text{STFT}$ , LHR$_\text{STFT}$ ). This analysis signifies that the LF power and LHR provide the best results for the noxious stimulation of endotracheal intubation. Meanwhile, the heart rate is represented the second best and the BIS index is represented the worst. The indices that are calculated from \textit{multitaper reassignment spectrogram} performs better than the corresponding indicators that are calculated from the \textit{STFT spectrogram}.

\section{Summary}
Several features on TF plane can be easily seen conveniently via the tvPS, including the ``rhythmic-to-non-rhythmic phenomenon", ``multiple component phenomenon", and ``LF surge phenomenon". The ``rhythmic-to-non-rhythmic phenomenon" and ``LF surge phenomenon" are apparent in the majority of my study cases. This universal consistency foreshadows the subsequent positive statistical results.

The quantitative analysis proves that the NRR index indicates the first motor movement reaction better than other clinical indices\cite{Lin_Wu_Tsao_Yien_Hseu:2013}. BIS index is the best indiator for LOC and ROC.  NRR index correlates with the concentration of sevoflurane. Furthermore, LF power indicates noxious stimulation best, and better than the heart rate. It is also shown the statistical benefit of the multitaper reassignment spectrogram over the classical STFT spectrogram.


\chapter{Implications to Anesthesiology and Physiology} 

\label{Chapter7} 

\lhead{Chapter 7. \emph{Implications}} 

\section{Main Findings}
There are two important clinical results in regards to the NRR index. First, the NRR index is able to predict first reaction during emergency period. Second, the NRR index correlates with the sevoflurance concentration during the patient's spontaneous breathing. Even so, the NRR index is not the best predictor for LOC and ROC. These findings suggest that there is a distinct physiological interpretation of the anesthetic depth level when compared to the hypnosis measured by surface EEG. Furthermore, tvLF correlates well with skin incision and endotracheal intubation. This momentary information is currently unavailable to clinical anesthesiologists as both the IHR and tvPS are not implemented in current monitor.

\section{Genuineness of NRR index}
The NRR index generally has a sharp increase nearly before the onset of the patient's first reaction. This then makes it highly unlikely that the motion artifacts interfere the NRR index. Furthermore, the observed NRR index performance in not confounded by the ventilation mode change. In particular, aside from the LOC, there are no transitions that occur from the mechanical ventilation to the spontaneous breathing in the data we have obtained. Also, there is no positive pressure ventilation for all the first reaction data. Furthermore, past literature on the association between "rhythmic-to-non-rhythmic" transition of the oscillatory pattern in RRI index and the different level of anesthesia has been mentioned but no quantitative works have been done. By using the classical power spectral analysis, Kato et al. has described that the high frequency component of RRI disperses when the patient awakens and concentrates when the patient is under anesthetics and controlled ventilation\cite{kato1992spectral}. Based on the above facts, I am more than confident to state that the NRR is convincing in its distinct and novel information that it is capable of providing us.

\section{Existing Findings}


The results we have obtained from the BIS (Fig.\ref{fig:spkloc}, Table \ref{table:corr}) are in agreement with previous studies showing that BIS is capable of monitoring awake status vs. anesthesia, the first reaction and the decrease of anesthetic gas concentration\cite{vakkuri2004time,soehle2008comparison,schmidt2004comparative}. Our BIS results (Fig.\ref{fig:spksi}, Fig.\ref{fig:spkint}) also agree with past studies that provide evidience that BIS or other EEG-derived indices cannot indicate the response to noxious stimulation\cite{storm2013nociceptive,velly2007differential,ekman2004comparison}.

The results from our analysis (Table \ref{table:represent}) are also consistent with previous reported association between anesthetic depth and traditional HRV parameters: HRV is decreased during general anesthesia and increased during recovery\cite{kato1992spectral,huang1997time,blues1998respiratory,galletly1994effect}. Lastly, our results are also consistent with the reported finding that combining classical HRV indices with BIS does not outperform BIS in discriminating awake from asleep during anesthetic induction\cite{sleigh1999comparison}.

It has been reported that classical HRV parameters do not outperform heart rate when predicting noxious stimulation\cite{seitsonen2005eeg,luginbuhl2007heart}. Contrarily, our results demonstrate that tvLF predicts skin incision better than HR. It seems that the noxious stimuli of the skin incision elicited a momentary increase of sympathetic activity, which leads to this finding. Although the study is not designed for noxious stimulation, the finding that tvLF correlates with skin incision shows the technical advantage of our approach and its potential as an index for noxious stimulation.

\section{Physiologic Interpretation of NRR}

The association between NRR and anesthetic depth can be partially explained by the cardiopulmonary coupling\cite{eckberg2009point} and the respiratory physiology. When the coupling effect is higher, the respiration pattern is reflected in the RRI index. The respiratory mechanism also contributes to the differential influence that occur from anesthetics. The neural respiratory control comprises of two systems, the involuntary automatic control system and the voluntary control system. The involuntary automatic control system is mainly controlled by the respiratory center in the pontomedullary area. The voluntary control system is mainly controlled by the forebrain\cite{mitchell1975neural}. These two systems are also distinct in their neural anatomy and their functions. Respiratory efferent signals from these two systems compete with each other and are integrated at the spinal level to control the respiratory motor neuron. However, during spontaneous respiration, studies have also shown that the involuntary automatic respiratory pattern generated in the pontomedullary center\cite{mitchell1975neural}, specifically the preB\"otzinger complex\cite{elsen2005postnatal,rekling1998prebotzinger,ramirez1998hypoxic}, is rhythmic.

 On the other hand, the breathing pattern of the voluntary respiratory motor control is non-rhythmic and involves the cortical processing and thalamic integration which responds to the peripheral inputs and the descending inputs\cite{mitchell1975neural,mcfarland2002thalamic,ramsay1993regional,mckay2003neural}. The thalamus also actively participates in different motor functions, including respiration\cite{ramsay1993regional,mckay2003neural}, speech\cite{murphy1997cerebral} and cough\cite{simonyan2007functional} to name a few\cite{tanaka2011contribution}. Our NRR index findings suggest that the non-rhythmic respiratory activity involves the nearly entire cerebral regions is more susceptible to sevoflurane than the rhythmic respiration generated in the medulla. This suggestion is supported by literature. Guedel made the classical observation on respiratory pattern\cite{Guedel:37}. Bimer et al denoted that the respiratory irregularities accompany electroencephalographic arousal reaction in spontaneous breathing\cite{bimar1977arousal}. Also, Studies have also shown that the respiratory activity of preB\"otzinger complex is less depressed by sevoflurane\cite{kuribayashi2008neural,takita2010effects}, and the effect of volatile anesthetics is less prominent to the brainstem compared to the cortical regions that are more abundant in synaptic transmissions.\cite{takita2010effects}.

The ability for the NRR index to predict first reaction suggests a connection between the ``rhythmic-to-non-rhythmic" phenomenon and motor reactions. Antognini et al. demonstrated that immobility to surgical stimulation by volatile anesthetics in goats is mainly modulated by the spinal cord, which also indirectly affects the thalamic response to noxious stimulation\cite{antognini1993exaggerated,antognini2000isoflurane}. Another study by Velly et al. also presented significant results\cite{velly2007differential}. He recorded the human subcortical electrophysiological activity and demonstrated that subcortical activity (possible the thalamic activicy) predicts suppression of movement to noxious stimuli, but not changes of consciousness, whereas cortical EEG predicts loss of consciousness, but not motor suppression. Based on the above, NRR might reflect the subcortical activity since it better reveals information about motor reaction, but not consciousness, compared with BIS. Since immobility to noxious stimulation is similar to, although not the same as the first reaction under surgical wound pain in the present study, this relationship suggests the possible role of the thalamus in ``rhythmic-to-non-rhythmic" phenomenon. Besides the NRR index, BIS also responded to first reaction in our serial $\PK$ analysis, which is in agreement with other studies\cite{schmidt2004comparative}. Therefore, it is possible that a cortical mechanism is also related to the first reaction.

From these results, it is possible to see that sevoflurane may affect non-rhythmic respiration level of the spinal cord, subcortical supraspinal regions, or the cortex in the human brain. Although we are currently unable to pinpoint the exact anatomic location, it is likely that the subcortial supraspinal area, possibly the thalamus, is the most plausible region. The above statement is sound as the NRR index generates different results in the serial $\PK$ analysis and is also closely connected to the human body's respiration responses. Thus, we hypothesize that the ``rhythmic-to-non-rhythmic" phenomenon in IHR exhibits the central respiratory activity via either central or peripheral mechanisms\cite{eckberg2009point}. The IHR also reflects the final integration of this "rhythmic-to-non-rhythmic" respiratory activity. Although more evidence is necessary to further clarify these hypotheses, we propose to use the NRR quantification methodology as a potential tool to evaluate the depth of anesthesia that differs from EEG-based monitoring.

\section{Clinical Application of the NRR Index}
The first strength of this study is a new quantitative approach, referred to as the NRR index, to analyze ``rhythmic-to-non-rhythmic" phenomenon. The NRR index has the potential to reflect different levels of anesthesia. Literature partially supports that there is an underlying physiological mechanism. Second, the signal processing technique, multitaper Synchrosqueezing transform, has been well studied in the literature, so we have adequate theoretical support to resolve the potential limitations of the PS. The momentary dynamics in IHR can be captured by the tvPS, which leads to the NRR index.

%

\section{Potential Index in Noxious Stimulation}

Although we did not design the study for noxious stimulation, from the data analysis result, we found that the indices we apply are potential in detecting pain reaction under anesthesia. In particular, the $tvLF$ index correlates with the noxious stimulation better than heart rate as an additional finding. More study will need to be conducted on the tvLF to provide sound results. It can also be helpful to compare this data with other well known indices like the Surgical Stress Index\textsuperscript{\textregistered} (SSI, GE healthcare) and Analgesia Nociception Index\textsuperscript{\textregistered} (ANI, Metrodoloris), which have been used on monitoring surgical stress {\cite{wennervirta2008surgical}}.
SSI is derived from the pulse wave amplitude and pulse beat interval of the photoplethysmography whereas ANI indicates the parasympathetic tone derived from ECG signal. 
It is possible that various indices can provide complementary information regarding noxious stimulation that is gathered from different modalities and concepts. 

NRR index measures the \textit{rhythmic-to-non-rhythmic} transition and quantifies a new kind of information different from the above two instruments in both physiologic and mathematical senses. Further investigations may be necessary to understand the clinical value and application of tvPS.

\section{Summary}
In conclusion, we are able to extract hidden information regarding the anesthetic levels from the routine ECG monitor. The quantitative results of the NRR index supports our initial hypothesis that the ``rhythmic-to-non-rhythmic" transition correlates with motor reactions during emergence period earlier than BIS index, and correlates with sevoflurance concentration. In addition, the potential of the time-varying HRV indices provides observations of the relationship between $\LFMR$ index and the noxious stimulation. The notion of these dynamics is rooted in the recently developed signal processing technique. Without tvPS, the above quantification cannot be easily achieved. Overall, my clinical study suggests that the ECG signal contains complementary information to the EEG-based depth-of-anesthesia index.
 

\chapter{Real-time Processing Using the Blending Operator} 

\label{Chapter8} 

\lhead{Chapter 8. \emph{Real-time Processing}} 


\section{Real-time Processing}
From the previous chapters, my study has concluded that the IHR signal during anesthesia contains a wealth of information when viewing with $tvPS$. It is possible to derive the IHR signal from ECG, the waveform of pulse oximetry, and also the invasive intra-arterial pressure waveform. Although we mainly use ECG signal in the present project, all possible sources of IHR are mandatory or frequently used functions in modern standard anesthesia monitors. A practical question is how do we apply the NRR index proposed in the present study to individual patients. 

The processing steps of NRR index starts from the identification of each heart beat from ECG signal. Next, an interpolation is used to convert the \textit{irregularly-sampled} heart beats data into \textit{regularly-sampled} IHR data. Next, the calculation of the $tvPS$ is based on the multitaper Synchrosqueezing tranform. And lastly the NRR index is computed based on the $tvPS$. Thus, the positive statistical results in previous chapter, are all obtained from \textit{off-line calculations}.

In anesthesia practices, the main purpose of anesthesia monitor is to provide real time monitoring of the patient's responses in order to provide timely responses. This purpose then turns ordinary measurements like heart rate, blood oxygen saturation, and blood pressure into vital information during anesthesia. The real-time processing of NRR index is necessary if we would like to ``steer" the anesthetic depth of patients based on their instantaneous information from IHR. The potential benefits include adequate dose of anesthetics, possibility to divert  critical situation, reduce stress response, and improve long-term welfare.

In this chapter, my main focus is the real-time processing from the irregular heart beat sample to the $tvPS$. The goal is to interpolate the irregular data samples in real time. Meanwhile, the polynomials property of the interpolation also serves as the vanishing-moment and minimum-supported wavelets.

\section{Technique Obstacles for Real-time Processing}
Due to the irregular interval of each heartbeat, the heartbeat data is irregularly sampled and this irregularity is unavoidable. Almost all continuous information are digitized in equally-spaced samples in time or in space before further analysis for convenience, and for efficiency. However, some situations provide non-uniform sampling and could be either unavoidable or intended\cite{Gunnarsson_Gustafsson_Gunnarsson:2004,Vio_Strohmer_Wamsteker:2000}. In order to derive the continuous instantaneous heart rate from the distinct heartbeats (Fig.\ref{fig:RRI}) is an example  of the rich IHR information that entirely relies on the irregularity of the heartbeats. Hence, the initial challenge is the real-time processing of the non-uniform data samples.

When analyzing data the scope of continuous wavelet transform (CWT), the theoretical background of Synchrosqueezing transform presents the optimal performance when using compactly supported windows in frequency domain\cite{Daubechies_Lu_Wu:2011}. That means we need whole data converted in frequency domain for subsequent calculation. This is inefficient and also highly unlikely for real-time data processing. Moreover, for real-time processing, an wavelet transform (or STFT) better has a short, and compact support window in time. This type of window may hamper the benefit of wavelet transform: an adequate vanishing moment can eliminate trend component in terms of corresponding polynomial order to preserve the oscillatory components we desire more efficiently.

A possible workaround is by cutting down a segment of data, windowing it, and then choosing a suboptimal wavelet function empirically. Next, we would observe the effect of suboptimal window and then adjust the parameters empirically until we finally accept the suboptimal effect in real-time display.

There is a better solution to consider before we proceed with the above workaround. The solution is a special spline interpolation, which has the interpolating ability to handle the unequally-spaced heart beat sample, and the polynomial property to match the vanishing moment property of wavelet transform. The resulting time-scale analysis is compactly supported in time and provides better performance due to the vanishing moment property. Based on this scheme, the real-time tvPS can help remove the above mentioned technical obstacles. 

\section{B-spline}

 We start from a brief introduction of the spline function. For digital samples $g(t_n)$ are sampled at time points
\begin{equation}\label{nonuniform:definition:t_knot}
\ut:\, a=t_0<t_1<t_2<\ldots,
\end{equation}
where $\{t_k\}$ may be irregular, or non-uniform signal, which means that we allow $t_{k+1}-t_k\neq t_{j+1}-t_j$ for $k\neq j$. Let $m\geq 3$ be any desired integer and $\Pi_{m-1}$ denote the space of all polynomials of degree less than $m$. 

The spline space is then defined as: 
\begin{equation}\label{nonuniform:definition:s_knot}
\us:\,s_{-m+1}=s_{-m+2}=\ldots=a=s_0<s_1<s_2<\ldots,
\end{equation}
and the notation $S_{\us,m}:=S_{\us,m}[a,\infty)$ denotes the spline space of order $m$ with the knot sequence $\us$.

To formulate a locally supported basis of $S_{\us,m}$, we now introduce the {\it truncated powers} 
\[
x_+^{m-1}:=(\max\{0,x\})^{m-1}.
\]
The truncated power functions $(s_k-t)^{m-1}_+$, $k=0,1,\ldots$, are in $S_{\us,m}$. Since $(x_k-t)_+^{m-1}$, $k\geq 1$, have global support, we apply the $m$-th order divided differences to form locally supported functions. The {\it divided differences} are
\[
[\,u,\ldots,u\,]f:=\frac{f^{(l)}(u)}{l!}
\]
if there are $l+1$ entries in $[u,\ldots,u]$, and 
\[
[\,u_0,\ldots,u_n\,]f:=\frac{[\,u_1,\ldots,u_n\,]f-[\,u_0,\ldots,u_{n-1}\,]f}{u_n-u_0}
\]
if $u_0\leq u_1\leq\ldots\leq u_n$ and $u_n>u_0$.

Now we consider the knot sequence $\us$ of the spline space $S_{\us,m}$. The {\it normalized B-splines}\cite{deBoor:1978} are the divided difference of the truncated power $(s_k-t)^{m-1}_+$, namely,
\begin{equation}\label{definition:Nmktuniform}
N_{m,k}(t)=N_{\us,m,k}(t)=(s_{m+k}-s_k)[\,s_k,\ldots,s_{m+k}\,](\cdot - t)_+^{m-1},
\end{equation}
for $k=-m+1,\ldots,0,1,2,\ldots$.

We would apply the recursive formula to implement $N_{m,k}$  \cite[page 143 (6.6.12a)]{Chui:1997}. The B-splines are a locally supported.


However, if we choose the irregularly-spaced sequence $\ut$ of time positions in (\ref{nonuniform:definition:t_knot}) as the knot sequences $\us$ in (\ref{nonuniform:definition:s_knot}) by attaching $t_{-m+1}=\ldots=t_0=a$ to $\ut$ as in (\ref{nonuniform:definition:s_knot}), it can be difficult to compute, or the computation cost will be higher for the non-uniform $\{t_j\}$ and large values of $n$.
\section{Quasi-interpolation}
On the other hand, de Boor and Fix \cite{deBoor_Fix:1973} proposed a spline representer $f(t)$, which approximates the data instead of providing exact interpolating the target data function $g(t)$ at $t=t_j$, $j=0,1,\ldots,n$ (that is, if $f(t_j)\neq g(t_j)$ is allowed). This \textit{quasi-interpolation} has a ``polynomial preservation" property:  $g(t)=p(t)\in \Pi_{m-1}$.

This polynomial preservation is important to improve the performance of the continuous wavelet transform (CWT), and hence the Synchrosqueezing transform (SST), to reveal the oscillatory components. Specifically, if the analysis wavelet is compactly supported by the spline of order $m$, then the $m$-th order vanishing moment annihilates the $m$-th order Taylor polynomial expansion of $g(t)$. This then facilitates the trend removal of the the signal $g(t)$ as a polynomial. However, the quasi-interpolation scheme de Boor proposed \cite{deBoor_Fix:1973,deBoor:1978} requires derivative data values of $g(t)$. 
 Others \cite{Lyche_Schumake:1975,Schumaker:1981} proposed to replace the derivatives of $g(t)$ by divided differences of $\{g(t_i)\}$. However, these quasi-interpolations\cite{deVilliers:2012} do not address the real-time issue at stake. In addition, quasi-interpolation introduced the error $g(t_i)-f(t_i)$ for $i=0,1,\ldots,n$.
 
Chui et al. proposed a real-time quasi-interpolation\cite{Chen_Chui_Lai:1988}. Chui and his college also introduced a ``local interpolation"\cite{Chui_Diamond:1991} to correct the error of ``quasi-interpolation".
 

The real-time quasi-interpolant \cite{Chen_Chui_Lai:1988} is introduced as follow: for each $k=0,1,\ldots$, consider the Vandermonde determinant
\[
D(t_k,\ldots,t_{k+m+1}):=\mbox{det}\left[
\begin{array}{cccc}
1 & 1 & \ldots & 1\\
t_k & t_{k+1} & \ldots & t_{k+m-1}\\
\vdots & \vdots & \vdots &\vdots \\
t^{m-1}_k & t^{m-1}_{k+1} & \ldots & t^{m-1}_{k+m-1}
\end{array}
\right],
\]
and the determinant $D(t_k,\ldots, t_{k+j-1},\xi_j,t_{k+j+1},\ldots,t_k+m-1)$ obtained by replacing the $(j+1)$-st column in the definition of $D(t_k,\ldots,t_{k+m+1})$ by the column vector
\[
\xi_j:=[\,\xi^0(j,m),\ldots,\xi^{m-1}(j,m)\,]^T,
\]
where $\xi^0(j,m)=1$ and 
\[
\xi^i(j,m)=\frac{\sigma^i(t_{j+1},\ldots,t_{j+m-1})}{\left(\begin{array}{c}m-1\\ i\end{array}\right)},
\]
for $i=1,\ldots,m-1$ with $\sigma^i(r_1,\ldots,r_{m-1})$ being the classical symmetric functions defined by $\sigma^0(r_1,\ldots,r_{m-1})=1$ and for $i=1,\ldots,m-1$,
\[
\sigma^i(r_1,\ldots,r_{m-1})=\sum_{1\leq l_1<\ldots<l_i\leq m-1} r_{l_1}\ldots r_{l_i}.
\]

Using the determinants introduced above, we apply the below spline coefficients
\begin{equation*}
a_{k,l}:=\frac{D(t_k,\ldots, t_{k+j-1},\xi_j,t_{k+j+1},\ldots,t_{k+m-1})}{D(t_k,\ldots,t_{k+m-1})} ,
\end{equation*}
to formulate the compactly supported spline function
\begin{equation*}
M_{\ut,m,k}(t):=\sum_{l=m-1}^{2m-2}a_{k,l-m+1}N_{\ut,m,k+l-m+1}(t) ,
\end{equation*}
with $\mbox{supp}M_{\ut,m,k}=[t_{k-m+1},t_{k+m}]$. These basis functions provide a real-time implementation of the {\it quasi-interpolation operator}
\begin{equation}\label{nonuniform:Q_m:definition}
(\mathsf{Q}_mg)(t)=\sum_{k}g(t_k)M_{\ut,m,k}(t).
\end{equation}

The quasi-interpolation operator $\mathsf{Q}_m$  (\ref{nonuniform:Q_m:definition}) possesses the polynomial preservation property.

For any $m\geq 1$, 
\[
(\mathsf{Q}_mp)(t)=p(t) ,
\]
for all $t\geq a$ and for all $p\in\Pi_{m-1}$, provided that the summation (\ref{nonuniform:Q_m:definition}) account for all non-negative integers $k=0,1,\ldots$. Furthermore, in view of the support of $M_{\ut,m,k}$, it follows that
\[
\sum_{k=v-m+1}^{v+m-2}p(t_k)M_{\ut,m,k}(t)=p(t),\quad t\in[t_u,t_v]
\]
for all $p\in \Pi_{m-1}$. 

This local polynomial preservation property allows the CWT, and hence the SST, to annihilate the $(m-1)$-th degree Taylor polynomial approximation of the signal at $t_j$, where $u<j<v$.

\section{Local Interpolation Operator}

The local interpolation operator, denoted by $\mathsf{R}_m$, satisfies the interpolation property. To define $\mathsf{R}_m$, we will insert knots to $\ut$ by considering a new knot sequence $\us\supset\ut$  (\ref{nonuniform:definition:s_knot}). For simplicity, we will introduce even order variable $m$ :

For even $m\geq 4$, we set
\[
s_{mk/2}=t_k,\quad k=0,1,2,\ldots.
\]
That is, we insert $(m/2-1)$ knots in between two consecutive knots in $\ut$. For convenience, we will choose the new knots to be equally spaced between every pair of two consecutive knots.

Fix even $m\geq 4$. Let $N_{\us,m,j}$ be the $m$-th order B-spline with knot sequence $\us$. Then the {\it completely local spline basis function} can be defined by
\begin{equation*}
L_{\us,m,j}(t):=\frac{N_{\us,m,m(j-1)/2}(t)}{N_{\us,m,m(j-1)/2}(t_j)}.
\end{equation*}

Since $t_j=s_{mj/2}$ is the ``centered'' knot and
\[
\mbox{supp}N_{\us,m,m(j-1)/2}=[\,s_{m(j-1)/2},s_{m(j+1)/2}\,]=[\,t_{j-1},t_{j+1}\,],
\]
we have
\begin{equation}\label{nonuniform:property:Lmsj}
\left\{
\begin{array}{l}
\displaystyle L_{\us,m,j}(t_j)=1\\ \\
\displaystyle\mbox{supp}L_{\us,m,j}=[\,t_{j-1},t_{j+1}\,].
\end{array}
\right.
\end{equation}
It is clear that $L_{\us,m,j}(t_k)=\delta_{j-k}$, where the Kronecker delta notation is used (\ref{nonuniform:property:Lmsj}). 
\vspace{0.4cm}

The above preparation provides a real-time implementation of the {\it local interpolation operator}, which satisfies the interpolation property (\ref{nonuniform:property:Lmsj}).

Fix $m\geq3$. For a given function $g\in C(\RR)$, the local interpolation operator $\mathsf{R}_m$ is defined as
\begin{equation}\label{nonuniform:R_m:definition}
(\mathsf{R}_mg)(t):=\sum_k g(t_k) L_{\us,m,k}(t).
\end{equation}

\section{Blending Operator}
We are now ready to apply (\ref{nonuniform:Q_m:definition}) to obtain the blending operator, denoted by $\mathsf{R}_m\oplus \mathsf{Q}_m$.  

Fix $m\geq 3$ and $g\in C(\RR)$. The \textit{blending operator} is defined as $\mathsf{P}_m:=\mathsf{R}_m\oplus \mathsf{Q}_m$, where
\begin{equation*}
\mathsf{R}_m\oplus \mathsf{Q}_m:=\mathsf{Q}_m+\mathsf{R}_m(\mathsf{I}-\mathsf{Q}_m)=\mathsf{Q}_m+\mathsf{R}_m-\mathsf{R}_m\mathsf{Q}_m,
\end{equation*}
and $\mathsf{I}$ is the identity operator. In particular, we have 
\begin{equation}\label{nonuniform:definition:Pm}
(\mathsf{P}_mg)(t):=\sum_{k}g(t_k)M_{\ut,m,k}(t)+\sum_k\big[g(t_k)-\sum_j g(t_j)M_{\ut,m,j}(t_k)\big]L_{\us,m,k}(t).
\end{equation}

We remark that in the definition of $\mathsf{P}_m$, the two operators $\mathsf{R}_m$ and $\mathsf{Q}_m$ are not commutative. Let us summarize the two key properties of the blending operator in the following theorem.
\begin{enumerate}
\item  The blending operator $\mathsf{P}_m$ possesses both the polynomial preservation property of $\mathsf{Q}_m$ and the interpolatory property of $\mathsf{R}_m$.
\item The error of spline interpolation by the blending operator is both small and bounded. 
\end{enumerate}

In conclusion, the blending operator, as a local spline interpolation operator, achieves the optimal interpolation error rate when compared with the traditional spline interpolation operator. In addition, the error depends only on the local data profile, which allows real-time implementation with optimal error rate.

\begin{algorithm}[h!]
  \begin{algorithmic}
\STATE First, pre-compute the B-spline values $N_{\us,m,l}(t_j)=:n_{l,j}$. Then for each $k$, since $M_{\ut,m,k}\in S_{\us,m}$, there exists a finite sequence $\{b_{k,l}\}$ in the formulation of  
\[
M_{\ut,m,k}(t)=\sum_{l}b_{k,l}N_{\us,m,l}(t).
\]
Also pre-compute $d_{k,j}=M_{\ut,m,k}(t_j)$.

Now, while the data sequence $\{g(t_k)\}$ is acquired, compute
\[
\tilde{g}_l=\sum_k b_{k,l}g(t_k)
\]
and simultaneously compute
\[
g^*_l=\frac{g(t_l)-\sum_{j} d_{l,j}g(t_j)}{n_{m(l-1)/2},l},
\]
and then up-sample $\{g^*_l\}$ by $m(l-1)/2$; that is, set
\[
g^{\#}_{m(k-1)/2}=g^*_k,\quad\mbox{and }g^{\#}_l=0\mbox .
\]
Then, we have an on-line computational scheme for the quasi-interpolation spline interpolation:
\begin{equation*}
f(t)=\sum_{l=-m+1}^n(\tilde{g}_l-g^{\#}_l)N_{\us,m,l}(t)
\end{equation*}
for increasing number of samples from $g(t_n)$ to $g(t_{n+1})$, $\ldots$; and this can be implemented for real-time D/A conversion. 
\end{algorithmic}
\caption{Real-time implementation of the blending operator}
\label{alg:Blending}
\end{algorithm}

\section{Real-time tvPS}
We provide a real-time computational algorithm to compute $f(t)$ for the upcoming data samples $g(t_0),g(t_1),\ldots$ etc. For convenience, we only consider even order $m$. The formulation for the odd order is similar but slightly more complicated.

%


The blending operator leads to the real-time spline interpolation. For real-time issue, the wavelets have to be minimum supported in time. We can design the wavelets with the same number of vanishing moments as the spline. Meanwhile the wavelets have minimum support and maximum order of vanishing moments.

Let $m,n\geq 1$ be arbitrary integers, and $\ux$ an arbitrary knot sequence. The spline basis functions is then as follows:
\begin{equation}\label{definition:psi_uxmnx}
\psi_{\ux,m;n,k}(x):=N^{(n)}_{\ux,m+n,k}(x),\quad k\in\ZZ,
\end{equation}
to be called {\it VM wavelets on $\ux$}, where $N_{\ux,m+n,k}(x)$ is defined in (\ref{definition:Nmktuniform}), which satisfies the moment conditions
\begin{equation}
\left\{\begin{array}{l}
\displaystyle\int_{-\infty}^\infty x^l\psi_{\ux,m;n,k}(x)\ud x=0,\quad l=0,\ldots,n-1\\ \\
\displaystyle\int_{-\infty}^\infty x^n\psi_{\ux,m;n,k}(x)\ud x\neq 0.
\end{array}
\right.
\end{equation}

That means the arbitrary chosen $n$ in the interpolator leads to the vanishing moment order of the wavelets if we use this spline basis as the wavelet function. The wavelet transform can successfully eliminate the $n$ order polynomial as a trend component.

Let $m,n\geq 1$ be arbitrary integers. Then the derivative of the wavelet $\psi_{\ux,m;n,k}$ on an arbitrary knot sequence $\ux$, is given by
\begin{equation}\label{formula:psi_uxmnxDerivative}
\psi'_{\ux,m;n,k}(x)=\psi_{\ux,m-1;n+1,k}(x).
\end{equation}

This property permits us to conveniently calculate the derivative of the wavelet, which is required to calculate the frequency reassignment in Synchrosqueezing transform.(see equation {\ref{freassign}})

Then, we can proceed to the Synchrosqueezed CWT to obtain the real-time time-frequency analysis as real-time tvPS. Using this tvPS, we can apply the research in previous chapters to quantify the ``rhythmic-to-non-rhythmic" phenomenon in real-time. Using the clinical data to ``emulate" the real-time calculation of NRR index, we obtained a comparable performance of the correlation with sevoflurane concentration during spontaneous respiration: The $\PK$ value of real-time NRR index is $0.711\pm 0.021$ ($\PK$ of off-line NRR index:$0.732 \pm 0.018$)(\ref{table:corr}).

\chapter{Conclusion} 

\label{Chapter9} 

\lhead{Chapter 9. \emph{Conclusion}} 

\section{Research Findings}
Thanks to the advancement of science in anesthesiology and neural physiology, I can propose a theory and modeling to investigate the ``rhythmic-to-non-rhythmic" phenomenon. Furthermore, because of the advancement of signal processing technique, I am able to use multitaper Synchrosqueezing technique to compute the NRR index. Anesthesia creates dynamic changes to the human body, and due to the time-varying ability and good TF resolution, the NRR index is a particularly suitable tool reflecting the constant changes during anesthesia\cite{Lin_Hseu_Yien_Tsao:2011,Auger_Flandrin_Lin_McLaughlin_Meignen_Oberlin_Wu:2013}.}

The numeric result from clinical database is positive \cite{Lin_Wu_Tsao_Yien_Hseu:2013}. NRR index performs better and faster than BIS index when detecting motor movement reaction. This also supports the fact that different brain regions mediating various functions, with differential susceptibility to anesthetics may require multi-modal assessment of these anesthetic effects. NRR index is therefore able to provide additional value. In addition, LF index provides good results as an indicator for noxious stimulation, for both skin incision and endotracheal intubation. ECG waveform is standard information in current anesthesia, while the IHR is not. My study shows that these new indices derived from ECG waveform can provide additional information that could be beneficial to patients undergoing anesthesia in the future.

\begin{figure}[htbp]
\centering
\includegraphics[width=1\textwidth]{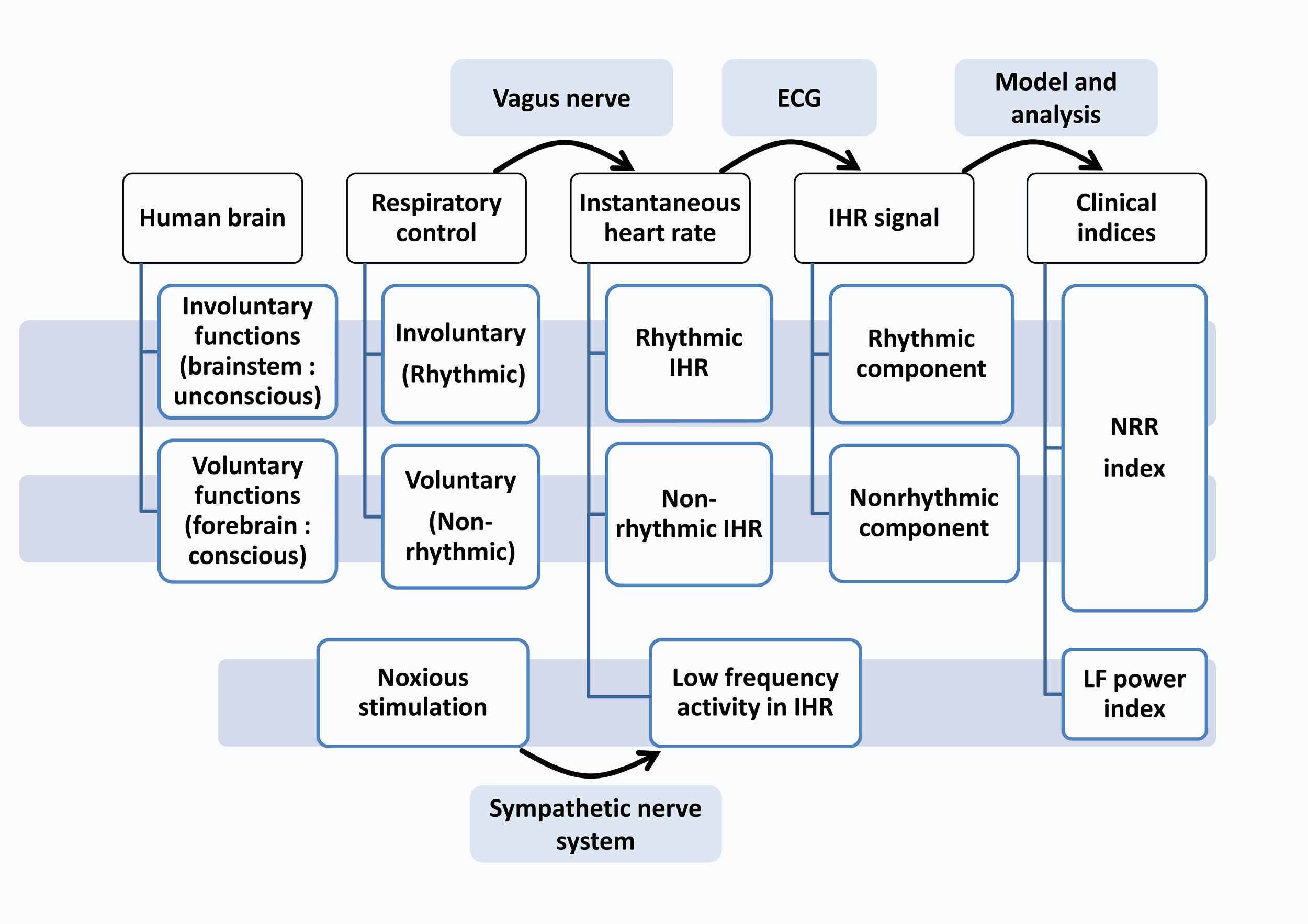}
\caption[Diagram of ``rhythmic-to-non-rhythmic phenomenon"]{The diagram represents the overall framework from the human brain to signal analysis.}
\label{fig:nrrdiag}
\end{figure}

\section{Accomplishments}

\begin{enumerate}
\item The \textit{adaptive harmonic model} is a method of modeling the rhythmicity of the ``rhythmic-to-non-rhythmic" phenomenon
\item A time-varying power spectrum that is based on multitapering Synchrosqueezed spectrogram.
\item Methodology of the NRR index
\item A methodology of serial $\PK$ analysis was developed to evaluate the dynamic performance of anesthetic depth index as \textit{time-varying} $\PK$ value.
\item Positive clinical value of the NRR index was obtained from a clinical database
\item Improved performance of classical HRV parameters was obtained using $tvPS$
\item A real-time scheme that provides real-time $tvPS$ and real-time NRR index.
\end{enumerate}
 
\section{Future Directions}
The results of my project indicates several future research possibilities.

In clinical anesthesiology, more studies should be conducted to answer questions regarding the value of NRR index in different kinds of anesthetics (i.e. propofol, ketamine, dexmedetomidine or xenon), in different types of patient (i.e. children, lying-in women, elderly patient, or patient with severe illness). Since the NRR index may indicate the activity of subcortical areas, this poses questions on if the application of the NRR index can lead to less stress response and better long-term outcomes, and if it is possible, how can we obtain those results. On the other hand, the IHR is derived from the ECG signal, I wonder whether it is possible to calculate the NRR index from photoplethsmography or intra-arterial pressure waveform. It also needs more clinical study to better understand the clinical role of NRR index and the $tvLF$ index.

It is possible that a thorough understanding of the basic mechanism can lead to better clinical benefit. The origin of ``rhythmic-to-non-rhythmic" phenomenon should be in the brain. However, the neural mechanism has not been addressed according to my best knowledge. For example, where exactly is the anatomical location that creates this phenomenon? What could cause the \textit{multiple component} phenomenon as previously mentioned? Lastly, how do the neurons coordinate with one another during this phenomenon? It is foreseeable that the answers of these questions can directly contribute to the brain science. Thus, the neurophysiological study based on the animal model is a potential future research.

The signal processing technique directly contributes to the performance of these clinical indices. The current result of tvPS is helpful in revealing various features in the TF plane, but there are still a lot of room for improvements. The time-frequency analysis is based mostly on the \textit{Fourier transform}. I am interested in whether it is possible to discard the idea of \textit{frequency} in order to have a better understanding of the signal's features. The real-time calculation and display of the clinical indices provides benefit to individual patients. Technique developments in real-time processing is imperative.

\section{Summary}

My study focuses on the qualifications of the ``rhythmic-to-non-rhythmic" phenomenon. I have verified that my proposed model and quantitative index through using clinical database. The positive results support my proposal on using the adaptive harmonic model and time-varying power spectrum based on the multitaper Synchrosqueezing transform. The NRR index also presents unique values in clinical anesthesia. The achievements in this thesis fill a gap in the fields of medicine and provide a new perspective on many of the issues that could be understood in the future.



\addtocontents{toc}{\vspace{2em}} 

\appendix 



\addtocontents{toc}{\vspace{2em}} 

\backmatter


\label{Bibliography}

\lhead{\emph{Bibliography}} 

\bibliographystyle{unsrtnat} 

\bibliography{dspmath,nrrmedical}

\end{document}